\documentclass[aps, reprint, onecolumn, superscriptaddress,amsmath,amssymb,noeprint,nofootinbib]{revtex4-2}
\usepackage[T1]{fontenc}
\usepackage[utf8]{inputenc}
\usepackage{lmodern}
\expandafter\let\csname equation*\endcsname\relax
\expandafter\let\csname endequation*\endcsname\relax
\usepackage{amsthm,amsmath,amssymb,amsfonts,amscd,mathrsfs}
\allowdisplaybreaks
\usepackage{float}
\usepackage{graphicx}
\usepackage{xcolor}
\usepackage{natbib}
\usepackage{bm}
\usepackage{dsfont}
\usepackage{soul}
\usepackage{bbold}
\usepackage[mathscr]{euscript}
\usepackage[cal=boondoxo]{mathalfa}
\usepackage{comment}
\usepackage{enumerate}
\usepackage[all]{xy}
\usepackage{booktabs}
\usepackage{braket}
\usepackage{subcaption}
\usepackage{bbm}
\usepackage{pgfgantt}
\usepackage[bottom]{footmisc}
\usepackage{comment}
\usepackage{tikz-cd}
\usepackage{stmaryrd}
\usepackage[normalem]{ulem}
\usepackage[textwidth=17cm]{geometry}
\usepackage[counter,user,hyperref]{zref}
\usepackage{notes2bib}
\bibnotesetup{
note-name = ,
use-sort-key = false
}
\graphicspath{ {Figures/} }
\renewcommand\footnotemark{}
\makeatletter
\renewcommand{\footnoterule}{%
  \kern -3pt
  \hrule \@width .4\columnwidth
  \kern 5pt
}
\makeatother

\newcommand{\II}{\mathbb{I}}

\def\hge{\hat{\sigma}_{ge}}  
\def\heg{\hat{\sigma}_{eg}} 
\def\ii{{\rm i}} 
\def\ra{\hat{\rho}}
\usepackage[colorlinks = true,
            linkcolor = blue,
            urlcolor  = blue,
            citecolor = blue,
            anchorcolor = blue]{hyperref}

\usepackage{ragged2e} 
\makeatletter
\long\def\@makecaption#1#2{%
  \vskip\abovecaptionskip
  \noindent
  \begin{minipage}{\linewidth}
    \justifying
    \parindent=0pt
    {#1.} #2
  \end{minipage}
  \vskip\belowcaptionskip
}
\makeatother
\makeatletter
\AtBeginDocument{%
\@ifpackageloaded{hyperref}{%
}{
  
  \providecommand{\hyperlink}[2]{#2}
}
\zref@newlist{sectionprop}
\zref@newprop{partinfo}[1]{\thepart}
\zref@newprop{sectioninfo}[1]{\number\value{section}}
\newcommand{\zclabel}[1]{%
  \zref@labelbyprops{#1}{partinfo,sectioninfo,anchor,counter}%
}
\newcommand{\countercref}[1]{%
  \expandafter\csname cref@\zref@extract{#1}{counter}@name\endcsname\ \hyperlink{\zref@extract{#1}{anchor}}{\zref@extract{#1}{sectioninfo}}%
}
\newcommand{\Countercref}[1]{%
  \expandafter\csname Cref@\zref@extract{#1}{counter}@name\endcsname\ \hyperlink{\zref@extract{#1}{anchor}}{\zref@extract{#1}{sectioninfo}}%
}
\newcommand{\secref}[1]{%
  \zref@ifrefundefined{#1}{}{%
    \edef\@tmp@a{\zref@extract{#1}{partinfo}}%
    \edef\@tmp@b{\thepart}%
    \ifx\@tmp@b\@tmp@a\relax
    \Countercref{#1} in this \partname%
    \else
    \Countercref{#1} in \partname\ \zref@extract{#1}{partinfo}%
    \fi
  }%
}
}
\makeatother
\usepackage{hyperref}
\usepackage{cleveref}
\renewcommand{\thepart}{\Alph{part}}

\begin{document}
\begin{abstract}
Many-body quantum-optical systems, where a collection of emitters interacts through a common electromagnetic reservoir, exhibit rich out-of-equilibrium behavior and hold promise for applications in quantum technologies. However, exact numerical simulations of their dynamics quickly become unfeasible due to the exponential growth of the Hilbert space with system size. Semiclassical, phase-space approaches -- such as the Truncated Wigner approximation (TWA) -- provide computationally efficient alternatives by capturing leading-order quantum fluctuations. In this paper, we present a comprehensive overview of how to tackle problems in many-body quantum optics using phase-space methods. We derive the exact partial differential equation governing many-body dissipative evolution in any phase-space representation and discuss the approximations that yield the dissipative TWA proposed by Mink and Fleischhauer~\cite{Mink23}. We find that $P$ and $Q$ distributions are generally suboptimal for many-body quantum optics. Additionally, we extend the formalism to calculate multi-time correlation functions, thereby broadening the scope of phase-space simulations of open spin systems to include coherence and spectral properties, as well as directional correlations of collectively radiating emitters. These developments provide valuable tools for investigating exotic light sources driven by collective dissipation, driven-dissipative phase transitions, and a wealth of many-body phenomena arising in state-of-the-art experimental platforms.
\end{abstract}

\onecolumngrid
\title{Phase-Space Methods for Many-Body Quantum Optics}
\author{Edgar Guardiola-Navarrete}
\affiliation{Department of Physics, Columbia University, New York, New York 10027, USA}
\author{Silvia Cardenas-Lopez}
\affiliation{Department of Physics, Columbia University, New York, New York 10027, USA}
\author{Ana Asenjo-Garcia}
\email{ana.asenjo@columbia.edu}
\affiliation{Department of Physics, Columbia University, New York, New York 10027, USA}
\maketitle
\vspace{0.5em}
\begingroup
  \setlength{\parskip}{0pt}
  \tableofcontents
\endgroup
 \section{Introduction}
 Ensembles of atoms (or other quantum emitters such as molecules, superconducting qubits, or color centers) interacting with each other via fluctuations of the electromagnetic field vacuum constitute a many-body open quantum system. These light-matter interfaces are of interest for observing out-of-equilibrium physics~\cite{Baumann10,Ferioli23,Goncalves24,Agarwal24}, many-body effects~\cite{Greiner02,Bakr10}, and hold promise for applications in quantum information~\cite{Bluvstein24}, quantum simulation~\cite{Bernien17,Semeghini21}, metrology~\cite{Bloom14,Safronova18,Norcia19,Bothwell22}, and the creation of novel light sources~\cite{Meiser09,Bohnet12,Perarnau20}. The interplay between dipole-dipole interactions and correlated decay gives rise to a myriad of effects that have attracted much attention, including spin-squeezing~\cite{Leroux10,Bornet23,Eckner23}, collective frequency shifts~\cite{Chang04,Bromley2016,Glicenstein20,Hutson24},  spatial self-organization~\cite{Domokos02,Black03,Baumann10,Ho24}, and many-body superradiance~\cite{Dicke54,Robicheaux21-2,Masson22,Sierra22,Rubies22,Rubies23,Cardenas23,Mok23,Masson24} and subradiance~\cite{Guerin16,Albrecht19,Rui2020,Ferioli21,Holzinger22,Zanner22,Rubies23-2}. Experimental platforms consisting of emitter arrays, either in free space~\cite{Endres16,Barredo16,Kim16,Kaufman21,Kumar18} or coupled to dielectric structures~\cite{Solano17,Mirhosseini19,Gonzalez24}, have rapidly matured.

Recent experimental progress in the field has spurred a growing demand for numerical simulations that faithfully capture the essential physics while remaining computationally efficient. Exact computations, whether through numerical solutions of the master equation or quantum trajectories~\cite{Wiseman09,Daley14}, are infeasible for large systems due to the exponential growth of the Hilbert space. Various strategies to circumvent this barrier include leveraging symmetries~\cite{Sarkar87,Xu13,Bolanos15} or adopting approximate methods, such as cumulant expansions~\cite{Robicheaux21,Plankensteiner22,Rubies-Bigorda23} and tensor network ans\"{a}tze~\cite{Verstraete08,Schollwock11}. Symmetry-based methods, however, are limited to systems exhibiting permutational symmetry, which narrows their applicability.  Approximation schemes risk introducing non-physical instabilities or fail to converge uniformly, thereby necessitating case-by-case validation. 

The quest for approximate computational methods in many-body physics remains an active area of research. Among these approaches, phase-space methods constitute a distinct category by representing quantum states as quasiprobability distributions, and recasting the dynamics as a partial differential equation (PDE) acting on the distribution. Initially envisioned as a framework to shed light on the interplay between classical dynamics and quantum fluctuations, phase-space methods have proven highly effective for developing efficient semiclassical simulations. While the PDE governing the dynamics is often of high order, in many cases it can be approximated as a Fokker-Planck equation (FPE)~\cite{Carmichael13}, which can then be unraveled into stochastic differential equations (SDEs) that are computationally efficient to simulate. Phase-space representations have been successfully used in the past by the quantum optics and cold atoms communities to capture the behavior of systems with semiclassical dynamics. Applications include modeling lasers~\cite{Carmichael13,Haken84}, Bose-Einstein condensates~\cite{Steel98,Sinatra02}, and closed spin systems~\cite{Schachenmayer15,Schachenmayer15_2,Ng13}. More recently, they have been applied to study open-system dynamics~\cite{Huber22,Mink22,Mink23,Hosseinabadi25}.

The aim of this paper is to provide a review of phase-space methods applied to the numerical simulation of open quantum systems composed of many two-level atoms that share a common environment. The paper is organized as follows: Section~\ref{Section PhaseSpace} provides an overview of previous work on phase-space methods and introduces the Stratonovich-Weyl correspondence, a general framework for deriving a phase-space formulation equivalent to the Hilbert-space approach. We also introduce the Truncated Wigner Approximation (TWA) as a semiclassical method for approximating exact coherent dynamics. Section~\ref{sc:phase_space_spins} reviews phase-space methods for two-level systems, including the use of the TWA for spins~\cite{Schachenmayer15}. Section~\ref{TWA section} focuses on many-body quantum optics. We review the spin model describing atoms interacting with a common electromagnetic bath and derive the exact PDE governing their phase-space evolution in any representation. By neglecting certain terms, we recover both the TWA for closed systems and the TWA for open dynamics proposed in Ref.~\cite{Mink23}. Additionally, we present a method for computing multi-time correlation functions and extend the analysis of errors introduced by these approximations. Finally, Section~\ref{sc:generalized_p_repre} expands the analysis of many-body quantum optics in phase space to an alternative quasiprobability distribution, the Positive $P$ representation.

\section{Phase-space methods: key concepts}\label{Section PhaseSpace}
In this section, we provide an overview on the history of phase-space methods and introduce a framework based on the Stratonovitch-Weyl correspondence, which can be applied to obtain the phase-space description of a general quantum system. To keep the discussion concise, we omit many details of this method but refer the interested reader to Refs.~\cite{Brif98, Brif99} for further details. To illustrate some of the concepts introduced in this section, we apply this formalism to a bosonic mode. Finally, we introduce the TWA, which incorporates quantum fluctuations only at the lowest order by sampling initial conditions from the initial Wigner function, while subsequent dynamics remains classical~\cite{Polkovnikov10}.

\subsection{Overview}
The interest in phase-space methods originated from efforts to formulate quantum mechanics as a statistical theory in phase space~\cite{Wigner32}. This formulation is achieved through Moyal quantization~\cite{Moyal49}, which maps a density matrix operator in Hilbert space to a quasiprobability distribution in phase space. The term quasiprobability reflects that this function is not necessarily positive.  Likewise, Moyal quantization transforms the von Neumann or master equation into a PDE for the quasidistribution, and maps quantum observables to functions in phase space. Once the evolution of the quasiprobability distribution is obtained, expectation values can be computed as statistical averages. Importantly, the phase-space representation of a quantum system is not unique~\cite{Carmichael13}. As operators may not commute, the ordering by which products in phase space map back into Hilbert matters. In bosonic systems, normal, anti-normal, and  symmetric order yield the Husimi $Q$~\cite{Husimi40},  Glauber-Sudarshan $P$~\cite{Glauber63,Sudarshan63}, and Wigner $W$ representations~\cite{Wigner32}, respectively.

The quantum optics community has extensively employed phase-space methods for the study of bosonic systems, such as lasers~\cite{Haken84}, masers~\cite{Gordon67}, and nonlinear processes such as parametric oscillators~\cite{Carmichael13}, among others. When the resulting PDE governing the dynamics in phase space is a FPE~\cite{Risken96}, an initially positive quasiprobability distribution remains positive, allowing efficient simulations via SDEs~\cite{Carmichael13,Risken96}. However, some systems of interest yield PDEs with non-positive semidefinite diffusion matrices, for which the unraveling in terms of SDEs is not possible. To address this issue, extensions such as the positive $P$~\cite{Drummond80_2} and gauge $P$ representations~\cite{Deuar02} were developed, enabling the study of nonclassical light and interacting Bose gases~\cite{Chaturvedi77,Drummond99,Drummond04,Kheruntsyan05}.

Phase-space representations can also be formulated for spins~\cite{Agarwal81,Varilly89}, and have been applied to the study of collective phenomena, including superradiance~\cite{Haake72,Glauber76} and optical bistability~\cite{Gronchi78,Drummond81}. A general framework based on the Stratonovich-Weyl correspondence~\cite{Stratonovich57,Varilly89} was introduced in Refs.~\cite{Brif98,Brif99}, defining mappings using phase-point or kernel operators and system symmetries, thereby generalizing phase-space methods to arbitrary quantum systems. Moreover, systems with finite-dimensional Hilbert spaces, such as ensembles of spins, allow for discrete phase-space representations~\cite{Wooters87}.

 Exact phase-space mappings of quantum dynamics often yield non-linear PDEs that are difficult to interpret or simulate efficiently ~\cite{Klimov02_2,Polkovnikov10}. Advancing the use of phase-space methods for simulations requires addressing two key questions: how to approximate the PDE by neglecting specific terms, and how to quantify the extent to which quantum fluctuations are captured by the chosen approximation.  The lowest-order approximation, the Truncated Wigner Approximation (TWA)~\cite{Blakie08,Polkovnikov10}, accounts for quantum fluctuations only through the initial quasiprobability distribution, with the subsequent evolution treated classically. In practice, this involves sampling initial conditions from the distribution and evolving them using mean field dynamics. Higher-order quantum fluctuations can be incorporated via stochastic quantum jumps~\cite{Polkovnikov10}. For spin systems, the Discrete Truncated Wigner Approximation (DTWA)~\cite{Schachenmayer15} applies the TWA by sampling the initial state in a discrete representation and evolving it according to classical spin equations of motion. The DTWA is effective for describing dynamics generated by spin Hamiltonians with high-coordination-number~\cite{Schachenmayer15}, and has been successfully applied to study spin squeezing~\cite{Qu19,Zhu19} and quantum quenches, though it fails to capture dissipation~\cite{Mink22}.

 \begin{figure}[!ht]
    \begin{center}
    \includegraphics[width = 0.9\textwidth]{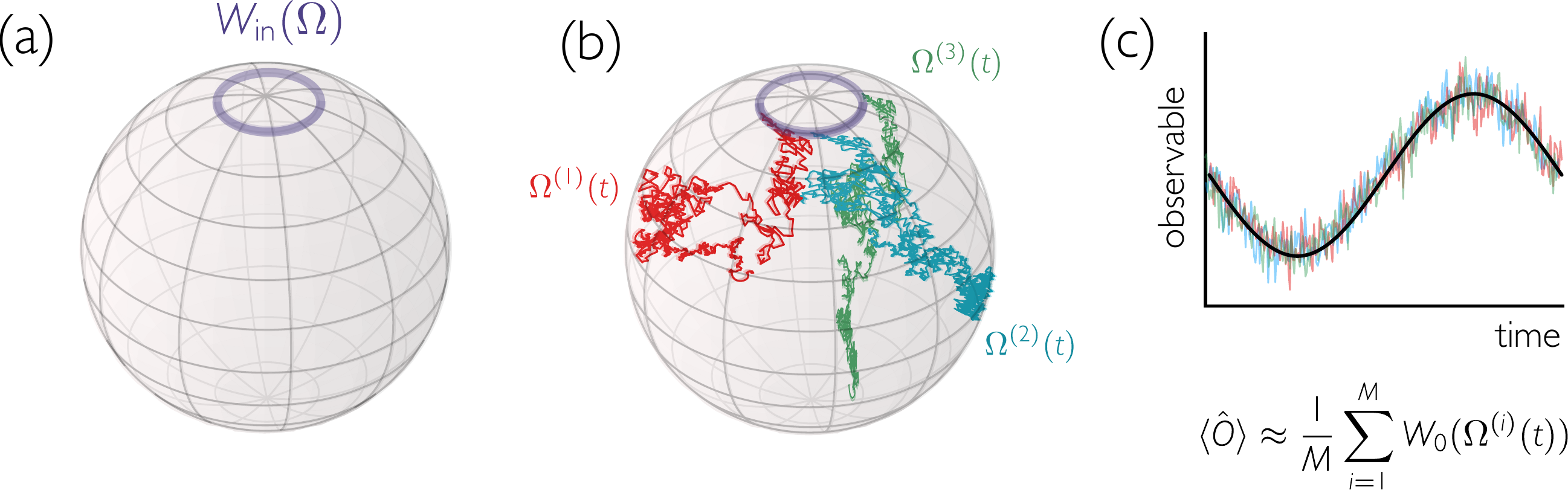}
    \caption{\textbf{Truncated Wigner approximation:} The method consists on  (a) obtaining the probability distribution for the initial state, $W_\text{in}(\Omega)$; (b)  sampling this initial distribution to generate initial conditions and evolving them using the SDE system that describes the approximated dynamics; and (c) computing expectation values $\langle\hat{O}\rangle$ by statistically averaging the phase-space representation of $\hat{O}$, $W_O(\Omega)$, over $M$ different realizations.}
    \label{fg:sketch_dtwa_trajectories}
    \end{center}
\end{figure}

 Because the DTWA cannot describe open systems, the Hybrid Discrete-Continuous Truncated Wigner Approximation (DCTWA)~\cite{Mink22} was recently developed to incorporate individual dissipative terms. In this approach, the continuous correspondence maps the spin dynamics to a PDE for the Wigner function. The resulting equation is then approximated as a FPE, and SDEs are used for simulation (see Fig.~\ref{fg:sketch_dtwa_trajectories}), making the computational cost scale linearly with the number of atoms.  Crucially, stochastic terms arise from dissipative dynamics, so quantum fluctuations are incorporated both in the initial state and during evolution.  If the dissipation terms are neglected and some of the drift terms are discarded, the approach reduces to the DTWA of Refs.~\cite{Schachenmayer15,Huber22}. 

If the master equation can be expressed in terms of a few collective spin operators, mapping to Schwinger bosons and applying TWA in bosonic phase space enables efficient simulations of dissipative many-body dynamics~\cite{Huber21}. A more general approach was proposed in Ref.~\cite{Mink23}, where a semiclassical approximation based on the exact correspondence rules for the master equation was used to derive SDEs where the noises driving different atoms are correlated. This approach effectively simulates collective dissipative dynamics populating states with large collective angular momentum, including superradiance~\cite{Mink23} and cascaded systems~\cite{Tebbenjohanns24,Bach24}, but seems to fail in subradiant sectors, where entanglement and strong correlations are expected to play a central role.

Another approach to achieving efficient simulations of spin systems using phase-space representations is to generalize the bosonic positive $P$ representation to spin systems, thereby ensuring that the resulting PDE takes the form of a FPE. This has been achieved either by mapping spin systems to Schwinger bosons and applying the bosonic positive $P$~\cite{Ng11}, or by extending the bosonic formulation to spins using coherent spin states~\cite{Ng13}. This formalism can be directly applied to both closed and open spin systems. However, these approaches are limited to short times due to instabilities in the SDEs, which cause some stochastic trajectories to diverge and distort statistical averages.

\subsection{Stratonovich-Weyl correspondence}
Historically, phase-space formulations for simple systems were first developed using the coherent states of the system at hand (see e.g. Ref.~\cite{Carmichael13} for bosonic modes and Ref.~\cite{Agarwal81} for spins). Here, we adopt a different approach by constructing the phase-space representation using the Stratonovich-Weyl (SW) correspondence~\cite{Stratonovich56,Brif98,Brif99}.  This method offers two advantages: 
(1) it generalizes the treatment of bosonic modes and spins to arbitrary quantum systems and (2) it establishes a notation that will be used throughout the paper.

Consider an arbitrary quantum system with Hilbert space $\mathcal H$. The first step towards a phase-space formulation is defining the corresponding phase space, denoted as $X$. The choice of phase space is often guided by the classical analogue of the quantum system. When no direct classical counterpart exists, the phase space can be determined by generalizing the notion of coherent states to a generic quantum system with an arbitrary symmetry group~\cite{Brif98,Brif99}.

The SW correspondence defines a mapping from the operator space of the quantum system, $ L(\mathcal{H})$, to a functional space over phase space, transforming an operator $\hat O\in L(\mathcal{H})$ into a function in phase space $F_O(\Omega;s)$, where $\Omega$ is a point in phase space and the parameter $s$ distinguishes different possible mappings. While $s$ can, in principle, be continuous, we focus on the discrete choices $s=\{-1,0,1\}$. The resulting distributions are referred to as the $Q$, Wigner, and $P$ representations, respectively, since for bosonic modes they reduce exactly to those well-known distributions~\cite{Carmichael13}. The mapping must satisfy a set of minimal and physically motivated rules:
\begin{enumerate}
    \item \textbf{Linearity}: The mapping $\hat O\rightarrow F_O(\Omega;s)$ is linear and bijective.
    \item \textbf{Reality}: $F_{O^{\dagger}}(\Omega;s) = F_O^*(\Omega;s)$, therefore if $\hat{O}$ is Hermitian, $F_O(\Omega;s)$ is real.
    \item  \textbf{Standardization}: 
    \begin{equation}
        \int_X d\mu(\Omega) F_O(\Omega;s) = \text{Tr} \{\hat O\},
    \end{equation}
    where $d\mu(\Omega)$ is the appropriate integration measure in phase space.
    \item \textbf{Covariance}:
    \begin{equation}
        F_{O(g)}(\Omega;s)=F_O(g\cdot\Omega;s),\; \text{with }\;\hat O(g)=\hat T(g^{-1})\hat O\hat T(g),
    \end{equation}
    where $g$ is an element of the quantum system's symmetry group and $\hat T(g)$ represents its action on the Hilbert space. 
    \item \textbf{Tracing}: For any two operators $\hat A$ and $\hat B$,
    \begin{equation}\label{eq:tracing_SW_rules}
        \int_X d\mu(\Omega)\, F_A(\Omega;s)F_B(\Omega;-s) = \text{Tr}\{\hat A\hat B\}.
    \end{equation}
\end{enumerate}

If a correspondence satisfies the previous rules, then $F_O(\Omega;s)$ is called the SW (or Weyl) symbol of the operator $\hat O$. In particular, the Weyl symbol of a density matrix $\hat \rho$, denoted as $F_\rho(\Omega;s)$, is a normalized real function due to the standardization  and real properties. According to the tracing property, the expectation value of any observable $\hat O$ can be calculated as
 \begin{equation}
        \int_X d\mu(\Omega) F_{\rho}(\Omega;s)F_O(\Omega;-s) = \text{Tr}\{\hat\rho\hat O\}=\braket{\hat O}.
        \label{expValue}
    \end{equation}
Thus, $F_{\rho}(\Omega;s)$ possesses some properties characteristic of a probability distribution. However, $F_{\rho}(\Omega;s)$ is not necessarily positive everywhere in phase space, making a strict probabilistic interpretation inaccurate. For this reason,  $F_{\rho}(\Omega;s)$ is commonly referred to as a \emph{quasi}distribution function.

As explained in Refs.~\cite{Brif98,Brif99}, constructing the Weyl symbol of any operator can be reduced to defining a  kernel operator $\hat\Delta(\Omega;s)\in L(\mathcal H)$ at each point in phase space. The correspondence is then implemented via
\begin{equation}\label{eq:CorrespondenceRuleKernel}
    F_O(\Omega;s) = \text{Tr}\{\hat\Delta(\Omega;s)\hat O\}, \quad \hat O = \int_X d\mu(\Omega) F_O(\Omega;s)\hat\Delta(\Omega;-s).
\end{equation}

The SW correspondence rules then translate into a series of properties for the kernel. In particular, the reality, standarization and covariance properties imply
\begin{eqnarray}
    \hat\Delta(\Omega;s)&&=\hat\Delta^\dagger(\Omega;s) \quad \forall \quad \Omega\in X,\\
    \int_X d\mu(\Omega)\hat\Delta(\Omega;s)&&=\mathbb 1,\\
    \hat\Delta(g\cdot\Omega;s)&&=\hat T(g)\hat \Delta(\Omega;s)\hat T(g^{-1}).
    \end{eqnarray}
A systematic method for constructing these operators while satisfying the SW rules was introduced in Refs.~\cite{Brif98,Brif99}.

\subsection{Moyal product and quantum evolution in phase space}
Unlike operators in $L(\mathcal{H})$, functions in phase space always commute. To preserve the structure of quantum mechanics, we must demand that $F_{AB}(\Omega,s)\neq F_{A}(\Omega,s)F_{B}(\Omega,s)$ if $\hat A$ and $\hat B$ do not commute. To formulate quantum mechanics in phase space in a way that remains fully equivalent to its formulation in Hilbert space, we must introduce an alternative, noncommutative product of real functions known as the Moyal product,
\begin{equation}
    F_A(\Omega,s_1)\star F_B(\Omega,s_2) = F_{AB}(\Omega,s_3).
\end{equation}
The explicit form of the Moyal product depends on the symmetry group of the specific quantum system~\cite{Brif98,Brif99,Zueco07}. Since the resulting general expression is involved, we do not reproduce it here. However, the Moyal products for a bosonic system [Eq. \eqref{MoyalProductBoson}] and a spin-$1/2$ [Eq. \eqref{eq:moyal_product_spins_definition}] are provided in later sections.

Once the Moyal product is defined, a formulation of quantum mechanics that is completely equivalent to that in the Hilbert space $\mathcal H$ is achieved. For example, consider a system with density matrix $\hat{\rho}$ evolving under a Hamiltonian $\hat{H}$. The evolution equation in phase space follows from the mapping
\begin{equation}
    \dot{\ra}=-\frac{\ii}{\hbar}\left[\hat{H},\ra\right] \; \rightarrow \; \dot{F_\rho}(\Omega;s)=-\frac{\ii}{\hbar}\left(F_H(\Omega;s) \star F_\rho(\Omega;s)- F_\rho(\Omega;s) \star F_H(\Omega;s)\right). 
    \label{eq:closedEvolution}
\end{equation}
Once the solution to this equation is obtained, Eq.~\eqref{expValue} can be used to compute expectation values of observables.

Beyond providing an alternative formulation of quantum mechanics, phase-space methods enable an interpretation of quantum dynamics as a classical evolution corrected by quantum fluctuations of increasing order~\cite{Wigner32,Polkovnikov10}. As we discuss in the following sections, this formulation is particularly valuable for many-body simulations, where neglecting higher-order quantum corrections allows for efficient semiclassical approximations.

\subsection{Example: bosonic system}\label{sc:harmonic_osc_introduction}
As a first example of this formalism, we review the phase-space representation of a bosonic mode~\cite{Carmichael13}. We consider a mode with lowering and raising operators $\hat a,\hat a^{\dagger}$, damped at a rate $\gamma$, so that its density matrix evolves according to
\begin{equation}\label{ExampleBoson}
    \dot{\hat  \rho} = \frac{\gamma}{2}\left(2\hat a\hat \rho \hat a^\dagger- \{\hat a^\dagger \hat a,\hat \rho\}\right).
\end{equation}
Following Refs.~\cite{Brif98,Brif99}, the phase space for this quantum system is the complex plane $X=\mathbb C$. Each point of $X$ corresponds to a coherent state $\ket\alpha$, which satisfies $\hat a\ket\alpha=\alpha\ket\alpha$. 

As a specific example, we focus on the $P$ representation ($s=1$). To express $\hat\rho$ in terms of the Weyl symbol $F_{\rho}(\alpha,1)$ as in Eq.~\eqref{eq:CorrespondenceRuleKernel}, we need the kernel with $s=-1$, which reduces to~\cite{Brif99}
\begin{equation}\label{Kernel P}
    \hat\Delta(\alpha;-1) = \ket{\alpha}\bra\alpha.
\end{equation}
Applying the correspondence rule [i.e., Eq.~\eqref{eq:CorrespondenceRuleKernel}], the $P$ representation of the density matrix $\hat \rho$ is given by
\begin{equation}
    \hat \rho = \int_{\mathbb C} d^2\alpha \ P(\alpha)\ket{\alpha}\bra\alpha,
\end{equation}
where we have defined $P(\alpha)\equiv F_{\rho}(\alpha;1)$. This representation interprets the state as an incoherent superposition of coherent states. The expectation value of an operator is then obtained via Eq.~\eqref{expValue} as
\begin{equation}
    \langle \hat O\rangle = \int d^2\alpha\ P(\alpha) \text{Tr}\{\hat O\ket\alpha\bra\alpha\},
\end{equation}
where $F_O(\alpha;s=-1)=\text{Tr}\{\hat O\ket\alpha\bra\alpha\}$ is the $Q$ representation of the operator. In particular, the calculation of the Weyl symbol for operators with normal order is straightforward. For example, the expectation value for $a^{\dagger p} \hat a^q$ is
\begin{equation}
    \langle \hat a^{\dagger p} \hat a^q\rangle=\text{Tr}\{\hat a^{\dagger p} \hat a^q\rho\} = \int d^2\alpha \ P(\alpha)\,\alpha^{*p}\alpha^q.
\end{equation}
Similar results hold for the $Q$ and Wigner representations with the anti-normal and symmetrized order, respectively~\cite{Carmichael13}.

To study the dynamics described in Eq.~\eqref{ExampleBoson}, we have to translate the master equation to phase space. This can either be done by writing down the Moyal product [Eq.~\eqref{MoyalProductBoson} below], or by using the action of creation and annihilation operators on the kernel along with integration by parts. We take the latter approach, for the sake of discussing the behavior of quasiprobability distributions at the phase-space boundary, which becomes crucial for understanding the failure of certain approximations discussed in Section \ref{sc:generalized_p_repre}.

The strategy consists in expressing the action of creation and annihilation operators on kernel operators in terms of derivatives. For example, for the $P$ representation,
\begin{equation}\label{Derivatives}
\hat a^\dagger\hat\Delta(\alpha;-1)=\left(\frac{\partial}{\partial\alpha}+\alpha^*\right)\hat\Delta(\alpha;-1),\ \ \hat\Delta(\alpha;-1)\hat a=\left(\frac{\partial}{\partial\alpha^*}+\alpha\right)\hat\Delta(\alpha;-1).\ \
\end{equation}
We apply this prescription (or its analogues for the other representations) on Eq.~\eqref{ExampleBoson} to recast the action of operators as derivatives, which then act on the quasidistribution  $F_{\rho}(\alpha,t;s)$ by integrating by parts. Denoting the master equation by  $\dot{\hat  \rho}=\mathcal L(\hat a,\hat a^\dagger)[\hat\rho]$ and its action on the kernel by the differential operator $D(\alpha,s)[\hat\Delta(\alpha;-s)]=\mathcal L(\hat a,\hat a^\dagger)[\hat\Delta(\alpha;-s)]$, we have
\begin{equation}
\begin{split}\label{IntegratingByParts}
    \int d^2\alpha\ \dot F_{\rho}(\alpha,t;s) \hat\Delta(\alpha;-s) = \dot\rho=\mathcal L(\hat a,\hat a^\dagger)[\hat \rho] = \int d^2\alpha\ F_{\rho}(\alpha,t;s) \mathcal L(\hat a,\hat a^\dagger)[\hat\Delta(\alpha;-s)]\\ = \int d^2\alpha\ F_{\rho}(\alpha,t;s) D(\alpha,s)[\hat\Delta(\alpha;-s)] = \int d^2\alpha\ D^\dagger(\alpha,s)[F_{\rho}(\alpha,t;s)]\hat\Delta(\alpha;-s)&,
\end{split}
\end{equation}
where $D^\dagger(\alpha,s)$ is the adjoint differential operator of $D(\alpha,s)$. In the last step, we have implicitly assumed that the quasidistribution vanishes at the boundaries, i.e., $\lim_{|\alpha|\rightarrow\infty} F(\alpha,t;s)=0$, so that the boundary terms produced by integrating by parts vanish. This assumption is reasonable for the purely dissipative dynamics in our example, but may not be true in general. As we discuss in section~\ref{sc:generalized_p_repre}, neglecting nonzero boundary terms can lead to instabilities in the resulting equations. A sufficient condition for Eq.~\eqref{IntegratingByParts} to be satisfied is that $F_{\rho}(\alpha,t;s)$ is solution to the PDE~\cite{Carmichael13} 
\begin{eqnarray}\label{FPEDamped}
    \dot F_{\rho}(\alpha,t;s) = D^\dagger(\alpha,s)[F_{\rho}(\alpha,t;s)]= \left(\frac{\gamma}{2}\frac{\partial}{\partial\alpha}\alpha +\frac{\gamma}{2}\frac{\partial}{\partial\alpha^*}\alpha^* +\gamma\lambda_s\frac{\partial^2}{\partial\alpha\partial\alpha^*}\right)F_{\rho}(\alpha,t;s),
\end{eqnarray}
where we have introduced the parameter $\lambda_s =(1-s)/2$ to account for the different representations. 

Therefore, the quasidistribution satisfies a FPE~\cite{Risken96}, which has the general form
\begin{equation}\label{eqFPE}
    \frac{\partial}{\partial t}P(\bm x, t) = -\sum_{i=1}^n \nabla_iA_i(\bm x) P(\bm x, t) + \frac{1}{2} \sum_{i,j=1}^n \nabla_i\nabla_j D_{ij}(\bm x) P(\bm x, t),
\end{equation}
where $\bm x \in \mathbb{R}^n$, and the vector $\bm A(\bm x)$ and the matrix $\bm D(\bm x)$ with elements $D_{ij}(\bm x)$ are the drift vector and the diffusion matrix, respectively. If $\bm D$ can be decomposed as $\bm D  = \bm B \cdot \bm B^T$, the FPE is equivalent to the system of \^Ito SDEs~\cite{Gardiner09,Carmichael13},
\begin{equation}\label{SDE}
    d\bm x = \bm A(\bm x)dt + \bm B(\bm x) d\bm W,
\end{equation}
where $d\bm W$ is a vector of independent infinitesimal Wiener increments drawn from a Gaussian distribution with zero mean and variance $dt$, i.e.,
\begin{eqnarray}
\mathbb E[dW_i]&=0,\\
    \mathbb E[dW_idW_j]&=\delta_{ij}dt.
\end{eqnarray}

The FPE has been extensively studied, with applications spanning many different fields~\cite{Risken96}. In the context of stochastic processes, it arises from the systematic truncation of the system-size expansion of the Chapman-Kolmogorov equation~\cite{Gardiner09,Kampen92} for a Markov process under certain regularity conditions. This expansion is characterized by a system-size parameter that enables a small-noise approximation of the equation. As a result, the FPE provides an alternative method to study complex stochastic processes, generally described by high-order PDEs, by capturing key features through a computationally efficient evolution governed by a system of SDEs with Gaussian noise.

For the damped bosonic mode, using $(\alpha,\alpha^*)\rightarrow(\text{Re}(\alpha),\text{Im}(\alpha)) \equiv (x,y)$, the SDEs corresponding to Eq.~\eqref{FPEDamped} take the form
\begin{equation}\label{CorrectSystem}
    \begin{cases} 
      dx = - \frac{\gamma}{2}xdt+\sqrt{\frac{\gamma\lambda_s}{2}}dW_x,\\
      dy = - \frac{\gamma}{2}ydt+\sqrt{\frac{\gamma\lambda_s}{2}}dW_y.
   \end{cases}
\end{equation}

\subsection{Truncated Wigner Approximation for bosons}

The PDE governing the evolution of a damped bosonic mode is a FPE in any representation. However, this is not always true for other systems, and one often requires methods to truncate the resulting PDE into a form that can be efficiently simulated. When a parameter exists that quantifies the magnitude of quantum fluctuations, it can serve as a system-size parameter, enabling the application of the truncation method described above~\cite{Carmichael13}. We now turn to one such method, the truncated Wigner approximation (TWA)~\cite{Blakie08,Polkovnikov10}, formulated for the Wigner representation $F_\rho(\Omega,0)$. This approach provides a semiclassical description of the dynamics of closed systems and has been successfully applied to the simulation of bosons, as we review here, spins (see next section), and fermions, among other systems.

The system's evolution can be derived either through the integration-by-parts method described earlier, or through a more direct application of the SW correspondence by computing the Moyal product -- an approach that can be implemented via Bopp operators~\cite{Polkovnikov10}. We follow the latter. For a bosonic mode, the Moyal product takes the form of an infinite series of differential operators acting on the Weyl symbols of the two operators involved,
\begin{equation}
    \label{MoyalProductBoson}
     \hat{A}\cdot\hat{B} \; \rightarrow \; F_A(\Omega,0)\star F_B(\Omega,0) =F_A(\Omega,0)e^{-\frac{\ii\hbar}{2}\Lambda}F_B(\Omega,0),
\end{equation}
where $\Lambda$ is the symplectic operator defined as
\begin{equation}
    \Lambda=\frac{\overset{\leftarrow}{\partial}}{\partial p}\frac{\overset{\rightarrow}{\partial}}{\partial x}-\frac{\overset{\leftarrow}{\partial}}{\partial x}\frac{\overset{\rightarrow}{\partial}}{\partial p},
\end{equation}
and $(x,p)$ are the position and momentum variables associated to the bosonic mode. The notation $\frac{\overset{\leftarrow}{\partial}}{\partial x}$ ($\frac{\overset{\rightarrow}{\partial}}{\partial x}$) indicates that the differential operator is applied to the function on the left (right). 

We consider a bosonic mode with purely coherent dynamics described by a Hamiltonian $\hat H$, evolving according to von Neumann's equation, i.e., Eq.~\eqref{eq:closedEvolution}. Denoting the Wigner representation of an operator $\hat A$ by $W_A(\Omega)\equiv F_A(\Omega,0)$, the time evolution of the Wigner function is given by
\begin{equation}
    \dot W_\rho(\Omega)=-\frac{\ii}{\hbar}\left(W_H(\Omega) \star W_\rho(\Omega)- W_\rho(\Omega) \star W_H(\Omega)\right).
\end{equation}

The Moyal product can often  be expanded in terms of a parameter that quantifies the strength of quantum fluctuations. In the case of a bosonic system, this parameter is $\hbar$. The truncated Wigner approximation consists of taking the limit  $\hbar\rightarrow 0$ in Eq.~\eqref{MoyalProductBoson} and keeping only the lowest-order term. In this limit, the evolution of the Wigner function for a bosonic system reduces to the classical Liouville equation~\cite{Polkovnikov10}, 
\begin{equation}
    \dot{W}_{\rho}(\Omega)\approx\{W_{H},W_{\rho}\}_\text{PB},
\end{equation}
where the Poisson bracket is defined as $\{A,B\}_\text{PB} \equiv \frac{\partial A}{\partial x}\frac{\partial B}{\partial p} - \frac{\partial A}{\partial p}\frac{\partial B}{\partial x}$.

According to Liouville's theorem, the Wigner function is conserved along the trajectories dictated by classical evolution. This implies that the Wigner distribution at any time can be obtained by sampling the initial distribution and evolving each sample according to classical equations of motion. More explicitly, if we sample $M$ initial conditions $\{\alpha_0^{(i)}\}_{i=1}^M$ from the initial distribution, we can approximate the Wigner distribution at any time by
\begin{equation}
    W_{\rho}(\alpha,t)\approx \frac{1}{M}\sum_{i=1}^M\delta(\alpha - \alpha^{(i)}(t)), 
    \label{eq:TWA}
\end{equation}
where $\{\alpha^{(i)}(t)\}_{i=1}^M$ are the solutions of the classical evolution. The expectation value of an observable can then be computed as
\begin{equation}\label{eq:exp_value_TWA}
  \langle \hat O \rangle=\int_X d\mu W_O(\Omega)W_{\rho}(\Omega)\approx \frac{1}{M}\sum_{i=1}^M W_O(\alpha^{(i)}(t)).
\end{equation}

Sampling the initial distribution and evolving according to the classical dynamics governed by $W_{H}$ thus provides an efficient approximate method for evolution.

\section{Phase-space formulation for spins}\label{sc:phase_space_spins}
Having discussed the main aspects of the phase-space formulation, we now develop the formalism to describe the collective dynamics of atoms using phase-space methods. We model the atoms as two-level systems or pseudo spins, and apply the SW correspondence to formulate the phase-space representation of $N$ spin-$1/2$ particles. Next, we discuss how the TWA has been adapted to study the coherent dynamics of many interacting atoms~\cite{Schachenmayer15}.

\subsection{Phase space and kernel}\label{Formalism Spins}
We consider a two-level system or pseudo spin-$1/2$, where we denote the ground and excited states as $\ket{g}$ and $\ket{e}$, respectively. The phase space is the unit sphere $X =\mathbb S^2$, parametrized using two angles $\Omega = (\theta,\phi)\in [0,\pi)\times[0,2\pi)$, and with invariant measure $d\mu(\theta,\phi) = \frac{\sin\theta}{2\pi}d\theta d\phi \equiv d\Omega$. The coherent states of the system are given by~\cite{Arecchi72}
\begin{equation}\label{eq:spin_coherent_state}
    \ket\Omega = \sin(\theta/2)e^{\ii\frac{\phi}{2}}\ket g + \cos(\theta/2)e^{-\ii\frac{\phi}{2}}\ket e = \begin{pmatrix}
    \cos(\theta/2)e^{-\ii\frac{\phi}{2}} \\
        \sin(\theta/2)e^{\ii\frac{\phi}{2}}
    \end{pmatrix}.
\end{equation}
As detailed in Refs.~\cite{Brif98,Brif99}, the kernel operators for a spin-$1/2$ that satisfy all the SW rules can be constructed from these states, and read
\begin{equation}\label{eq:kernel_general_spin}
    \hat\Delta (\Omega;s) = \frac{1}{2}\II - \frac{(-\sqrt{3})^{1+s}}{2}\bm m    \cdot\hat{\bm\sigma} = \frac{1}{2}
    \begin{pmatrix}
1-(-\sqrt{3})^{1+s}\cos\theta & -(-\sqrt{3})^{1+s}e^{-i\phi}\sin\theta\\
-(-\sqrt{3})^{1+s}e^{i\phi}\sin\theta & 1+(-\sqrt{3})^{1+s}\cos\theta 
\end{pmatrix}.
\end{equation}
Here, $\bm m = (\cos\phi\sin\theta,\sin\phi\sin\theta,\cos\theta)$, $\hat{\bm \sigma} = (\hat \sigma_x,\hat \sigma_y,\hat \sigma_z)$ are the Pauli matrices, and $s\in\mathbb Z$ labels the possible representations. Note that choosing the parametrization  $(\theta,\phi)\rightarrow (\pi-\theta,-\phi)$ would yield an equivalent formulation. However, we adopt the prior convention for consistency with the existing literature~\cite{Mink23}. 

With the kernel operators defined, we can now compute the Weyl symbols of operators and states using the correspondence rule in Eq.~\eqref{eq:CorrespondenceRuleKernel}. For instance, the Weyl symbols of the excited state $\hat\rho = \ket{e}\bra{e}$ in the three representations are 
\begin{equation}
    F_{\rho}(\Omega;s)=\frac{1}{2}\left(1-(-\sqrt{3})^{1+s}\cos\theta\right).
\end{equation}
More generally, since $\{\hat\II,\hat{\bm\sigma}\}$ forms a basis for the linear operators of a spin-$1/2$, and the correspondence is linear, we obtain
\begin{equation}
    L(\mathcal{H})=\text{span}\left\{\hat\II,\hat{\bm\sigma}\right\}\longmapsto\text{span}\left\{1,-(-\sqrt{3})^{1+s}\bm m(\theta,\phi)\right\}=\text{span}\{Y_{00},Y_{10},Y_{11},Y_{1-1}\}\equiv \mathcal U,
\end{equation}
where $Y_{\ell m}$ denote the spherical harmonics. Hence, the SW correspondence establishes a bijection between the original space of linear operators, $L(\mathcal{H})$,  and a certain functional space $\mathcal U$ defined over the unit sphere $\mathbb S^2$. This showcases the equivalence between formulations, since the phase representation is also given in terms of a complex 4-dimensional vector space.

The representation induced by the kernel $\hat\Delta (\Omega;s)$ for fixed $s$ is not unique~\cite{Mink22}. Specifically, if $\hat\rho$ is mapped to a distribution function $F_\rho(\theta,\phi,s)$, the function
\begin{equation}\label{Gauge}
    F_\rho'(\theta,\phi,s)=F_\rho(\theta,\phi,s)+\sum_{\ell=2}^\infty\sum_{m=-\ell}^\ell c_{\ell,m} Y_{\ell m}(\theta,\phi)
\end{equation}
with arbitrary coefficients $\{c_{\ell,m}\}$ is an equivalent representation for $\hat\rho$. This equivalence arises because, when integrating against the kernel, all spherical harmonics with $\ell>1$ are suppressed, as all the matrix elements of $\hat\Delta $ belong to $\mathcal U$. Thus, the spin phase space exhibits a gauge freedom in the sense that two distributions are equivalent, $F_\rho^{(1)}(\Omega,s)\sim F_\rho^{(2)}(\Omega,s)$, if and only if 
\begin{equation}
    F_\rho^{(1)}(\Omega,s)- F_\rho^{(2)}(\Omega,s)\in \mathcal{U}^C,
\end{equation}
where $\mathcal U^C$ consists of all functions spanned by spherical harmonics with $\ell>1$.

Finally, let us note that extending this phase-space formalism to systems of $N$ spin-$1/2$ particles is straightforward. If $\hat\Delta_i(\Omega_i;s)$ is the kernel of the phase-space correspondence for the Hilbert space $\mathcal H_i$ of the $i$th spin, then the tensor product 
$$\hat\Delta(\bm\Omega;s)=\bigotimes_{i=1}^N\hat\Delta_i(\Omega_i;s)$$
serves as the kernel for the Hilbert space $\bigotimes_{i=1}^N \mathcal H_i$, with the phase space given by 
$$\bm \Omega = (\Omega_1,\dots,\Omega_N)\in X_1\times\dots\times X_N.$$

\subsection{Mapping dynamics to phase space}
We now consider a single spin-$1/2$ evolving under some dynamics $\dot{\hat  \rho}=\mathcal L [\rho]$. As for a single bosonic mode discussed in Section~\ref{sc:harmonic_osc_introduction}, we can map this evolution to phase space using two different approaches. The first approach involves defining the Moyal product, which provides an exact mapping of the dynamics in phase space. The second approach involves identifying the action of the Pauli matrices on the kernel operators with differential operators that can be transferred to the quasiprobability distribution via integration by parts. We review both below.

\subsubsection{Moyal product}
Since any operator can be expressed in terms of $\{\hat\II,\hat{\bm\sigma}\}$, it suffices to compute the Moyal products of the Pauli matrices $\hat{\bm\sigma}$ and an arbitrary operator $\hat O$, that is, the actions $\bm{\mathcal S}_s[F_O(\Omega;s)] = F_{\bm \sigma}(\Omega;s)\star F_O(\Omega;s)$ and $\bm{\mathcal S}_s^*[F_O(\Omega;s)] = F_O(\Omega;s)\star F_{\bm \sigma}(\Omega;s)$. Using Bopp operators~\cite{Zueco07}, it can be shown that these products are given by

\begin{subequations}\label{eq:moyal_product_spins_definition}
\begin{equation}
\hat{\bm \sigma}\cdot \hat O \longmapsto  \bm{\mathcal S}_s[F_O(\Omega;s)] = \left(2\bm m\eta_1(L^2;s) + 2\ii(\bm m \times \bm L)\eta_2(L^2;s) + \bm L\right)F_O(\Omega;s),
\end{equation}
\begin{equation}
 \hat O\cdot\hat{\bm \sigma} \longmapsto \bm{\mathcal S}_s^*[F_O(\Omega;s)] =\left(2\bm m\eta_1(L^2;s) + 2\ii(\bm m \times \bm L)\eta_2(L^2;s) - \bm L\right)F_O(\Omega;s),
 \end{equation}
\end{subequations}
where the angular momentum operator is defined as
\begin{equation}\label{AngularOpDef}
 \bm L = \ii \begin{pmatrix}
        \sin \phi \frac{\partial}{\partial\theta} + \cot \theta\cos \phi\frac{\partial}{\partial\phi}\\
        -\cos \phi \frac{\partial}{\partial\theta} + \cot \theta\sin \phi\frac{\partial}{\partial\phi}\\
        -\frac{\partial}{\partial\phi}.
    \end{pmatrix}.
\end{equation}
The action of the operators $\eta_i$ on the basis of $\mathcal U$ of spherical harmonics is
\begin{equation}
    \eta_i(L^2;s)Y_{\ell m} = \eta_i(\ell;s)Y_{\ell m} = \left(\eta_i(0;s) + \frac{\eta_i(1;s) -\eta_i(0;s) }{2}L^2\right)Y_{\ell m}, \; \text{for}\; i=\{1,2\},
\end{equation}
with the coefficients $\eta_i(\ell;s)$ listed in Table~\ref{ZuecoCoeffs}. Our numerical values for these coefficients differ from those presented in Ref.~\cite{Zueco07}, which we suspect may contain a typographical or minor derivation error leading to the discrepancy.
\begin{center}
\begin{table}[!ht]
    \begin{tabular}{||c ||c| c||} 
 \hline
  $\eta_1(\ell;s)$ & $\ell=0$ & $\ell=1$ \\ [0.5ex] 
 \hline\hline
  $s=1 $& $-{3}/{2}$ & $-1/6$ \\  
 \hline
 $s=0$ & ${\sqrt{3}}/{2}$ & ${\sqrt{3}}/{6}$ \\ 

 \hline
 $s=-1$ & $-{1}/{2}$ &$ -{1}/{2} $ \\
 \hline
\end{tabular}
\begin{tabular}{||c ||c | c||} 
 \hline
  $\eta_2(\ell;s)$ &  $\ell=0$ & $\ell=1$ \\ [0.5ex] 
 \hline\hline
 $s=1 $& ${1}/{6}$& ${1}/{6}$ \\ 
 \hline
 $s=0$ & $-{\sqrt{3}}/{6}$& $-{\sqrt{3}}/{6}$ \\ 
 \hline
 $s=-1$ &$1/2 $&$1/2 $ \\
 \hline
\end{tabular}
\caption{\centering Coefficients for the Moyal product of Eq.~\eqref{eq:moyal_product_spins_definition}.}
\label{ZuecoCoeffs}
\end{table}
\end{center}

These results can be readily extended to systems with multiple spins. The Moyal products of Pauli matrices of the $n$-th spin and an arbitrary operator, $F_{\bm \sigma^n}(\bm\Omega;s)\star F_O(\bm\Omega;s)$ and $F_O(\bm\Omega;s)\star F_{\bm \sigma^n}(\bm\Omega;s)$, are given by Eqs.~\eqref{eq:moyal_product_spins_definition} and~\eqref{AngularOpDef}, with the partial derivatives replaced by $\frac{\partial}{\partial\theta_n},\frac{\partial}{\partial\phi_n}$. Additionally, the Moyal product is known for angular momentum $j>1/2$ ~\cite{Zueco07}, which opens the door to the use of a collective spin basis.

\subsubsection{Integration by parts}
The second approach is more direct and relies on the set 
$$\mathcal{B}=\left\{\hat\Delta(\Omega;s),\frac{\partial}{\partial\theta}\hat\Delta(\Omega;s),\frac{\partial}{\partial\phi}\hat\Delta(\Omega;s),\frac{\partial^2}{\partial\phi^2}\hat\Delta(\Omega;s)\right\}$$
forming a basis for the linear operators of a two-level system~\cite{Mink22}. Consequently,  the product of any Pauli matrix and the kernel, $\hat \sigma_j\hat\Delta(\Omega;s) = D_j[\hat\Delta(\Omega;s)]$ for $j=\{x,y,z\}$, can be written as a linear combination of the elements of $\mathcal{B}$, i.e.,
$$
\hat \sigma_j\hat\Delta(\Omega;s) =\left(\alpha_j^{0}+\alpha_j^{\theta}\frac{\partial}{\partial\theta}+\alpha_j^{\phi}\frac{\partial}{\partial\phi}+\alpha_j^{\phi \phi}\frac{\partial^2}{\partial\phi^2}\right)\hat\Delta(\Omega;s)\equiv D_j[\hat\Delta(\Omega;s)].
$$

An alternative definition of the Moyal product  $F_{\sigma_j}(\Omega;s) \star F_{\rho}(\Omega,t;s)$ can be obtained by integrating by parts and transferring the derivatives to the distribution~\footnote{ Explicitly, \begin{equation}
\begin{split}
    \hat \sigma_j\cdot \hat \rho &= \int_Xd\Omega \  F_{\rho}(\Omega,t;s)\ \hat \sigma_j \hat\Delta(\Omega;s) = \int_Xd\Omega \  F_{\rho}(\Omega,t;s)D_j[\hat\Delta(\Omega;s)]\\
    &= \int_X\frac{d\theta d\phi}{2\pi}\sin\theta \csc\theta \ D_j^{\dagger}[F_{\rho}(\Omega,t;s)\sin\theta]\hat\Delta(\Omega;s) = \int_Xd\Omega \  \tilde {D}_j^{\dagger} [F_{\rho}(\Omega,t;s)]\hat\Delta(\Omega;s),\\
    &\text{therefore }\  \hat \sigma_j\cdot \hat \rho \longmapsto \tilde {D}_j^{\dagger} [F_{\rho}(\Omega,t;s)].
\end{split}
\end{equation}}. The product is then mapped to the action of the adjoint operator of $D_j$ (denoted as $ {D}_j^{\dagger}$), i.e.,
\begin{equation}\label{eq:integration_by_parts_wigner}
     \hat \sigma_j\cdot \hat \rho \longmapsto \tilde {D}_j^{\dagger} [F_{\rho}(\Omega,t;s)].
\end{equation}

Here, $\tilde {D}_j^{\dagger}[\ \cdot\ ] = \csc\theta \ D_j^{\dagger}[\sin\theta\ \cdot\ ]$ corresponds to the same expression as $D_j^\dagger$ but with all partial derivatives replaced by covariant derivatives $\nabla_x = \csc\theta\frac{\partial}{\partial x}\sin\theta$ with $x=(\theta,\phi)$. Explicit examples can be found in Ref.~\cite{Mink22}. As for the bosonic case, we have implicitly neglected boundary terms. The boundary for $\mathbb S^2$, given by the chosen parametrization, is
\begin{equation}\label{Boundary sphere}
    \delta\mathbb S^2 = \{0,\pi\}\times(0,2\pi)\cup(0,\pi)\times\{0,2\pi\}.
\end{equation}
The operators $D_j^\dagger$ and the kernels $\hat\Delta(\Omega;s)$ are $2\pi$-periodic in $\phi$, ensuring that the boundary terms vanish in that coordinate. However, there is no physical or rigorous argument ensuring that $F_{\rho}(\Omega,t;s)$ vanishes at $\theta=0$, and $\theta=\pi$. Hence, neglecting boundary terms and using the alternative Moyal product is an approximation that should be validated case by case.  For example, Ref.~\cite{Mink22} shows that this approximation is suitable for modeling spontaneous decay and coherent driving.

\subsection{Example: Ising model}
As an example of this formalism, we analyze the Ising model, of Hamiltonian
\begin{equation}
    \hat{H}_I=\frac{1}{2}\sum_{n,m=1}^N J_{nm}\hat \sigma_n^z\hat \sigma_m^z,
\end{equation}
with $J_{nm}=J_{mn}$ and $J_{nn}=0$.  Using the Moyal product from Eq.~\eqref{eq:moyal_product_spins_definition}, the evolution of the Wigner distribution (taking $\hbar=1$) is given by
\begin{equation}\label{eq:ising_model_equation_exact}
\begin{split}
\begin{aligned}
\dot{\hat \rho} = -\ii[ \hat H_I,\hat \rho]\longmapsto \dot W_{\rho} = -\frac{2}{\sqrt 3}\sum_{n,m=1}^N J_{nm}\Big( 
    &\nabla_{\phi_n}3\cos\theta_m - \nabla_{\phi_n}\nabla_{\theta_m}\cos2\theta_m\csc\theta_m\\
    &+\nabla_{\phi_n}\nabla_{\theta_m}^2\cos\theta_m + \nabla_{\phi_n}\nabla_{\phi_m}^2\cot\theta_m\csc\theta_m\Big)W_{\rho}.\\
    \end{aligned}
\end{split}
\end{equation}

The resulting evolution equation includes third order derivatives and a non positive diffusion matrix. Consequently, the phase-space formulation does not yet provide any clear physical insight or computational advantage over its Hilbert-space counterpart. As discussed in Section~\ref{Section PhaseSpace}, accessing semiclassical and computationally efficient simulations requires truncating Eq.~\eqref{eq:ising_model_equation_exact} to the lowest order in quantum fluctuations. However, unlike for bosons, Eq.~\eqref{eq:ising_model_equation_exact} lacks a parameter quantifying the strength of quantum fluctuations, making the choice of approximation less evident. 

\subsection{Truncated Wigner Approximation for spins}\label{TWA for spins}
The approximation used for spin systems  was originally developed in Ref.~\cite{Schachenmayer15} to study many-body coherent spin dynamics. To identify the terms to be neglected, let $D$ denote the right-hand side of Eq.~\eqref{eq:ising_model_equation_exact}, and let $D^{\dagger}$  be its adjoint. We can then describe the evolution in a Heisenberg-like formalism by putting the time dependence of the state in the kernel
\begin{equation}\label{eq:Heisenberg}
    {\hat \rho}(t) = \int_X d\bm\Omega \  W_{\rho}(\bm\Omega,t){\hat\Delta}(\bm\Omega) = \int_X d\bm\Omega \ W_{\rho}(\bm\Omega,t=0)\hat\Delta(\bm\Omega,t),
\end{equation}
which now satisfies the differential equation
\begin{equation}
    \begin{cases}
        \dot \Delta(\bm\Omega,t) &= D^{\dagger}\Delta(\bm\Omega,t),\label{EqKernelEv}\\
        \Delta(\bm\Omega,t=0) &= \Delta(\bm\Omega)=\bigotimes_{n}\Delta(\Omega_n).
    \end{cases}
\end{equation}
Note that in order to obtain this equation, we have assumed that $W_{\rho}(\bm\Omega,t=0)$ vanishes at the boundaries so that we can integrate by parts to pass the time propagator $e^{Dt}$ in $W_{\rho}(\bm\Omega,t)=e^{Dt}W_{\rho}(\bm\Omega,t=0)$ to the kernel. Such boundary conditions can always be enforced by exploiting the gauge freedom to eliminate any contributions of $W_{\rho}(\bm\Omega,t=0)$ at the boundaries (using e.g. the discrete sampling defined below in Eq.~\eqref{Eq:discreteSampling}).

The exact solution to Eq.~\eqref{EqKernelEv} would in general lead to kernels $\Delta(\bm\Omega,t)$ whose dimension scales exponentially with atom number. To achieve the linear complexity scaling that would arise if we knew which terms to neglect to implement the TWA, we introduce the following ansatz for the kernel operators~\cite{Mink22}
\begin{equation}\label{ansatz}
    \hat\Delta(\bm\Omega,t) = \bigotimes_n \hat\Delta(\Omega_n,t),
\end{equation}
i.e., the kernels remain separable \emph{at all times} throughout the evolution. Examining Eq.~\eqref{eq:ising_model_equation_exact}, we conclude that this ansatz holds only if we neglect cross derivatives and approximate the evolution by
\begin{equation}\label{eq:LiuvilleSpins}
     \dot W_{\rho} \approx -2\sqrt{3}\sum_{n,m=1}^N J_{nm}\nabla_{\phi_n}\cos\theta_m W_{\rho},
\end{equation}
which (by Eqs.~\eqref{eqFPE} and~\eqref{SDE}) is equivalent to the system of ordinary differential equations (ODEs)
\begin{equation}\label{eq:ising_model_equation_dtwa}
    \begin{cases}
        \dot\theta_n = 0,\\
        \dot \phi_n = 2\sqrt{3}\sum_{m=1}^N J_{nm}\cos\theta_m.
    \end{cases}
\end{equation}

Equations~\eqref{eq:LiuvilleSpins} and~\eqref{eq:ising_model_equation_dtwa} are, respectively, the Liouville equation and the classical evolution equations corresponding to the Weyl symbol of the Hamiltonian, $W_H(\bm \Omega) = \frac{3}{2}\sum_{n,m}J_{nm}\cos\theta_n\cos\theta_m$~\cite{Sudarshan2016}. Thus, by neglecting cross terms, we can invoke Liouville's theorem to approximate the solution of Eq.~\eqref{eq:ising_model_equation_exact} by
\begin{equation}
    W_{\rho}(\bm{\Omega},t)\approx \frac{1}{M}\sum_{i=1}^M\delta(\bm\Omega - \bm\Omega^{(i)}(t)), 
    \label{eq:TWA_spins}
\end{equation}
where $\{\bm\Omega^{(i)}(t)\}$ are classical trajectories of the sampled initial conditions. Just as in Eq.~\eqref{eq:exp_value_TWA}, expectation values are computed as
\begin{equation}\label{eq:exp_value_TWA2}
  \langle \hat O \rangle\approx \frac{1}{M}\sum_{i=1}^M W_O(\bm\Omega^{(i)}(t)).
\end{equation}

\subsection{Discrete Truncated Wigner Approximation}\label{section:DTWA}
Here we discuss practical implementation of the TWA to the numerical simulation of multiple interacting  atoms undergoing purely coherent dynamics~\cite{Schachenmayer15}. The main challenge in implementing the ideas presented before is that some states of interest have Wigner functions that are not completely positive, making it nontrivial to sample them for initial conditions. For example, the fully excited state is mapped to 
\begin{equation}\label{eq:excitedW}
    \hat \rho = \bigotimes_{n=1}^N \ket{e_n}\bra{e_n}\longmapsto W_{\rho}(\bm \Omega) = \prod_{n=1}^N \frac{1+\sqrt{3}\cos\theta_n}{2},
\end{equation}
which is negative for some combinations of angles.

To circumvent this issue, we adopt an alternative \emph{discrete} phase-space representation for systems with a finite Hilbert space~\cite{Wooters87}. When applied to two-level systems, this representation is completely analogous to the continuous representation used so far. The key observation is that the kernel operators associated with the set of points
\begin{equation}\label{Eq:WootersPoints}
    X_d =\{\Omega_1,\Omega_2,\Omega_3,\Omega_4\}= \{(\theta_0,\pi/4),(\pi - \theta_0,3\pi/4),(\theta_0,5\pi/4),(\pi-\theta_0,7\pi/4)\},
\end{equation}
with $\theta_0=\arccos(1/\sqrt 3)$, are sufficient to establish a mapping that satisfies the SW correspondence rules. These kernel operators have unit trace, are orthogonal, and form a basis for single-spin operators, so they satisfy all the correspondence rules of Section~\ref{Section PhaseSpace}. As a result, the Wigner function of a single spin can be expressed using just four real numbers,
\begin{equation}\label{Eq:discreteSampling}
   \ra= \sum_{\alpha=1}^4  W_\alpha \hat{\Delta}(\Omega_\alpha),
\end{equation}
with $W_\alpha=\frac{1}{2}\text{Tr}[\ra\hat{\Delta}(\Omega_\alpha)]$. For example, alternative representations for an atom in the excited state and in the ground state are, respectively,
\begin{equation}\label{examples_discrete}
    \ket e \bra e \longmapsto \frac{1}{2}(\delta(\Omega-\Omega_1) + \delta(\Omega-\Omega_3)),\quad  \ket g \bra g \longmapsto \frac{1}{2}(\delta(\Omega-\Omega_2) + \delta(\Omega-\Omega_4)),
\end{equation}
where $\delta(\Omega-\Omega_0) = 2\pi\delta(\cos\theta-\cos\theta_0)\delta(\phi-\phi_0)$ is the Dirac delta that accounts for the manifold measure.  Thanks to this discrete sampling method, the TWA can be used to numerically simulate spin systems, leading to the discrete truncated Wigner approximation (DTWA) introduced in Ref.~\cite{Schachenmayer15}. After sampling, initial conditions are evolved using the mean-field equations of motion, and expectation values are computed using Eq.~\eqref{eq:exp_value_TWA2}.

Since the initial states in Eq.~\eqref{examples_discrete} are sampled from only four points, initial conditions exhibit strong non-physical correlations.  At its core, the discrete sampling method exemplifies how the gauge freedom introduced in Section~\ref{Formalism Spins} allows for different representations of a state. The gauge freedom implies that we can add spherical harmonics with $\ell>1$ to Eq.~\eqref{eq:excitedW} to construct an alternative representation. In particular, it can be shown that  the representations in Eqs.~\eqref{eq:excitedW} and~\eqref{examples_discrete} are equivalent under this transformation.

We can further exploit this gauge freedom to modify Eq.~\eqref{examples_discrete} and mitigate the effect of initial non-physical correlations.  For example, it can be seen that the value for $\phi$ is irrelevant for these states, so any rotation  around the z axis of the sample points in Eqs. \eqref{examples_discrete} is a valid alternative sampling. In particular, this means that we can sample the continuous lines~\cite{Mink22}
\begin{equation}\label{eq:infinite_sampling}
    \ket e \bra e \longmapsto \delta(\cos\theta-\cos\theta_0),\quad  \ket g \bra g \longmapsto \delta(\cos\theta-\cos(\pi-\theta_0))
\end{equation}
for initial conditions.

\section{Phase-space methods for many-body quantum optics}\label{TWA section}

In this section, we present the phase-space formulation of many-body quantum optics. We begin by reviewing the spin model governing the dynamics of an ensemble of atoms that interact via a common electromagnetic reservoir and discuss collective phenomena that typically appear in these systems. Using the SW correspondence, we map the spin model from Hilbert space to phase space. 

The resulting  PDE describes collective dissipative dynamics in any representation, fully retaining the effect of quantum fluctuations, and thus encoding the same information as the spin model. However, numerical solution of this PDE becomes intractable for large particle numbers. As in previous sections, a truncation method is needed to reduce the equation to a more tractable form, preserving quantum fluctuations only to leading order. We discuss a recently proposed approximation~\cite{Mink23} that enables efficient simulations of open many-body spin dynamics by truncating the exact PDE for the Wigner distribution into a FPE that can be unraveled into SDEs. We analyze the accuracy of this approximation and present new sampling techniques for cases where the Wigner function cannot be directly interpreted as a probability distribution. 

Exploiting the linearity of the dynamics, we extend the method to compute multi-time correlation functions, enabling the calculation of coherence properties of the emitted electromagnetic field with polynomial scaling in atom number. Finally, we examine the truncation procedure in Ref.~\cite{Mink23}, outline the key criteria a valid truncation should satisfy, and explore analogous approaches for the $P$ and $Q$ representations. We find that the Wigner representation is the only case that yields a valid FPE with minimal ingredients.

\subsection{Spin model for many-body quantum optics}\label{section:spinModel}
 \begin{figure}[!ht]
    \begin{center}
    \includegraphics[width = 0.9\textwidth]{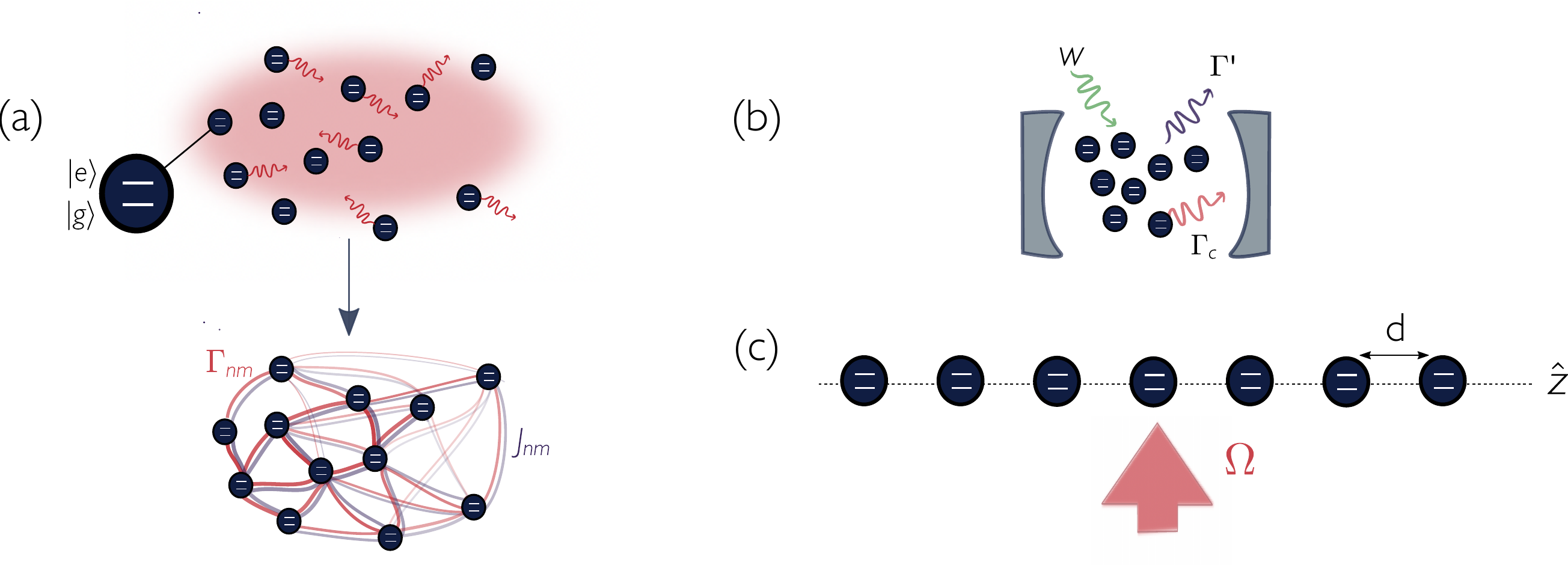}
    \caption{\textbf{Spin model for many-body quantum optics.} (a) An ensemble of atoms in contact with a Markovian electromagnetic environment can be described as an open spin model by integrating out the electromagnetic degrees of freedom. The field effectively mediates coherent and dissipative interactions among the atoms. Examples of typical systems include: (b) Atoms decaying at rate $\Gamma_c$ into a ``bad'' cavity, while being subjected to incoherent pumping and parasitic decay with rates $w$ and $\Gamma'$, respectively. (c) Atoms in free space arranged in a 1D array with lattice constant $d$ subjected to identical coherent pumping with Rabi frequency $\Omega$.}\label{fg:some_sample_configurations}
    \end{center}
\end{figure}
Here, we expand on the discussion from the first section regarding the many-body systems we aim to study. Many experiments described in the introduction, as well as other contemporary quantum optics scenarios, involve a collection of $N$ two-level atoms interacting with a continuum of electromagnetic field modes, such as those in free space or near a dielectric structure (as shown in Fig.~\ref{fg:some_sample_configurations}). A complete description of the atom-field system is infeasible due to the infinite number of field degrees of freedom.

A common approach, valid when the environment has a large bandwidth, is to trace out the field to obtain a master equation describing only the atomic degrees of freedom. While the derivation of the master equation can be carried out for specific cases~\cite{Lehmberg70,Pichler15}, an elegant and general approach, applicable to any linear isotropic medium, is given in Refs.~\cite{Gruner96,Dung02}. In this approach, a key quantity is the propagator of the electromagnetic field, i.e., the Green's tensor, which satisfies
\begin{equation}
   \left[\nabla \times \nabla -\frac{\omega^2}{c^2} \epsilon(\textbf{r},\omega)\right]\textbf{G}(\textbf{r},\textbf{r}',\omega)=\delta(\textbf{r}-\textbf{r}')\mathbb{1},
\end{equation}
where $\epsilon(\textbf{r},\omega)$ is the relative permittivity of the medium, and the boundary conditions are dictated by the geometry of the problem. The Green's tensor encodes the field propagation information, as $\textbf{G}(\textbf{r},\textbf{r}',\omega)\cdot \textbf{p}(\omega)$ is the field at  $\textbf{r}$ radiated by a dipole $\textbf{p}(\omega)$ at $\textbf{r}'$~\cite{Buhmann12}. The key physical insight is that while the quantum properties of the electromagnetic field are dictated by correlations and fluctuations, its propagation obeys the wave equation and is thus captured by the Green's tensor.

Following Refs.~\cite{Gruner96,Dung02}, after tracing out the field degrees of freedom under the Born-Markov approximation, the atomic system is described by the master equation
\begin{equation}
\dot{\ra}=\mathcal L[\rho] \equiv -\frac{\ii}{\hbar}\left[\hat{H},\ra\right]+\mathcal{L}_\text{dis}[\ra], \label{eq:master}    
\end{equation}
where
\begin{eqnarray}
\hat{H}&&=\hbar \sum_{n,m=1}^N J_{nm}\heg^n\hge^m ,\\
\mathcal{L}_\text{dis}[\ra]&&=\sum_{n,m=1}^N\frac{\Gamma_{nm}}{2}\left(2\hge^m\ra\heg^n-\ra\heg^n\hge^m-\heg^n\hge^m\ra\right),
\end{eqnarray}
and $\heg^n\equiv \ket{e_n}\bra{g_n}$ is the atomic coherence operator of atom $n$. The coherent and dissipative interactions mediated by the electromagnetic environment can be expressed in terms of the Green's tensor as 
 \begin{eqnarray}
     J_{nm}&=-\frac{\mu_0\omega_0^2}{\hbar} \mathcal{P}^*\cdot \text{Re}\left[\textbf{G}(\textbf{r}_n,\textbf{r}_m,\omega_0)\right]\cdot \mathcal{P},\label{GeneralFormJ}\\
     \Gamma_{nm}&=\frac{2\mu_0\omega_0^2}{\hbar} \mathcal{P}^*\cdot  \text{Im}\left[\textbf{G}(\textbf{r}_n,\textbf{r}_m,\omega_0)\right]\cdot \mathcal{P}\label{GeneralFormG}.
 \end{eqnarray}
Here, $\mu_0$ is the vacuum magnetic permeability, and ${\textbf{r}_n}$ and $\mathcal{P}$ denote the atomic positions and the dipole matrix element of the two-level transition, respectively. 

The dissipator in the master equation can readily be written in terms of collective operators. This is achieved by diagonalizing the $\mathbb{\Gamma}$ matrix (of elements $\Gamma_{nm}$) in Eq.~\eqref{GeneralFormG} to obtain $N$ collective jump operators $\{\hat{O}_\nu\}$ and collective decay rates $\{\Gamma_\nu\}$, yielding
\begin{equation}\label{CollectiveOps}
    \mathcal{L}_\text{dis}[\ra]=\sum_{\nu=1}^N\frac{\Gamma_\nu}{2}\left(2\hat{O}_\nu\ra\hat{O}_\nu^\dagger-\ra\hat{O}_\nu^\dagger\hat{O}_\nu-\hat{O}_\nu^\dagger\hat{O}_\nu\ra\right).
\end{equation}
Here, collective decay rates describe the rate at which each jump operator acts on the system. One can also unravel the master equation in terms of directional operators to extract information on the spatial distribution of the emitted light~\cite{Carmichael00}. 
 
In an experiment, the state of individual atoms may not be directly accessible. Instead, many-body dynamics is often characterized by analyzing different correlation functions of the field radiated by the system~\cite{Solano17,Glicenstein22,Ferioli23,Ferioli24}. For instance, the atomic decay rate or, equivalently, the photon emission rate reads
\begin{equation}\label{rate}
    R=\sum_{n,m=1}^N\Gamma_{nm}\langle\heg^n\hge^m\rangle\equiv\sum_{\nu=1}^N \Gamma_\nu \langle\hat{O}_\nu^\dagger\hat{O}_\nu\rangle.
\end{equation}

Coherence and spectral properties can also be extracted by tracking atomic degrees of freedom. Once Eq.~\eqref{eq:master} is solved, an input-output relation allows us to retrieve the (positive-frequency component of) the field at position $\textbf{r}$ emitted by the atoms~\cite{Caneva15,Xu15}, i.e.,
 \begin{equation}
     \hat{\textbf{E}}^+(\textbf{r})=\mu_0\omega_0^2\sum_{n=1}^N \textbf{G}(\textbf{r},\textbf{r}_n,\omega_0)\cdot \mathcal{P} \hge^n.
     \label{Eq:field}
 \end{equation}
From here, we obtain the spectrum of the emitted light, as well as the second-order correlation function between different components of the field, i.e.,
 \begin{eqnarray}
     S(\omega,\textbf{r},t)&&=2\,\text{Re}\left[\int_0^\infty d\tau\, e^{-\ii\omega \tau}\langle{\hat{\textbf{E}}^{-}(\textbf{r},t+\tau)\cdot\hat{\textbf{E}}^{+}(\textbf{r},t)}\rangle\right],\label{EqSpectrum}\\
      g^{(2)}_{\alpha\beta}(t,\tau,\textbf{r},\textbf{r}')&&=\frac{\langle \hat{E}_{\alpha}^-(t,\textbf{r})\hat{E}_{\beta}^-(t+\tau,\textbf{r}') \hat{E}_{\beta}^+(t+\tau,\textbf{r}')\hat{E}_{\alpha}^+(t,\textbf{r})\rangle}{\langle \hat{E}_{\alpha}^-(t,\textbf{r}) \hat{E}_{\alpha}^+(t,\textbf{r})\rangle\langle \hat{E}_{\beta}^-(t+\tau,\textbf{r}') \hat{E}_{\beta}^+(t+\tau,\textbf{r}')\rangle}.
 \end{eqnarray}

Beyond its elegance, the spin model in Eq.~\eqref{eq:master} is general, describing atoms decaying into an arbitrary Markovian environment. This generality allows for broad conclusions about dissipative dynamics without requiring explicit knowledge of the system's full evolution~\cite{Masson22,Mok24}. To illustrate the application of phase-space methods in the following sections, we consider two specific situations: (i) Atoms decaying into a single-mode ``bad'' cavity and (ii) Atoms forming a one-dimensional (1D) array in free space.

\textbf{(i) Atoms decaying into a single-mode bad cavity}

In the first scenario, atoms decay into a single-mode bad cavity, as shown in Fig.~\ref{fg:some_sample_configurations}(b). If the atoms and cavity are on resonance, the interaction coefficients are given by~\cite{Bonifacio71}:
 \begin{equation}\label{Dicke}
      J_{nm}=0,  \qquad \Gamma_{nm}=\Gamma_c  \qquad \forall \quad n,m,
 \end{equation}
where $\Gamma_c$ denotes the single-atom decay rate into the cavity mode. Additionally, atoms may be subject to parasitic individual decay and incoherent pumping occurring at respective rates $\{\Gamma'_n\}$ and $\{w_n\}$. The interaction coefficients in Eq.~\eqref{Dicke} correspond to the Dicke limit, where a single bright collective jump operator governs the decay of the system. The evolution of the density operator obeys the master equation
\begin{equation}
\dot{\ra}= \mathcal{L}_\text{dis}[\ra]+\mathcal{L}_{\Gamma'}[\ra]+ \mathcal{L}_{w}[\ra],
\end{equation}
with 
\begin{equation}
  \mathcal{L}_\text{dis}[\ra]=\frac{N\Gamma_c}{2}\left(2\hat{J}^-\ra\hat{J}^+-\ra\hat{J}^+\hat{J}^--\hat{J}^+\hat{J}^-\ra\right),  
\end{equation}
 where $\hat{J}^-=\frac{1}{\sqrt{N}}\sum_{n=1}^N \hge^n$ and   
\begin{eqnarray}
    \mathcal{L}_{\Gamma'}[\ra]&&=\sum_{n=1}^N \frac{\Gamma'_n}{2}\left(2\hge^n\ra\heg^n-\ra\heg^n\hge^n-\heg^n\hge^n\ra\right),\\
     \mathcal{L}_{w}[\ra]&&=\sum_{n=1}^N \frac{w_n}{2}\left(2\heg^n\ra\hge^n-\ra\hge^n\heg^n-\hge^n\heg^n\ra\right).\label{InCohPumping}
\end{eqnarray}
If the pumping is set to zero and atoms are initially inverted, the system undergoes Dicke superradiance, emitting a burst of light whose maximum intensity scales as $\sim N^2$, rather than the $\sim N$ scaling expected for independent atomic decay~\cite{Dicke54}. For nonzero pumping, steady-state superradiance is achieved~\cite{Meiser09,Bohnet2012} for a range of pumping rates, and the system constitutes a superradiant laser.

\textbf{(ii) 1D atomic array in free space}

In the second scenario, a coherently-driven 1D array of atoms decays into free space, as shown in Fig.~\ref{fg:some_sample_configurations}(c). The coherent drive has Rabi frequencies
$\{\Omega_n\}$, and is included in the dynamics by adding the Hamiltonian term
\begin{equation}
\hat{H}_\text{coh}=\hbar\sum_{n=1}^N\left(\Omega_n\hat\sigma_{eg}^n+\Omega_n^*\hat\sigma_{ge}^n\right)\label{CohPumping}
\end{equation}
to Eq.~\eqref{eq:master}. The interaction coefficients are determined by Eqs.~\eqref{GeneralFormJ} and~\eqref{GeneralFormG}. Since the vacuum is isotropic, the Green's tensor mediating the interactions depends only on the relative positions of the atoms and reads~\cite{Asenjo17}
 \begin{equation}
\textbf{G}_0(\textbf{r}_{n},\textbf{r}_{m},\omega_0)\equiv\textbf{G}_0(\textbf{r}_{nm},\omega_0)=\frac{e^{\ii k_0 r_{nm}}}{4 \pi k_0^2r_{nm}^3}\left[\left(k_0^2r_{nm}^2+\ii  k_0 r_{nm}-1\right)\mathbb{1}+(-k_0^2 r_{nm}^2-3\ii k_0r_{nm}+3)\frac{\textbf{r}_{nm}\otimes \textbf{r}_{nm}}{r_{nm}^2}\right].\label{GfreeSpace}
 \end{equation}
Here, $\textbf{r}_{nm}=\textbf{r}_{n}-\textbf{r}_{m}$, $r=|\textbf{r}|$, and $k_0=\omega_0/c$ is the wave number of the atomic transition. For a single atom, the spontaneous emission rate into the vacuum,  $\Gamma_0=\frac{\omega_0^3|\mathcal{P}|^2}{3\pi\hbar \epsilon_0c^3}$, is readily found from Eqs.~\eqref{GeneralFormG} and~\eqref{GfreeSpace} taking the limit $r_{nm}\rightarrow 0$.

Unlike in the Dicke limit, where a single bright jump operator dominates, the free-space scenario involves multiple bright jump operators~\cite{Masson22}. For many-body decay, this leads to deviations from the quadratic scaling of the maximum emitted intensity~\cite{Mok24}, competition effects, and potential quenching of superradiance.

\subsection{Exact many-body open quantum dynamics in phase space}

We now  apply the phase-space formalism to the master equation for the spin model, i.e., Eq.~\eqref{eq:master}. Using the Moyal product expressions from Eq.~\eqref{eq:moyal_product_spins_definition} for each term in the master equation, we find the PDE for the Weyl symbol
$F_{\rho}(\bm \Omega,t;s)$ in any representation, which reads
\begin{equation}\label{eq:exact_pde}
\frac{\partial F_{\rho}(\bm \Omega,t;s)}{\partial t}=\left(-\mathcal L_1+\frac{1}{2}\mathcal L_2+\mathcal L_3\right)F_{\rho}(\bm \Omega,t;s),
\end{equation}
where
\begin{align}
    \mathcal L_1   = \sum_{n=1}^N&\frac{\Gamma_{nn}}{2}\nabla_{\theta_n}\csc \theta_n\left(\cos\theta_n+\frac{\nu_{s}^{(1)}}{2}(3^s-1)\right)+ \nu_{s}^{(1)} \sum_{n,m=1}^N\left(\nabla_{\theta_n}\sin\theta_m A_{mn}+\nabla_{\phi_n}\cot\theta_n\sin\theta_mB_{mn}\right),\notag\\
    \mathcal L_2   = \sum_{n,m=1}^N&\big\{\nabla_{\theta_n}\nabla_{\theta_m}\left(\nu_{s}^{(2)}\cos\theta_mA_{mn}+\Gamma_{mn}\cos\phi_{mn}\right)
        -\nabla_{\theta_n}\nabla_{\phi_m}\csc\theta_m\left(\tilde{\nu}_{s}^{(2)} B_{mn}+\Gamma_{mn}\cos\theta_m\sin\phi_{mn}\right)\notag\\
        +&\nabla_{\phi_n}\nabla_{\theta_m}\cot\theta_n\left(\nu_{s}^{(2)}\cos\theta_mB_{mn}+\Gamma_{mn}\sin\phi_{mn}\right)
        +\nabla_{\phi_n}\nabla_{\phi_m}\cot\theta_n\csc\theta_m\left(\tilde{\nu}_{s}^{(2)} A_{mn}+\Gamma_{mn}\cos\theta_m\cos\phi_{mn}\right)\big\}\notag\\
        +&2\nu_{s}^{(3)}\sum_{n=1}^N\nabla_{\phi_n}^2\Gamma_{nn}\cot\theta_n\csc\theta_n,\notag\\
    \mathcal L_3 = \nu_{s}^{(3)}&\sum_{n,m=1}^N\big\{\nabla_{\theta_n}\nabla_{\theta_m}^2\sin\theta_m A_{mn}+  \nabla_{\theta_n}\nabla_{\phi_m}^2\csc\theta_m A_{mn} \notag\\
    &+\nabla_{\phi_n}\nabla_{\theta_m}^2\cot\theta_n\sin\theta_m B_{mn} +\nabla_{\phi_n}\nabla_{\phi_m}^2\cot\theta_n\csc\theta_m B_{mn}\big\}.
    \end{align}
    In the above expressions, 
\begin{align*}
\begin{array}{rl@{\hspace{1em}}rl}
    \nu_s^{(1)} &= (-1)^s\,3^{\frac{1 - s}{2}}, 
    & \nu_s^{(2)} &= \dfrac{\nu_s^{(1)}}{3}\left(3^{2 + s} -5\right), \\[2pt]
    \tilde{\nu}_s^{(2)} &= \dfrac{2\nu_s^{(1)}}{3}\left(3^{1 + s} -2\right), 
    & \nu_s^{(3)} &= -\dfrac{\nu_s^{(1)}}{6}\left(3^{1 + s} -1\right),
\end{array}
  \end{align*} 
and
    \begin{align}
    A_{mn} &= J_{mn}\sin\phi_{mn}+\frac{\Gamma_{mn}}{2}\cos\phi_{mn} \equiv \text{Im}\left\{g_{nm}e^{-\ii\phi_{mn}}\right\},\\
    B_{mn} &=-J_{mn}\cos\phi_{mn}+\frac{\Gamma_{mn}}{2}\sin\phi_{mn}\equiv \text{Re}\left\{g_{nm}e^{-\ii\phi_{mn}}\right\},
    \end{align}
where $g_{mn}=\frac{\mu_0\omega_0^2}{\hbar} \mathcal{P}^*\cdot \textbf{G}(\textbf{r}_n,\textbf{r}_m,\omega_0)\cdot \mathcal{P}\equiv-J_{mn}+i\frac{\Gamma_{mn}}{2}$, $\phi_{mn}=\phi_m-\phi_n$, and $\nabla_{x_n}=\csc\theta_n\frac{\partial}{\partial x_n}\sin\theta_n$ are covariant derivatives with $x_n=(\theta_n,\phi_n)$. The derivation of Eq.~\eqref{eq:exact_pde} is lengthy but straightforward, so it will not be reproduced here. 

The fundamental equation~\eqref{eq:exact_pde} provides an exact description of the dynamics in phase space, and is one of the central results of this paper. The solution of the PDE, $F_{\rho}(\bm \Omega,t;s)$, contains the same information as the solution of the master equation in Eq.~\eqref{eq:master}. However, the phase-space representation does not offer any numerical advantages over the Hilbert-space formulation because the above equation cannot be reduced to a FPE, meaning that no direct unraveling into SDEs is possible for any representation. In particular:  
\begin{itemize}
    \item The $P$ and the $W$ representations evolve according to a third-order PDE, which cannot be expressed as a FPE.
    \item In the $Q$ representation ($s=-1$), all third-order derivatives vanish (as $\nu_{s=-1}^{(3)}=0$). However, numerical observations indicate that the region in phase space where the diffusion matrix is positive semidefinite has zero measure for $N>1$. 
\end{itemize}

A naive numerical solution of Eq.~\eqref{eq:exact_pde}, using a finite-volume method, would require memory scaling as $\mathcal O(M^N)$, where $M$ is the number of partitions used in the phase space of each atom. A more refined approach would involve analyzing the action of the right-hand side of the PDE on the basis elements $\{Y_{00},Y_{10},Y_{11},Y_{1-1}\}^N$, and expressing the evolution exactly as a linear combination of these elements. However, this method ultimately results in a computational cost of $4^N$, which is the same as directly solving the master equation. Thus, no computational advantage is achieved.

The exact evolution in phase space is thus intractable for every representation, and a systematic approximation is not feasible because of the lack of a system size parameter quantifying quantum fluctuations. The goal of the of the rest of the section is to leverage mathematical tools from stochastic methods and functional analysis to develop approximations of Eq.~\eqref{eq:exact_pde} that enable efficient numerical simulation of many-body quantum optics problems. In particular, we will see that recent approximations introduced in the Truncated Wigner approximation formalism can be obtained by appropriately truncating some of the terms in Eq.~\eqref{eq:exact_pde}. For coherent evolution (with coefficients $\Gamma_{mn}=0\,\,\forall \,m,n$), neglecting all second- and third-order derivatives in Eq.~\eqref{eq:exact_pde} allows us to recover a purely deterministic evolution described by a system of ordinary differential equations in phase space. This system of equations corresponds to the semi-classical approximation of dynamics initially presented by Schachenmayer, Pikovski, and Rey~\cite{Schachenmayer15} reviewed in Section~\ref{TWA for spins}. 

\subsection{Dissipative TWA}

A natural first avenue towards truncating the fundamental equation~\eqref{eq:exact_pde} in open systems is to follow the approach outlined in Section~\ref{TWA for spins}, i.e., to assume that kernel operators remain unentangled during the evolution. Heuristically, the validity of the TWA discussed in past sections relies on Liouville's theorem for Hamiltonian dynamics. However, no direct analogue of Liouville's theorem exists for non-conservative systems. Consequently, when a system is coupled to an environment and dissipative terms are included in the density matrix evolution, the DTWA introduced in Ref~\cite{Schachenmayer15} is not expected to remain valid. 

The first step in developing an approximation to the exact evolution is to identify terms to be neglected. Here, we focus on the Wigner representation and discuss the approximation recently introduced by Mink and Fleischhauer~\cite{Mink23}, while postponing a more general discussion to Section~\ref{PQGeneralization}. Our goal is to recast Eq.~\eqref{eq:exact_pde} as a FPE for $s=0$. To achieve this, we neglect third-order derivatives by setting $\nu_{s=0}^{(3)}=0$, and ensure that the resulting diffusion matrix is positive semi-definite -- for instance, by imposing $\nu_{s=0}^{(2)}=0$. With these choices, Eq.~\eqref{eq:exact_pde} reduces to
\begin{equation}\label{eq:dtwa_pde}
\begin{split}
    \frac{\partial}{\partial t}W_{\rho}(\bm \Omega,t) =& \left(-\mathcal L_1 + \frac{1}{2}\mathcal L_2\right)W_{\rho}(\bm \Omega,t),\\
    \mathcal L_1 =& \sum_{n=1}^N \nabla_{\theta_n}\left[\frac{\Gamma_{nn}}{2}\cot\theta_n + \sqrt{3}\sum_{m=1}^N\sin\theta_m\left(J_{mm}\sin\phi_{mn} + \frac{\Gamma_{mn}}{2}\cos\phi_{mn}\right)\right] \\
    &+\sum_{n=1}^N \nabla_{\phi_n} \sqrt{3}\cot\theta_n\sum_{m=1}^N\sin\theta_m\left(-J_{mm}\cos\phi_{mn} + \frac{\Gamma_{mn}}{2}\sin\phi_{mn}\right),\\
    \mathcal L_2 =& \sum_{n,m=1}^N\Gamma_{mn}\Big(\nabla_{\theta_m}\nabla_{\theta_n}\cos\phi_{mn} - \nabla_{\theta_m}\nabla_{\phi_n}\cot\theta_n\sin\phi_{mn} \\
    &+\nabla_{\phi_m}\nabla_{\theta_n}\cot\theta_m\sin\phi_{mn} + \nabla_{\phi_m}\nabla_{\phi_n}\cot\theta_n\cot\theta_m\cos\phi_{mn}\Big).
\end{split}\end{equation}

This is precisely the equation originally derived by Mink and Fleischhauer~\cite{Mink23}. Equation~\eqref{eq:dtwa_pde} can be mapped  
into a system of SDEs by decomposing the diffusion matrix $\bm D = \bm B \cdot \bm B^T$, with both matrices given explicitly by
\begin{equation}
    \bm D = \left(\begin{array}{c|c} 
	\Gamma_{mn}\cos\phi_{mn} & -\Gamma_{mn}\cot\theta_n\sin\phi_{mn}\\ 
	\hline 
	\Gamma_{mn}\cot\theta_m\sin\phi_{mn} & \Gamma_{mn}\cot\theta_n\cot\theta_m\cos\phi_{mn}
\end{array}\right)\Longrightarrow
    \bm B = \left(\begin{array}{c|c} 
	-\Upsilon_{nm}\cos\phi_{n} & \Upsilon_{nm}\sin\phi_{n}\\ 
	\hline 
	\Upsilon_{nm}\cot\theta_n\sin\phi_{n} & \Upsilon_{nm}\cot\theta_n\cos\phi_{n}
\end{array}\right),
\end{equation}
where $\bm \Upsilon$ (with matrix elements $\Upsilon_{nm}$) is the decomposition of the matrix $\bm \Gamma=\bm \Upsilon\cdot\bm \Upsilon^T $. The SDE system corresponding to Eq.~\eqref{eq:dtwa_pde} is hence
\begin{eqnarray}
        d\theta_n = & \left(\frac{\Gamma_{nn}}{2}\cot\theta_n +  \sqrt{3}\sum_{m=1}^N\sin\theta_m\left(J_{mn}\sin\phi_{mn} + \frac{\Gamma_{mn}}{2}\cos\phi_{mn}\right)\right)dt \nonumber\\
        &+\sum_{m=1}^N\Upsilon_{nm}(-\cos\phi_ndW_{\theta_m} + \sin\phi_ndW_{\phi_m}),\label{eq:FleischauerSDE}\\
        d\phi_n = & \sqrt{3}\cot\theta_n\sum_{m=1}^N\sin\theta_m\left(-J_{mn}\cos\phi_{mn} + \frac{\Gamma_{mn}}{2}\sin\phi_{mn}\right)dt\nonumber\\
        &+\sum_{m=1}^N\Upsilon_{nm}\cot\theta_n(\sin\phi_{n}dW_{\theta_m} +\cos\phi_{n}dW_{\phi_m}), \label{eq:FleischauerSDE2}
\end{eqnarray}
where $\{W_{\theta_n},W_{\phi_n}\}_{n=1}^N$ are independent Wiener noises. Qualitatively, dissipation introduces stochasticity into the trajectories through random noise. Studying the evolution of the kernels in the Heisenberg representation, as done in Eq.~\eqref{eq:Heisenberg}, reveals that the kernel is no longer separable for all times, i.e.,  $\hat\Delta(\bm\Omega,t) \neq \bigotimes_n \hat\Delta(\Omega_n,t)$.  This behavior is expected since atoms are subjected to correlated noise. Note however that the non separability is due only to dissipation since coherent interactions are treated at the same level as before.

The truncation of the fundamental equation [Eq.~\eqref{eq:exact_pde}] remains \emph{a priori} arbitrary and cannot be systematically implemented given the lack of a system size parameter that quantifies fluctuations. The truncation can alternatively be performed at the level of the Moyal product, rather than directly on the PDE in Eq.~\eqref{eq:exact_pde}. This is the approach taken in Ref.~\cite{Mink23}, where Eq.~\eqref{eq:moyal_product_spins_definition} is approximated as 
\begin{equation}\label{eq:fleischauer_approx_moyal_definition}
\hat {\bm \sigma}\cdot \hat \rho \longmapsto \bm{\mathcal S}[W_{\rho}] \approx\tilde{\bm{{\mathcal S}}}[W_{\rho}] = (\bm s + \bm L)[W_{\rho}],   
\end{equation}
where $\bm s(\theta,\phi)\equiv2\eta_1(\ell=0,s=0)\bm m(\theta,\phi)=\sqrt 3\bm m(\theta,\phi)$. We now note that any error induced by the approximation in Eq.~\eqref{eq:fleischauer_approx_moyal_definition} is eliminated by projecting back into $\mathcal U$. That is (denoting the composition of two functions by $\circ$), we have
\begin{equation}\label{eq:dtwa_moyal_exact_projection}
    (P_{|\mathcal U}\circ \tilde{\bm{{\mathcal S}}} )[Y_{\ell m}] =  \bm{\mathcal S} [Y_{\ell m}]
\end{equation}
for any $\ell \leq1, |m|\leq \ell $, and thus for any $f\in\mathcal U$. In the last equation, we have defined the projection operator as 
\begin{equation}\label{projector}
    P_{|\mathcal U} Y_{\ell m} = \begin{cases}
        0 \quad \;\;\;\,\text{ if  } \ell>1\\
        Y_{\ell m}\quad \text{ otherwise.  }
    \end{cases}
\end{equation}

Equations~\eqref{eq:FleischauerSDE} and ~\eqref{eq:FleischauerSDE2} allow for a numerical simulation with time and memory scaling linearly with atom number. The numerical implementation closely follows that of the DTWA presented in Section~\ref{section:DTWA}, with the modification that the initial conditions sampled from the initial distribution must be evolved using the SDEs above rather than the classical evolution equations. Once the evolution is complete, expectation values can be obtained by averaging the Weyl symbols of the observables of interest, which can be obtained by using the correspondence rule in Eq.~\eqref{eq:CorrespondenceRuleKernel}. For example, the Weyl symbols corresponding to the total excited state population, the decay rate and the total angular momentum are given by
\begin{eqnarray}
    &\hat \sigma_{ee}= \sum_{n=1}^N\ket{e_n} \bra{e_n} \longmapsto \sum_{n=1}^N\frac{1+\sqrt{3}\cos\theta_n}{2},\label{PopulationTWA}\\
    &\hat R = \sum_{n,m=1}^N\Gamma_{nm}\hat \sigma_n^+\hat \sigma_m^-\longmapsto  3\sum_{n\neq m}\Gamma_{nm}\sin\theta_n\sin\theta_me^{i\phi_{nm}} + \sum_n\frac{1+\sqrt{3}\cos\theta_n}{2},\label{rTWA}\\
    &\hat J^2=\sum_{\alpha=\{x,y,z\}}\left(\sum_{n=1}^N\hat \sigma_\alpha^n\right)^2 \longmapsto 3\left(N+\sum_{n\neq m} \sin\theta_n\sin\theta_m\cos\phi_{nm}+\cos\theta_n\cos\theta_m\right).\label{jTWA}
\end{eqnarray}

An important point to note is that a single realization of the dynamics described by Eqs.~\eqref{eq:FleischauerSDE} and~\eqref{eq:FleischauerSDE2} does not correspond to a single physical realization of the conditioned dynamics of the atomic system under any measurement scheme~\cite{Wiseman09}. This becomes evident when examining the kernel in Eq.~\eqref{eq:kernel_general_spin}, which is not a valid density matrix, as it may possess negative eigenvalues. Consequently, while the TWA scheme is useful for computing expectation values of observables associated with the full density matrix, $\hat \rho(t)$, it fails to capture features related to individual trajectories. 

\subsection{Multi-time correlation functions}\label{sc:multi_time_correlation}
Here, we introduce a method to calculate multi-time correlation functions using Eqs.~\eqref{eq:FleischauerSDE} and~\eqref{eq:FleischauerSDE2}. These correlation functions are crucial in quantum optics problems: as discussed in Section~\ref{section:spinModel}, spatio-temporal correlations in the field emitted by a collection of atoms provide valuable insights into the underlying many-body dynamics. 

Let $\hat O_1,\dots,\hat O_K$ be a set of atomic operators and consider $K$ evaluation times, $t_1<\dots<t_K$. The quantum regression formula~\cite{Carmichael13} states that
\begin{equation}\label{eq:quantumRegression}
    \langle \hat O_1(t_1)\dots \hat O_K(t_K)\rangle = \text{Tr} \left[e^{\mathcal L (t_K-t_{K-1})}\left\{\dots \left\{  e^{\mathcal L (t_2-t_{1})}\left\{\hat \rho(t_1)\hat O_1(0)\right\}\hat O_2(0)\right\}  \dots\right\}\hat O_K(0) \right],
\end{equation}
where $\mathcal L$ is the Lindbladian of the system in Eq.~\eqref{eq:master} and $e^{\mathcal L t}$ is the time propagator. For simplicity, we assume that operators are ordered according to the evaluation time, though the following discussion can be adapted to other orderings. Furthermore, we assume each operator $\hat O_j$ acts on a single atom. The input-output relation, Eq.~\eqref{Eq:field}, implies that calculating correlators of this form is sufficient to determine the coherence properties of the emitted field, as field correlators reduce to sums of correlators involving products of single-atom operators. While the generalization to operators with larger support is straightforward, it is omitted here for brevity.

\begin{figure}[!ht]
    \begin{center}
    \includegraphics[width = 0.6\textwidth]{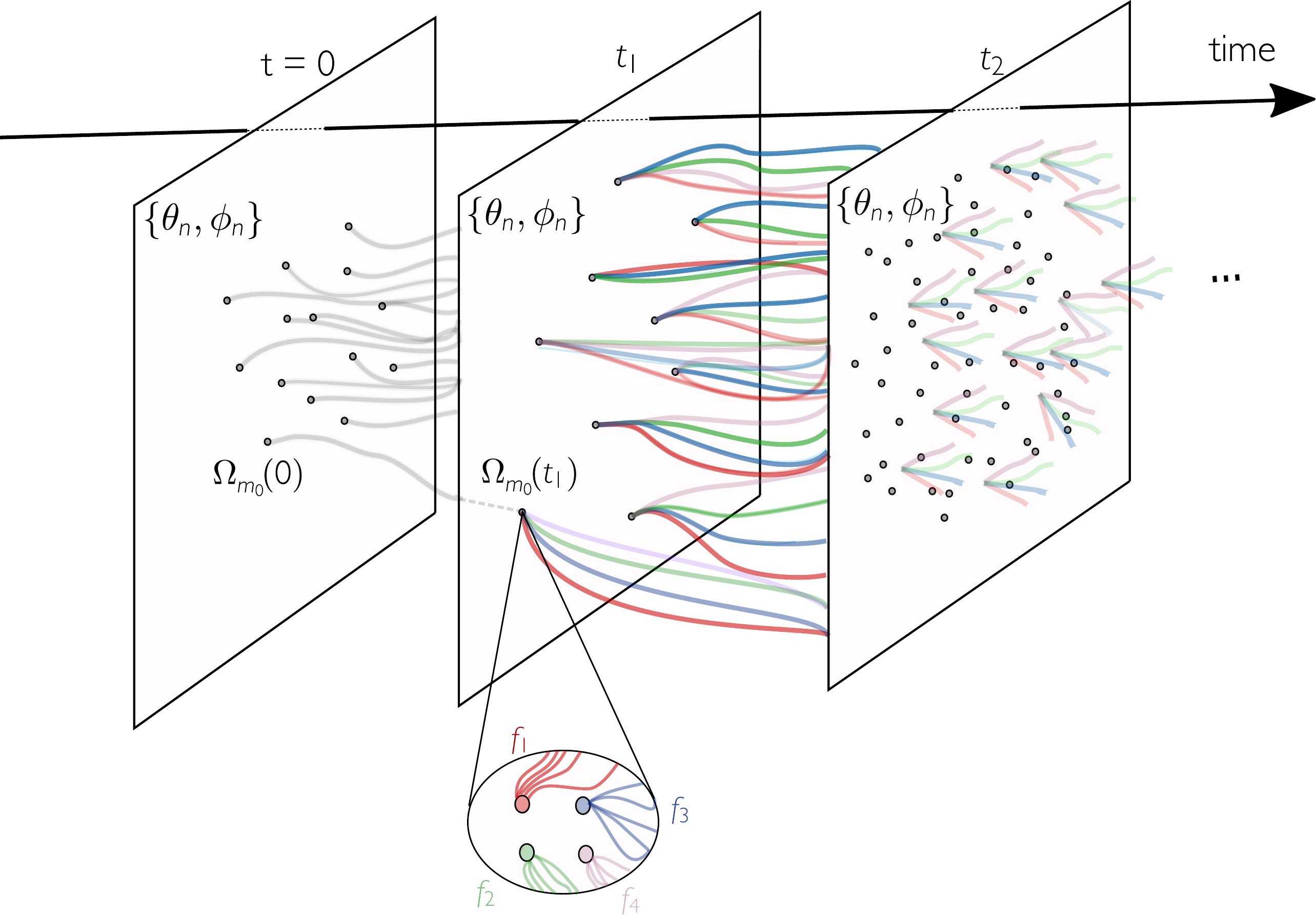}
    \caption{\textbf{Schematic representation of the calculation of multi-time correlation functions}. A set of $M_0$ samples from the initial distribution, $\Omega_{m_0}(0)$, is evolved to time $t_1$. Using the kernel of the discrete representation of a spin, the auxiliary operator is sampled. Since there are four kernel operators, each trajectory generates four distinct types of new initial conditions.  These new initial conditions are then evolved up to time $t_2$, and the process is iteratively repeated until reaching the final time $t_K$. }\label{fg:sketch_N_time_correlation_pipeline}
    \end{center}
\end{figure}

The phase-space methods presented earlier allow us to compute an approximation of the state at $t=0$. Specifically, by sampling $M_0$ initial conditions from the initial distribution, we approximate the Wigner function as
\begin{equation}
    W_\rho(\bm\Omega,t=0)\approx\frac{1}{M_0}\sum_{m_0=1}^{M_0}\delta(\bm\Omega -\bm \Omega_{m_0}(t=0)).
\end{equation}
We then evolve the sample points to a time $t=t_1$ (see Fig.~\ref{fg:sketch_N_time_correlation_pipeline}) and approximate the Wigner function at that time as
\begin{equation}\label{eq:Wignert1}
   W_\rho(\bm\Omega,t_1)\approx\frac{1}{M_0}\sum_{{m_0}=1}^{M_0}\delta(\bm\Omega -\bm \Omega_{m_0}(t_1)).
\end{equation}

Following the quantum regression formula, i.e., Eq.~\eqref{eq:quantumRegression}, we must now evolve the auxiliary operator $\hat \chi_1(0) =\hat \rho(t_1)\hat O_1(0)$ to $\hat \chi_1(t_2-t_1)$. This represents a twofold complication: first, we need to apply the correspondence rules to transform the Wigner function in Eq.~\eqref{eq:Wignert1} back into a density matrix, perform the operator product, and then transform back to phase space, which involves handling objects of dimension $2^N$. Second, $\chi_1(0)$ is generally not Hermitian, making the Wooters sampling method~\cite{Wooters87} unsuitable for obtaining initial conditions representing it.

The key to devising a new sampling method for $\chi_1(0)$ lies in recognizing that the problem is \emph{linear}, i.e., if $f_1(\bm \Omega,t)$ and $f_2(\bm \Omega,t)$ are solutions to the fundamental equation [Eq.~\eqref{eq:exact_pde}] with initial conditions $g_1(\bm \Omega)$ and $g_2(\bm \Omega)$, then $f_1(\bm \Omega,t)+f_2(\bm \Omega,t)$ is also a solution to Eq.~\eqref{eq:exact_pde} with initial condition $g_1(\bm \Omega)+g_2(\bm \Omega)$. This motivates analyzing the contribution of a single trajectory, which, combined with the discrete representation of phase space, enables sampling of the auxiliary operator without reverting to Hilbert space.

Consider thus a single trajectory $\bm \Omega_{m_0}(t)$ used to compute the distribution at time $t_1$ in Eq.~\eqref{eq:Wignert1}. This trajectory can be visualized as a delta function moving through the $2N$-dimensional phase space. Denoting its components by $\bm \Omega_{m_0}(t)=(\Omega_{{m_0};1}(t),\dots,\Omega_{{m_0};N}(t))$, the kernel operator corresponding to this moving delta function can be written as
\begin{equation}\label{eq:multi_kernel_separable}
    \hat\Delta(\bm \Omega_{m_0}(t))=\bigotimes_{n=1}^N \hat\Delta_n(\Omega_{{m_0};n}(t))\equiv\bigotimes_{n=1}^N \hat\Delta(\Omega_{{m_0};n}(t)), 
\end{equation}
where we introduce the notation $\hat\Delta(\Omega_{{m_0};n}(t))\equiv\hat\Delta_n(\Omega_{{m_0};n}(t))$ to indicate the kernel corresponding to the $n$-th spin to simplify notation. The kernel in the above equation is separable because it corresponds to a single trajectory, represented by a delta function. However, as discussed earlier, the full state -- obtained by averaging over many such trajectories -- is not separable. The Weyl symbol corresponding to $\hat\Delta(\bm \Omega_{m_0}(t))\hat O_1(0)$ is thus

\begin{equation}\label{WeylSybolt1}
    F_{\Delta(\bm \Omega_{m_0}(t_1))O_1(0)}(\bm \Omega) = \text{Tr}\{ \hat\Delta(\bm \Omega_{m_0}(t_1)) \, \hat O_1(0)\, \hat\Delta(\bm \Omega)\}.
\end{equation}

Our goal is to evolve the Weyl symbol in Eq.~\eqref{WeylSybolt1} for a time $\tau_1=t_2-t_1$. The function $F_{\Delta(\bm \Omega_{m_0}(t))O_1(0)}(\bm \Omega)$ is analytic, generally complex and non-positive.  We assume $O_1$ only addresses atom $n$ and use the discrete phase-space formalism~\cite{Wooters87} to write $\Delta(\bm \Omega_{m_0}(t))\,\hat O_1(0)$ in terms of the Wooters points defined in Eq.~\eqref{Eq:WootersPoints}, i.e.,

\begin{equation}\label{AuxOp}
    \hat\Delta(\Omega_{{m_0};n}(t_1))\, \hat O_1(0) = a_1^{(1)}\hat\Delta(\Omega_1)+a_2^{(1)}\hat\Delta(\Omega_2)+a_3^{(1)}\hat\Delta(\Omega_3)+a_4^{(1)}\hat\Delta(\Omega_4),
\end{equation}

since $\{\hat\Delta(\Omega_i)\}_{i=1,\dots,4}$ form a basis for the operators on the $n$-th atom. The expansion coefficients $a_{i}^{(1)}\in\mathbb C$ depend on the time $t_1$ and the trajectory $m_0$ considered, as they are functions of the $n$-th component $\Omega_{m_0;n}(t_1)$, but we have eliminated the dependence to lighten the notation. By substituting Eq.~\eqref{AuxOp} into Eq.~\eqref{WeylSybolt1}, and using the orthogonality of the kernel operators, we obtain

\begin{equation}
\begin{aligned}\label{IntermediateStep}
    F_{\Delta(\bm \Omega_{m_0}(t_1))O_1(0)}(\bm \Omega) &= (a_1^{(1)}\delta(\Omega_n-\Omega_1)+a_2^{(1)}\delta(\Omega_n-\Omega_2)+a_3^{(1)}\delta(\Omega_n-\Omega_3)+a_4^{(1)}\delta(\Omega_n-\Omega_4)) \prod_{r\neq n}^N\delta(\Omega_r-\Omega_{{m_0};r}(t_1))\\
    &\equiv \sum_{i=1}^4 a_i^{(1)}f_i(\bm\Omega),
    \end{aligned}
\end{equation}

where each function in the expansion is a product of deltas, i.e., $f_i(\bm\Omega) = \delta(\Omega_n-\Omega_i) \prod_{r\neq n}\delta(\Omega_r-\Omega_{{m_0};r}(t_1))$. By using the fact that the evolution is linear, the Weyl symbol in Eq.~\eqref{IntermediateStep} can now be straightforwardly evolved to

\begin{equation}\label{t1WeylSymbol}
    F_{\Delta(\bm \Omega_{m_0}(t_1))O_1(0)}(\bm \Omega,\tau_1)=\sum_{i=1}^4 a_i^{(1)}(t_1)f_i(\bm\Omega,\tau_1).
\end{equation}

Here, $f_i(\bm\Omega,\tau_1)$ is the product of deltas resulting from evolving the corresponding initial conditions for a time $\tau_1=t_2-t_1$, i.e.,
\begin{equation}
    f_i(\bm\Omega) = \delta(\Omega_n-\Omega_i) \prod_{r\neq n}\delta(\Omega_r-\Omega_{{m_0};r}(t_1))\rightarrow f_i(\bm\Omega,\tau_1) = \delta(\Omega_n-\Omega_i(\tau_1)) \prod_{r\neq n}\delta(\Omega_r-\Omega_{{m_0};r}(\tau_1)).
\end{equation}

The fact that $f_i(\bm\Omega)$ is a product of deltas simplifies the task significantly, as each term $f_i(\bm\Omega,\tau_1)$ involves only a single initial condition. However, in the case of open dynamics, the stochastic nature of the evolution still requires multiple trajectories to evolve each delta.

The process for computing a two-time correlator $\langle\hat O_1(t_1)\hat O_2(t_2)\rangle$ with $t_1<t_2$ thus involves the following steps:
\begin{enumerate}
    \item Running $M_0$ evolutions of the initial distribution.
    \item Constructing four different ``flavors'' of initial conditions from each final condition obtained in step 1. This is achieved by substituting the final phase-space coordinate of the $n$-th atom, i.e., $\Omega_{m_0;n}(t_1)$, by each of the points that define the Wooters kernel $\Omega_i$, and by computing the weights $a_i^{(1)}$ in Eq.~\eqref{AuxOp}.
    \item Running $M_1$ trajectories sampling from the initial conditions generated in step 2. 
    \item Computing the two-time correlator through a statistical average. By indexing the trajectories from steps 1 and 3 as $m_0$ and $m_1$, respectively, and assigning the $m_1$-th trajectory from step 2 to flavor $i$ based on the modulus $i=m_1\text{ mod }4+1$, the two-time correlator is given by

\begin{equation}
    \langle \hat O_1(t_1)\hat O_2(t_2)\rangle = \text{Tr}  \left\{ e^{\mathcal L (t_2-t_{1})}\left\{\hat \rho(t_1)\hat O_1(0)\right\}\hat O_2(0) \right\}=\frac{4}{M_0M_1}\sum_{{m_0}=1}^{M_0}\sum_{{m_1}=1}^{M_1} a_{m_0,m_1}^{(1)}W_{O_2(0)}(\bm\Omega_{{m_0},{m_1}}(\tau_1)).
\end{equation}

Here, $\bm\Omega_{{m_0},{m_1}}(\tau_1)$ represents the trajectory originating from sample $m_0$ of the initial Wigner distribution, and from the initial condition $m_1$ of step 2. Additionally, we use the shorthand notation $a_{m_0,m_1}^{(1)}=a_{m_1,\text{ mod }4+1}^{(1)}$ to condense all complex weight information for the $m_0$-th trajectory. The generalization to operators with support on more than a single atom can be worked out by considering that $\{\hat\Delta(\Omega_{i_1})\otimes\dots\otimes\hat\Delta(\Omega_{i_p})\}$, with $i_1,\dots,i_p\in\{1,2,3,4\}$, form a basis for the Hilbert space of $p$ atoms, $\mathcal H^p$.
\end{enumerate}

Identical reasoning applies to calculate correlators involving more than two times: once the $M_1\times M_0$ trajectories are completed to account for the evolution up to $t_1$, we consider a single trajectory, $\bm\Omega_{m_0,m_1}(\tau_1)$ and its corresponding operator $\hat\Delta(\bm\Omega_{m_0,m_1}(\tau_1))$. Just as in Eq.~\eqref{IntermediateStep}, the discrete phase-space sampling can be applied to write down the Weyl symbol of $\hat\Delta(\bm \Omega_{m_0,m_1}(\tau_1))\hat O_2(0)$,

\begin{equation}
    F_{\hat\Delta(\bm \Omega_{m_0,m_1}(\tau_1))O_2(0)}(\bm \Omega) = (a_1^{(2)}\delta(\Omega_n-\Omega_1)+a_2^{(2)}\delta(\Omega_n-\Omega_2)+a_3^{(2)}\delta(\Omega_n-\Omega_3)+a_4^{(2)}\delta(\Omega_n-\Omega_4))\prod_{r\neq n}\delta(\Omega_r-\Omega_{m_0,m_1;r}(\tau_1)).
\end{equation}

 We then sample this distribution with  $M_2$ points and evolve for a time $\tau_2=t_3-t_2$. This process is repeated until we reach time $t_K$ and operator $\hat O_K$ in Eq.~\eqref{eq:quantumRegression}. For clarity, we explicitly outline the procedure for the $k$-th step. Using the superindex notation $\bm m^{(k)} = (m_0,m_1,\dots,m_{k-1})$ to label trajectories, 

\begin{enumerate}
    \item Given the trajectory $\bm\Omega_{\bm m^{(k)}}(\tau_{k-1})$, apply the operator $\hat O_{k}(0)$ and calculate the Weyl symbol in the form of Eq.~\eqref{IntermediateStep}  to obtain the coefficients $a_i^{(k)}$ for $i=1,\dots,4$, which we denote by $a_{\bm m^(k+1)}^{(k)}$ as they depend on the choices of trajectories for all previous evolutions $m_0,\dots,m_{k-1}$ and they have four possible values labeled as $i=m_{k}\text{ mod }4+1$.
    \item Compute the evolution of the $M_k$ trajectories $\bm\Omega_{\bm m^{(k+1)}}(\tau_{k+1})$ with initial conditions $\bm\Omega_{\bm m^{(k+1)}}(0) = (\Omega_{\bm m^{(k)};1}(\tau_k),\dots,\Omega_{m_k\text{ mod }4+1},\dots,\Omega_{\bm m^{(k)};N}(\tau_k))$ with $m_k=1,\dots,M_k$.
    \item Once we have completed all evolutions with $k=K-1$, we can evaluate the multi-time correlation function as
    \begin{equation}
        \langle \hat O_1(t_1)\dots \hat O_K(t_K)\rangle = \frac{4^{K-1}}{M_0\dots M_{K-1}}\sum_{\bm m^{(K)}} W_{O_K(0)}\left(\bm\Omega_{\bm m^{(K)}}(\tau_{K-1})\right)\prod_{k=1}^{K-1} a_{\bm m^{k+1}}^{(k)},
    \end{equation}
    where $\bm m^{(K)}=(m_0,\dots,m_{N-1})$ and $m_k = 1,\dots,M_k$.
\end{enumerate}

An schematic pipeline of this procedure is shown in Fig.~\ref{fg:sketch_N_time_correlation_pipeline}, displaying the main steps for the calculations described above. This formalism is not restricted to the four point sampling method~\cite{Wooters87}, but it can also be implemented with the continuous version of the sampling or other alternatives presented in Appendix~\ref{sc:appendix_sampling_method_no_harmonics}, by appropriately changing the weights $a_i^{(k)}$. In the previously described scheme, strong correlations may build up between trajectories, since they are sampled only from four different phase-space points, and only differ through the noise in the SDEs. 

The dissipative TWA enables the computation of many-body dynamics with relatively low computational cost. In particular, the cost for calculating one time averages scales as $\mathcal O(N M)$, where $N$ is the number of spins and $M$ is the number of trajectories that are being sampled from the initial distribution. The previously described procedure to calculate multi-time correlators retains the computational advantage, but with a cost that scales as $\mathcal O(N M_0\dots M_K)$. This represents the best-case scenario, where the operators $\hat O_1(0),\dots,\hat O_K(0)$ each act only on a single spin. When the operators $O_k(0)$ are a linear combination of operators acting on each atom -- e.g. for the emitted field operator [Eq.~\eqref{Eq:field}] -- the computational cost increases to $\mathcal O(N^{(K+1)}M_0\dots M_K)$, as, for each of the possible $N^K$ combinations of single atom operators, we add a computation cost $\mathcal O(N M_0\dots M_K)$. Nevertheless, for few-time correlators, this remains significantly more efficient than the exponential scaling associated with exact evolution in the original Hilbert space.

\subsection{Numerical examples and validity of the approximation}

Natural questions regarding the dissipative TWA and its extension to open systems are: Is there a way to quantify the error it introduces? Under what conditions can we trust the approximation to capture the key aspects of a system's evolution? An initial attempt to address these questions was made in Ref.~\cite{Mink23}, where it was shown that if the dynamics predominantly populates states with high cooperativity (specifically, Dicke states  $\ket{J,M}$ with $J\sim \frac{N}{2}$) then the relative error of the approximation in Eq.~\eqref{eq:fleischauer_approx_moyal_definition} scales as $\mathcal O(1/\sqrt N)$ with atom number. The dissipative TWA is hence expected to perform well when the system evolves into such states.

However, no analogous bound exists for states with low angular momentum. In fact, similar arguments lead to a relative error bound that scales as $\mathcal O(N)$. Intuitively, we might expect the dissipative TWA to perform poorly in low-angular momentum subspaces, as they host highly entangled states~\cite{Toth10}. Nevertheless, this space also includes, for instance, the completely mixed state $\hat \rho=\frac{1}{2^N} \mathbb 1^{\otimes N}$, for which the approximation in Eq.~\eqref{eq:fleischauer_approx_moyal_definition} is exact. As a result, the accuracy of the approximation in these cases remains uncertain. 

In what follows, we aim to develop an intuitive understanding of how the approximation performs. We approach this first through numerical simulations of the spin model introduced in Section~\ref{section:spinModel}. We begin by applying the dissipative TWA to study the paradigmatic Dicke superradiance for atoms in a cavity~\cite{Dicke54}, the collective dynamics of an atom array in free space under coherent pumping, and the steady-state behavior of a superradiant laser~\cite{Meiser09}. These cases provide insights into when the approximation is expected to be valid.

\subsubsection{Dicke superradiance in a cavity}
\begin{figure}[!ht]
\begin{center}\includegraphics[width=1.0\linewidth]{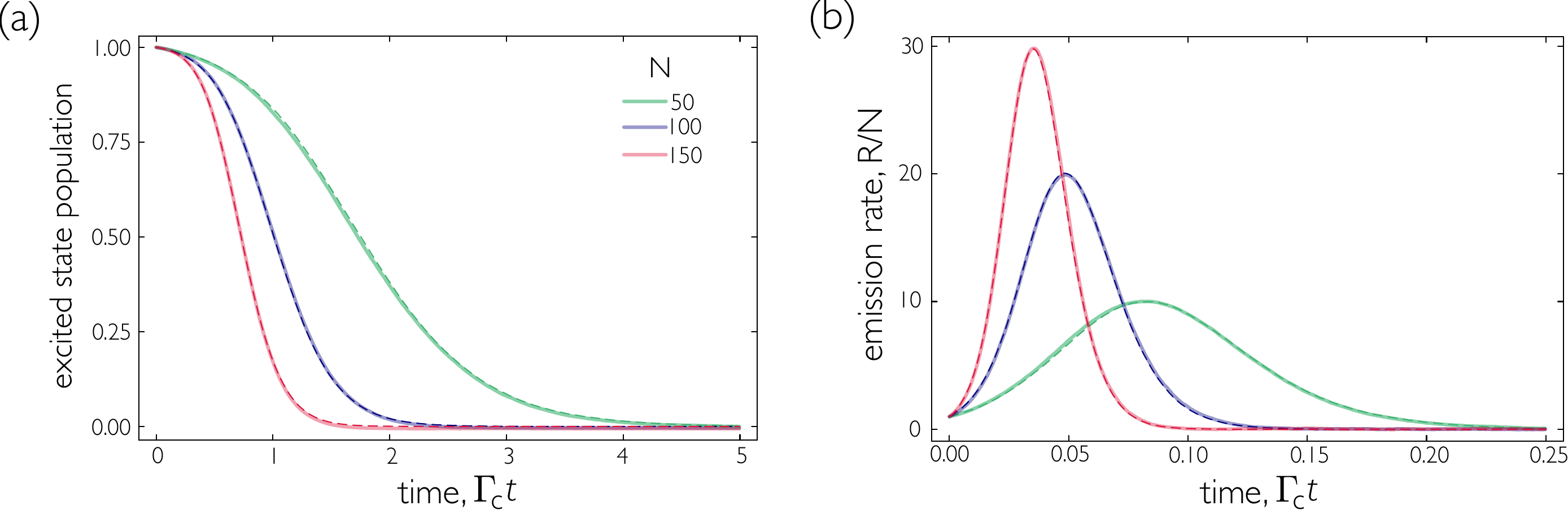}
    \caption{\textbf{Dicke superradiance in a cavity}: (a) Normalized total excited population $\sum_n\braket{\hat{\sigma}^n_{ee}
    }/N$ and (b) decay rate for $N$ atoms decaying collectively with a single bright jump operator obtained exactly (dashed lines), and using the dissipative TWA (solid lines) with $M_0=1000$ trajectories. }\label{fg:excited_population_dtwa_fleischauer_vs_dicke}
    \end{center}
\end{figure}
As discussed in Section \ref{section:spinModel}, inverted atoms decaying into a single-mode bad cavity develop spontaneous coherence during the decay process. We simulate this process initializing the system in the fully inverted state, $\rho(t=0)=\bigotimes_n\ket e\bra e$, and evolving up to a time $\Gamma_c t=5$, using both the dissipative TWA and exact evolution with the master equation. Figure~\ref{fg:excited_population_dtwa_fleischauer_vs_dicke} shows that the approximated dynamics given by Eqs.~\eqref{eq:FleischauerSDE} and \eqref{eq:FleischauerSDE2} is in excellent agreement with the exact evolution, capturing accurately the rapid decay of the population in the early stages of the evolution, as well as the burst of light resulting from it. Dicke superradiance populates only the $N+1$ states belonging to the Dicke ladder with highest angular momentum, $J=\frac{N}{2}$, so the success of the approximation in this example is in accordance with our understanding of the error, as discussed in Ref.~\cite{Mink23}.

\subsubsection{Coherent driving of a 1D atomic array in free space}
\begin{figure}[!ht]
   \begin{center}\includegraphics[width=1.0\linewidth]{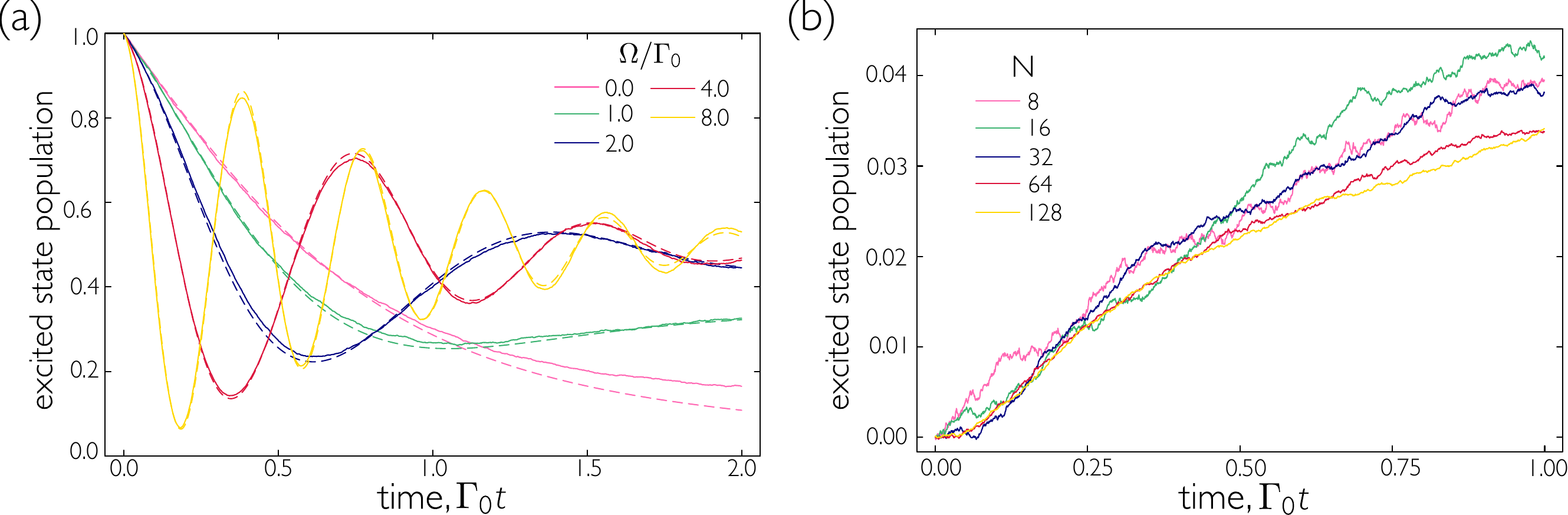}
    \caption{\textbf{Dissipative TWA evolution for a coherently driven 1D atomic array in free space}. (a)Excited state population $\sum_n\braket{\hat{\sigma}^n_{ee}}/N$ for a 1D array of $N=5$ atoms in free space with $d/\lambda_0=0.1$, polarization perpendicular to the chain, and different Rabi frequencies, calculated exactly by solving the master equation (dashed lines) and using the dissipative TWA (solid lines) with $M_0=1000$ trajectories. (b)~Evolution of excited state population if the system is initially in the ground state and $\Omega=0$, computed with the dissipative TWA and $M_0=1000$ trajectories.}   \label{fg:excited_population_dtwa_fleischauer_vs_free_space_n_4_pumping}
    \end{center}
\end{figure}
Next, we study a coherently-driven 1D array of atoms in free space, where the 
spin model is defined by the interaction coefficients in Eqs.~\eqref{GeneralFormJ} and~\eqref{GeneralFormG}. Using the approximation of Eq.~\eqref{eq:fleischauer_approx_moyal_definition}, the coherent pumping on the $n$-th atom contributes with the following term to the PDE and the subsequent SDE~\cite{Mink23},
\begin{equation}
    \begin{split}
        -\ii\left[\Omega_n \hat\sigma^n_{eg}+\Omega_n^* \hat\sigma^n_{ge},\hat\rho\right]\longmapsto &2\left[\nabla_{\theta_n}\text{Im}\left\{\Omega_n e^{i\phi_n}\right\}+\nabla_{\phi_n}\text{Re}\left\{\Omega_n e^{i\phi_n}\right\}\cot \theta_n\right]W_{\rho}(\bm \Omega)\\
        &\equiv \begin{cases}
        d\theta_n = -2\,\text{Im}\left\{\Omega_n e^{i\phi_n}\right\}dt,\\
        d\phi_n = -2\,\text{Re}\left\{\Omega_n e^{i\phi_n}\right\}\cot \theta_ndt.
    \end{cases}
    \end{split}
\end{equation}
This mapping coincides with the result obtained using the exact Moyal product in Eq.~\eqref{eq:moyal_product_spins_definition}, meaning that coherent pumping is treated exactly within this approximation. 

Figure~\ref{fg:excited_population_dtwa_fleischauer_vs_free_space_n_4_pumping}(a) shows the evolution of the excited-state population for different pumping strengths. The approximation performs well when the coherent pumping is strong, as a significant part of the evolution is captured exactly. However, it fails to accurately describe the system when the population remains low. This situation is further explored in Fig.~\ref{fg:excited_population_dtwa_fleischauer_vs_free_space_n_4_pumping}(b), where we initialize the system in the ground state, $\rho(t=0)=\bigotimes_n\ket g\bra g$, and set all Rabi frequencies to zero, so the system should remain de-excited. Instead, we observe an unphysical increase in the excited-state population -- an artifact of the approximation, which can destabilize fixed points of the exact dynamics, as we further explore in Section~\ref{Geometry}.

\subsubsection{Superradiant lasing in a cavity}
\begin{figure}[!ht]
    \begin{center}\includegraphics[width=1.0\linewidth]{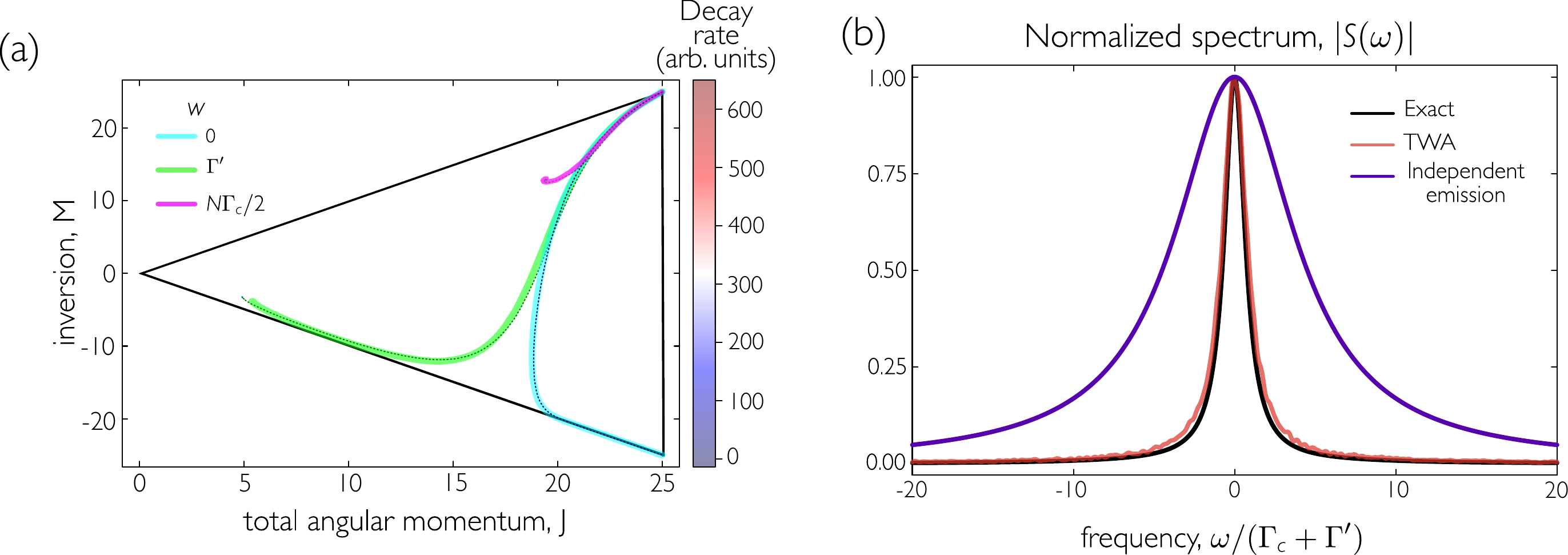}
\caption{\textbf{Superradiant lasing of 50 atoms in a bad cavity}. (a) Trajectories in the $(J,M)$ space for different pumping strengths $w$, calculated using an exact algebraic method~\cite{Xu13} (solid line) and dissipative TWA  with $M_0=2500$ trajectories (dashed). The two angular momentum numbers are defined as $M=\sum_n\langle\hat\sigma^n_{ee}\rangle - N/2$ and $J(J+1)=\langle\hat J^2\rangle$. The heatmap shows the collective decay rate into the cavity. (b) Normalized spectrum of the emitted light, for $w=25\Gamma_c$. In both plots, the decay rate into free space is twice that into the cavity, i.e., $\Gamma'=2\Gamma_c$. The total number of trajectories is $M_0=M_1=300$ for both the first and second evolution.}\label{fg:fleischauer_low_total_ang_momentum_error}
\end{center}
\end{figure}

To highlight our method for calculating multi-time correlators, we examine the spectral properties of the light emitted by a collection of $N$ atoms inside a bad cavity~\cite{Meiser09,Meiser10} that are driven incoherently. The atoms are subjected to parasitic decay and incoherent pumping at rates $\Gamma'$ and $w$, respectively, as shown in Fig.~\eqref{fg:some_sample_configurations}(b). Unlike Dicke superradiance, where decay only populates Dicke states with $J=\frac{N}{2}$, here the system is driven into states with lower angular momentum (see Fig.~\ref{fg:fleischauer_low_total_ang_momentum_error}). This occurs because individual decay and pumping break the full permutational symmetry that originally constrained the atoms to be in the maximum $J$ space. In particular, it is known that for $N\gg1$, there is a subradiant to superradiant phase transition at $\Gamma'=w$, and the system populates states with $J\sim 0$~\cite{Shankar21}.

Under the approximation of Eq.~\eqref{eq:fleischauer_approx_moyal_definition}, the individual decay and incoherent pumping terms for the $n$-th atom can be computed in the phase-space formalism. However, an exact mapping~\cite{Mink22} can be obtained by integrating by parts the action on the kernel as shown in Eq. \eqref{eq:integration_by_parts_wigner}, resulting in terms
\begin{equation}
\begin{aligned}
    \frac{\Gamma'}{2}\left(2\hge\ra\heg-\ra\heg\hge-\heg\hge\ra\right)\longmapsto&& \Gamma'\left[- \nabla_{\theta}\left(\cot\theta+\frac{\csc\theta}{\sqrt 3}\right)+\frac{1}{2}\nabla_{\phi}^2\left(1+2\cot^2\theta+\frac{2\cot\theta\csc\theta}{\sqrt 3}\right)\right]W_{\rho}(\bm\Omega), \\
    \frac{w}{2}\left(2\heg\ra\hge-\ra\hge\heg-\hge\heg\ra\right)\longmapsto&& w\left[- \nabla_{\theta}\left(\cot\theta-\frac{\csc\theta}{\sqrt 3}\right)+\frac{1}{2}\nabla_{\phi}^2\left(1+2\cot^2\theta-\frac{2\cot\theta\csc\theta}{\sqrt 3}\right)\right]W_{\rho}(\bm\Omega),
\end{aligned}
\end{equation}
for each atom.

Since the model exhibits full permutational symmetry at the level of the master equation, we can use algebraic methods for exactly simulating the system for large $N$~\cite{Xu13}. Figure~\ref{fg:fleischauer_low_total_ang_momentum_error}(a) compares the
trajectories in the $(J,M)$ space for three different choices of $(\Gamma',w)$, using both exact methods and the dissipative TWA. The approximation captures reasonably well the exact dynamics both qualitatively and quantitatively. For $w\in\left[\Gamma'+\Gamma_c,N\Gamma_c\right]$, the system reaches a bright steady state, yielding a continuous source of light known as the superradiant laser~\cite{Meiser09,Meiser10}. 

Due to collective decay, the system becomes robust against noise from the individual incoherent processes. This robustness is quantified by the narrow linewidth of the emitted light [as shown in the calculation of the spectrum, i.e., Eq.~\eqref{EqSpectrum}]. Specifically, in the limit $N\gg 1$, the linewidth approaches $\Gamma_c$, which can be order of magnitude narrower than state-of-the-art lasers when an ultra-narrow transition is used~\cite{Meiser09}. Figure~\ref{fg:fleischauer_low_total_ang_momentum_error}(b) compares the spectrum of a superradiant laser obtained using exact algebraic methods based on permutational symmetry~\cite{Xu13} with results from the dissipative TWA and its extension presented in Section~\ref{sc:multi_time_correlation}. The semiclassical approximation successfully captures the line-narrowing effect.

\subsection{Improvements and generalizations of the dissipative Truncated Wigner Approximation}\label{sc:beyond_dtwa}
The examples above seem to indicate that the approximation tends to perform well for large atom numbers, and for dynamics populating high angular momentum sectors. Nevertheless, this is far from a satisfying characterization of the error introduced by the approximation. Questions aimed at improving the approximation include identifying the origin of the error in more intuitive terms and determining whether there exist alternatives to the approximation in Eq.~\eqref{eq:fleischauer_approx_moyal_definition} that enable a simulation cost scaling polynomially with atom number, such as with the SDE system in Eqs.~\eqref{eq:FleischauerSDE} and~\eqref{eq:FleischauerSDE2}. Here we discuss these aspects.

\subsubsection{Geometric origin of the error}\label{Geometry}
We now provide a geometric interpretation of the error, by means of the projector onto the $\mathcal U$ [defined in Eq.~\eqref{projector}]. From the spin model in Eq.~\eqref{eq:master}, we observe that once mapped to phase space, all terms involve the application of two differential operators $\mathcal S^i_n$. Additionally, to compute any observable, we only need the portion of the Wigner distribution that belongs to  $\mathcal U$. 

\begin{figure}[!ht]
    \begin{center}
    \includegraphics[width = 0.6\textwidth]{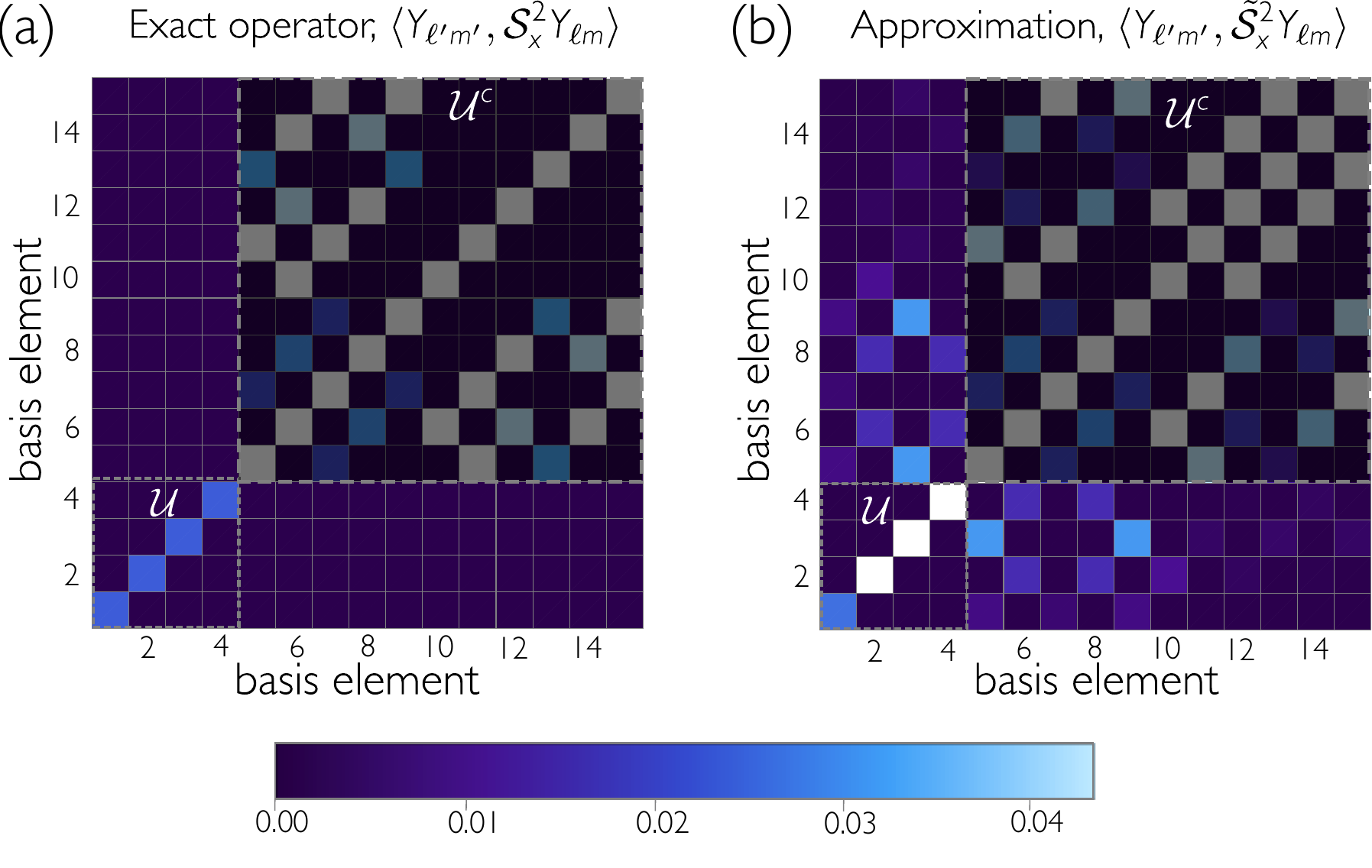}
    \caption{\textbf{Error in the approximated operators}.
    Matrix elements of the (a) exact $\mathcal S_x^2$ and (b) its approximation $\tilde{\mathcal S}_x^2$, as defined in Eq.~\eqref{eq:fleischauer_approx_moyal_definition}, in the first elements of the spherical harmonics basis. The exact operator is block diagonal, so $\mathcal U$, and its complement, $\mathcal U^c$, are completely decoupled. In contrast, the approximate operator introduces nonzero matrix elements between these two spaces, leading to population leakage from $\mathcal U$.}
    \label{fg:harmonic_leaking}
    \end{center}
\end{figure}

Therefore, to analyze the error in geometric terms, we study how two successive applications of the approximate differential operator, followed by a projection, act on the basis of $\mathcal U$ -- i.e., we examine $(P_{|\mathcal U}\circ\tilde{\bm{\mathcal S}}\circ\tilde{\bm{\mathcal S}})[Y_{\ell m}]$ with $\ell\leq 1$. After the application, $\tilde{\bm{\mathcal S}}\circ\tilde{\bm{\mathcal S}}[Y_{\ell m}]$ contains contributions both inside and outside $\mathcal U$, as illustrated in Fig.~\ref{fg:harmonic_leaking},  which compares the exact operator  ${\mathcal S^2_x}$ with its approximate counterpart $\tilde{{\mathcal S}}^2_x$. The components outside $\mathcal U$ generate nonzero contributions inside  $\mathcal U$ once the projection is applied a second time, introducing a source of error.

\begin{figure}[!ht]
    \begin{center}
    \includegraphics[width = 1\textwidth]{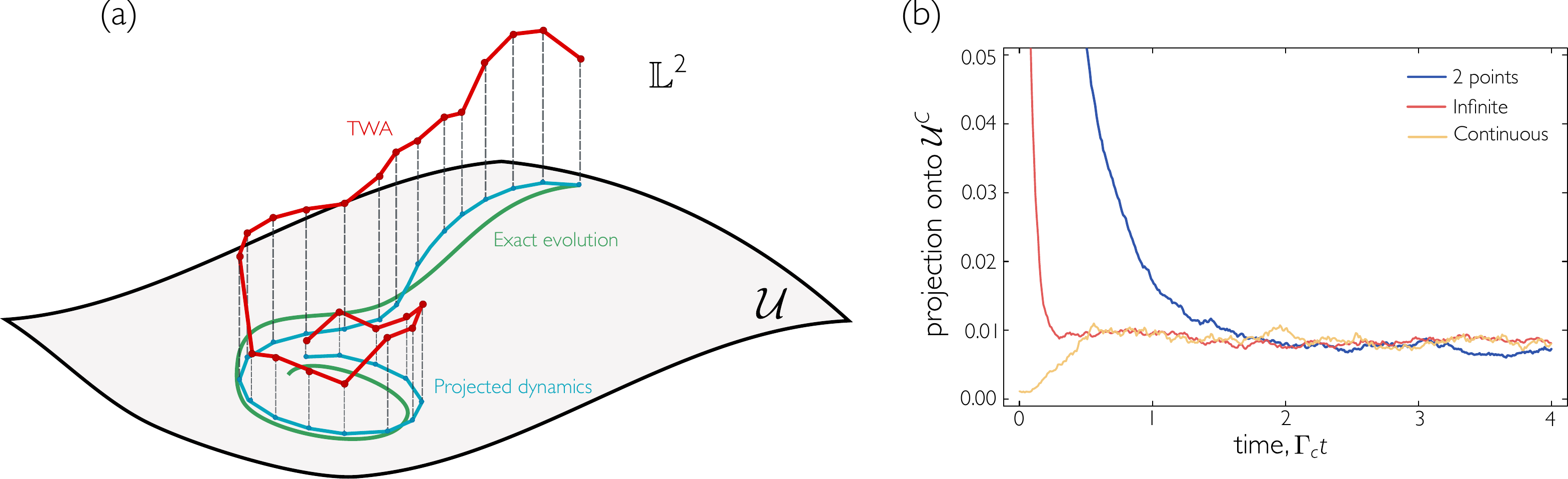}
    \caption{\textbf{Evolution out of $\mathcal U$}. (a) Sketch of the evolution set by the exact dynamics and by the dissipative TWA. (b) Evolution of the projection of the Wigner distribution onto the spherical harmonics with $1<\ell\leq L_{max} = 6$, i.e., $\sum_{\ell=2}^{L_{max}}\sum_{|m|\leq \ell}|\langle Y_{\ell m}(\Omega)\rangle|^2$, for a single initially excited atom in a cavity. Different plots belong to different sampling schemes with $M_0=10000$ trajectories.}
    \label{fg:manifold_sketch}
    \end{center}
\end{figure}

A similar phenomenon occurs for every term in the spin model. The approximated evolution described in Eq.~\eqref{eq:dtwa_pde} is thus not entirely contained in $\mathcal U$. At each time step, spherical harmonics $Y_{\ell m}$ with $\ell>1$ outside this space become populated, and the projection back to $\mathcal U$ introduces deviations from the exact evolution, which always remains within $\mathcal U$, as illustrated in Fig.~\ref{fg:manifold_sketch}(a). The accumulation of these errors leads to non-physical instabilities, such as the growth of excited state population without pumping shown in Fig.~\ref{fg:excited_population_dtwa_fleischauer_vs_free_space_n_4_pumping}(b). Additionally, if the initial distribution is sampled by exploiting the Wooters discrete representation, high-order $Y_{\ell m}$ are populated from the start, so the initial condition itself already lies outside $\mathcal U$.

A first approach to reducing the error is to develop a continuous sampling method that ensures the initial state remains strictly within
$\mathcal U$. This method is detailed in Appendix~\ref{sc:appendix_sampling_method_no_harmonics}. However, as Fig.~\ref{fg:manifold_sketch}(b) illustrates, numerical experiments suggest that this new sampling approach does not usually yield significant improvements compared to the two-point [Eq.~\eqref{examples_discrete}] and infinite sampling methods [Eq.~\eqref{eq:infinite_sampling}]. After a short transient, all three sampling approaches converge to the same level of population in $\mathcal U^C$.

\subsubsection{Uniqueness of the approximation}
Equipped with this geometric understanding, we now ask whether an improved approximation of Eq.~\eqref{eq:fleischauer_approx_moyal_definition} exists -- one that enables efficient numerical simulations in the form of a system of SDEs. Such an approximation must: 
\begin{itemize}
\item Yield a FPE when applied to the spin model in Eq.~\eqref{eq:master}.

\item Satisfy the condition in Eq.~\eqref{eq:dtwa_moyal_exact_projection}, i.e.,
$(P_{|\mathcal U}\circ \tilde{\bm{{\mathcal S}}} )[Y_{\ell m}] =  \bm{\mathcal S} [Y_{\ell m}]$. This guarantees that the action of all operators is exact after we project back to $\mathcal U$.
\end{itemize}

Since each term in the full quantum evolution involves the application of two differential operators, the approximate Moyal product must be further restricted to containing only first-order derivatives. Under these constraints, we propose an approximation of the form
\begin{equation}\label{eq:generalized_approx_alpha_beta}
    \tilde{\bm{\mathcal S}}=\bm s+\bm\alpha\frac{\partial}{\partial\theta}+\bm\beta\frac{\partial}{\partial\phi}+\bm L
\end{equation} 
with parameters $\alpha^i, \beta^i\in L^2(\mathbb S^2)$ for $i=\{x,y,z\}$. The term $\bm s=\sqrt 3 \bm m(\theta,\phi)$ is included to ensure that the approximation correctly reproduces the action of the exact Moyal product [Eq.~\eqref{eq:moyal_product_spins_definition}] on the Weyl symbol of the identity, $W_{\mathbb I} = 1$. The angular momentum term $\bm L$ is included separately, as it appears explicitly in the exact expression in Eq.~\eqref{eq:moyal_product_spins_definition}.

The coefficients $\bm\alpha, \bm\beta$ thus constitute free parameters. The goal is to determine values for these parameters that yield a second-order PDE with a positive diffusion matrix. If such solutions exist, one could then identify the approximation that minimizes the error. Because $\bm\alpha, \bm\beta$ belong to a large functional space, we rely on approximations to find solutions. By truncating the functional space $L^2(\mathbb S^2)$ to $\text{span}\{Y_{\ell m}\}_{\ell<\ell_{max},|m|<\ell}$, we approximate the coefficients as
$$\bm \alpha \approx \sum_{\ell<\ell_{max},|m|\leq \ell}\bm\alpha_{\ell m}Y_{\ell m},\quad \bm \beta \approx \sum_{\ell<\ell_{max},|m|\leq \ell}\bm\beta_{\ell m}Y_{\ell m},$$ 
with $\bm \alpha_{\ell m}, \bm \beta_{\ell m}\in\mathbb C$. This truncation further justifies including $\bm L$ explicitly in the approximation: from the definition of the angular momentum operator in Eq.~\eqref{AngularOpDef}, we see that $\bm L$ inherently contains higher-order spherical harmonics. A low-order truncation approximation would therefore fail to capture the $\bm L$ term appearing in the exact Moyal product expression.

The solution for the coefficients is $\bm\alpha=\bm\beta=\bm 0$ for any dimension of truncation $\ell_{max}$, corresponds to the dissipative TWA \eqref{eq:fleischauer_approx_moyal_definition}. Solutions with $\bm\alpha, \bm\beta\neq  0$ define alternative approximations that lead to a second-order PDE for the Wigner distribution. However, as detailed in Appendix \ref{sc:generalized_DTWA}, all the alternatives for low truncation order $\ell_{max}$ produce a PDE with a diffusion matrix that is  not positive semidefinite, thus preventing a reformulation of the problem in terms of a system of SDEs. Therefore, we conclude that the FPE in Eq.~\eqref{eq:dtwa_pde} is, in a sense, the only available approximation in the Wigner representation that minimally uses the basis of $\mathcal U$.

\subsubsection{Generalization to $P$ and $Q$ representations}\label{PQGeneralization}
Finally, a third potential avenue for improvement is to explore whether there is an alternative approximation for the Moyal product that transform the fundamental PDE governing the evolution of the $P$ and $Q$ representations [Eq.~\eqref{eq:exact_pde}] to a FPE. To identify such an approximation, we follow the same approach as with the Wigner function: we propose a form for the Moyal product with free parameters and then determine all solutions for these free functions that satisfy the condition in Eq.~\eqref{eq:dtwa_moyal_exact_projection}. The full details of this procedure are covered in Appendix~\ref{sc:generalized_DTWA}.

From the exact Moyal products for these representations, we observe that the direct analogue of Eq.~\eqref{eq:fleischauer_approx_moyal_definition},
\begin{equation}
    \tilde{\bm{\mathcal S}}_s\simeq\bm s_s+\bm L,
\end{equation}
with $\bm s_s = -(-\sqrt 3)^{1+s}\bm m(\theta,\phi)=\text{Tr}\{\hat{\bm\sigma}\cdot\hat\Delta(\Omega,s)\}$ being the Weyl symbol of the Pauli matrices in any representation, does not satisfy the condition of exactness after projection back into $\mathcal U$ [Eq.~\eqref{eq:dtwa_moyal_exact_projection}]. Therefore, we propose the alternative approximation
\begin{equation}
    \tilde{\bm{\mathcal S}}_s=\bm s_s+\bm\alpha_s\frac{\partial}{\partial\theta}+\bm\beta_s\frac{\partial}{\partial\phi}+\bm L,
\end{equation} 
and rely on the coefficients $\bm \alpha_s,\bm \beta_s$ to obtain a valid expression for the operator. As before, even with truncation, the resulting equations for $\bm \alpha_s$, $\bm \beta_s$ contain many degrees of freedom. We thus employ numerical approaches to find solutions and verify whether the resulting PDE has a positive definite diffusion matrix. Similarly to the Wigner case, we find that no approximation with $\bm\alpha_s, \bm\beta_s\neq  0$ exists that both satisfies condition~\eqref{eq:dtwa_moyal_exact_projection} and results in a valid FPE.  

The analysis in this and the previous subsection provides evidence suggesting that the approximation of the Moyal product~\eqref{eq:fleischauer_approx_moyal_definition} and the resulting FPE~\eqref{eq:dtwa_pde} represent the only available approximation that enables efficient simulations of collective dissipative dynamics in phase space when restricting the basis. As a result, the dissipative TWA introduced in Ref.~\cite{Mink23} remains the only case where the approximation has no errors in the first time step of the evolution and produces a PDE with a positive semidefinite diffusion matrix. However, we emphasize that this analysis does not constitute a formal proof that no alternative FPE exists beyond Eq.~\eqref{eq:dtwa_pde}, as we have only explored low-truncation examples. A more extensive investigation would require exploring truncations with higher $\ell_\text{max}$, finding alternative basis, and developing a deeper understanding of how different terms in the approximate Moyal product contribute to the non-positivity of the resulting diffusion matrix. Additionally, this analysis does not rule out the possibility of better-suited methods, such as approximating the Moyal product for angular momentum with $j>1/2$~\cite{Zueco07} to derive a FPE in terms of collective angular momenta. Yet another alternative, which we explore in the next section, is to introduce new representations that further exploit the non-uniqueness of the phase-space representation, and force the evolution to be described by a FPE. 

\section{Positive $P$ representation}\label{sc:generalized_p_repre}
As seen in the previous section, numerical simulations of dissipative spin dynamics using the Wigner representation effectively capture the dynamics that populate high-cooperativity states. However, this approach suffers from instabilities, such as the unphysical growth of the ground state population in the absence of pumping, and no bound of the error made by the approximation is known for a arbitrary states. Moreover, extending the FPE treatment to other representations is not straightforward, as the formalism leads to non-positive semidefinite diffusion matrices.

The latter issue has previously arisen in the context of bosonic systems, motivating the development of generalized representations such as the positive $P$ representation. This approach exploits the overcompleteness of coherent operators to construct an alternative expansion that ensures a positive semidefinite diffusion matrix in the resulting FPE, at the cost of doubling the number of phase-space variables. In this section, we apply the positive $P$ representation for $N$ atoms, developed in Ref.~\cite{Ng13} to the study of coherent dynamics. As in its predecessor in bosonic systems, the PDE governing the evolution of the spin positive $P$ distribution directly takes the form of a FPE. Consequently, there are no ambiguities or additional complications in transforming the evolution into a system of SDEs for efficient simulation. However, the derivation of the PDE implicitly assumes that the population near the phase-space boundary is negligible. As we have discussed repeatedly throughout this paper, there is no physical justification for this assumption for a general case. If the dynamics is such that the distribution at the boundaries becomes non negligible, the Positive $P$ SDE system exhibits divergent trajectories, leading to ``spikes'' that distort observable averages.

\subsection{Positive $P$ for bosonic systems}
To illustrate the underlying philosophy of the positive $P$ representation, we first discuss its version for bosonic systems by considering a two-photon absorption process~\cite{Chaturvedi77}. The master equation for this system is given by
\begin{equation}\label{two photon process}
    \dot{\hat \rho}=-[\epsilon^*\hat a-\epsilon \hat a^{\dagger},\hat \rho] + \frac{\Gamma}{2}(2\hat a^2\rho \hat a^{2\dagger} - \hat a^2 \hat a^{2\dagger}\rho - \rho  \hat a^2 \hat a^{2\dagger}),
\end{equation}
where $\Gamma$ is the photon absorption rate, and we assume the mode is coherently pumped at rate $\epsilon$. Mapping this equation to the $P$ representation, following the steps in Section~\ref{sc:harmonic_osc_introduction}, yields the PDE
\begin{equation}
\begin{split}\label{P two photon}
    \frac{\partial}{\partial t}P(\alpha) =& \left(-\frac{\partial}{\partial\alpha}(\epsilon^*-\Gamma\alpha^*\alpha^2) + \frac{1}{2}\frac{\partial^2}{\partial\alpha^2}(-\Gamma\alpha^2) + \text{h.c.}\right)P(\alpha).
\end{split}
\end{equation}
It is immediately evident that the diffusion matrix in the above equation is not positive semidefinite. Consequently, the standard mapping to SDEs, which enables, for example, the exact simulation of damping [Eq.~\eqref{CorrectSystem}], is not feasible. Still, naively translating the PDE in Eq.~\eqref{P two photon} to SDEs yields
\begin{equation}\label{SDEs Two photon}
    \begin{cases}
        d\alpha = (\epsilon^* - \Gamma\alpha^*\alpha^2)dt + \ii\sqrt{\Gamma}\alpha dW_1,\\
        d\alpha^* = (\epsilon - \Gamma\alpha\alpha^{*2})dt - \ii\sqrt{\Gamma}\alpha^* dW_2.\\
    \end{cases}
\end{equation}
This system of SDEs is problematic because evolution does not preserve the conjugacy relation, i.e., $(\alpha(t))^*\neq \alpha^*(t)$, since the Wiener increments $dW_1$, $dW_2$ are not related to each other. To resolve this issue, $\alpha^*$ can be promoted to a new independent variable, $\beta$. 

Following this idea, Drummond and Gardiner~\cite{Drummond80_2} exploited the overcompleteness of coherent states to introduce a generalization of the $P$ representation kernel from Eq.~\eqref{Kernel P}, i.e.,
\begin{equation}\label{kernel positive P}
    \hat\Delta(\alpha,\beta)= \frac{\ket\alpha \bra{\beta^*}}{\braket{\beta^*|\alpha}}=e^{-\alpha\beta}e^{\alpha \hat{a}^{\dagger}}\ket 0 \bra 0e^{\beta \hat{a}}.
\end{equation}
It can be shown~\cite{Drummond80_2} that all states can be expanded in terms of this kernel as 
\begin{equation}\label{Positive P}
    \rho = \iint d\mu(\alpha,\beta) P(\alpha,\beta)\hat\Delta(\alpha,\beta),
\end{equation}
where $P(\alpha,\beta)$ is a generalized $P$ representation and  $d\mu(\alpha,\beta)$ is the integration measure. Specifically, if we choose $d\mu(\alpha,\beta) = d^2\alpha \ \delta(\alpha-\beta^*)$, Eq.~\eqref{Positive P} reduces to the standard $P$ representation. Alternatively, selecting  $d\mu(\alpha,\beta) = d^2\alpha d^2\beta$ effectively promotes $\beta$ to an independent variable, extending the phase space to a double complex plane, $X=\mathbb C^2$. This generalized $P$ representation is called the positive $P$ because, using the correspondence rule in Eq.~\eqref{eq:CorrespondenceRuleKernel}, one can show that
\begin{equation}
    P(\alpha,\beta)=\frac{1}{4\pi^2}e^{-\frac{|\alpha-\beta^*|^2}{4}}\biggl< \frac{\alpha+\beta^*}{2}\bigg|\rho\bigg|\frac{\alpha+\beta^*}{2}\biggr>,
\end{equation}
and thus $P(\alpha,\beta)$ is guaranteed to be real and positive for any quantum state $\hat\rho$. The positive $P$ representation is another example of the non-uniqueness of the representation in phase space. Given the overcompleteness of coherent states, even the Positive $P$ representation of a state is not unique -- an observation that can be useful for sampling.  For example, if the $P$ representation of a state is $P(\alpha)$, then one possible positive $P$  representation for the same state is $P(\alpha,\beta)=P(\alpha)\delta(\alpha-\beta^*)$.

The derivation of the evolution equation for the positive $P$ corresponding to the master equation~\eqref{two photon process} follows the standard procedure outlined in Section~\ref{sc:harmonic_osc_introduction}. This involves translating the action of the creation and annihilation operators on the kernel into differential operators, which are then transferred to $P(\alpha,\beta)$ by integrating by parts. The action of $\hat a$, $\hat a^\dagger$ on the kernels is given by     
\begin{equation}
\hat a^\dagger\hat\Delta(\alpha,\beta) =\left(\frac{\partial}{\partial\alpha}+\beta\right)\hat\Delta(\alpha,\beta),\ \
\hat\Delta(\alpha,\beta)\hat a=\left(\frac{\partial}{\partial\beta}+\alpha\right)\hat\Delta(\alpha,\beta).
\end{equation}
These are equivalent to the identities in Eq.~\eqref{Derivatives} for the $P$ representation if we replace $\beta\longmapsto \alpha^*$, which is exactly the replacement suggested by the naive mapping from Eq.~\eqref{P two photon} to a SDE system. Hence, the SDE system for the positive $P$ representation is 
\begin{equation}\label{eq:two_photon_abs_positive_p_sde}
    \begin{cases}
        d\alpha = (\epsilon^* - \Gamma\beta\alpha^2)dt + \ii\sqrt{\Gamma}\alpha dW_1,\\
        d\beta = (\epsilon - \Gamma\alpha\beta^{2})dt - \ii\sqrt{\Gamma}\beta dW_2.\\
    \end{cases}
\end{equation}
Here, $\alpha$ and $\beta$ are no longer constrained to be complex conjugates. After solving the evolution, the expectation value of any observable is obtained using the correspondence rules from Section~\ref{Section PhaseSpace},
\begin{equation}
    \langle \hat O(t)\rangle =\iint d^2\alpha\ d^2\beta \ P(\alpha,\beta)\text{Tr}\{\hat O\hat\Delta(\alpha,\beta)\}.
\end{equation}

Equation~\eqref{eq:two_photon_abs_positive_p_sde} correctly reproduces the evolution of the exact master equation~\eqref{two photon process}\cite{Chaturvedi77}. In general, if the $P$ representation satisfies a second-order evolution equation, the Positive $P$ is guaranteed to evolve according to a FPE~\cite{Drummond80_2}, enabling efficient simulation via SDEs. Due to this property, the positive $P$ has been widely applied to quantum optics problems with non-positive diffusion matrices. Moreover, the Positive $P$ representation allows straightforward sampling of any initial state. 

Since no terms are neglected to obtain the FPE, the dynamics in this representation appears to be captured \emph{exactly} by simulations whose complexity scales linearly. This raises a natural question: where is the quantumness hidden? It seems highly unlikely that any physical evolution of a bosonic mode, starting from an arbitrary state, can be successfully captured by semiclassical-like simulations. The answer lies in that the FPE is not always equivalent to its master equation counterpart, as boundary terms arising from integration by parts were implicitly neglected. As a result, the FPE remains equivalent to the master equation only if $P(\alpha,\beta)$ decays sufficiently fast at the boundaries. This approximation is reasonable in bosonic systems when damping effects dominate over nonlinearities~\cite{Gilchrist97}. However, when damping and nonlinearities are of comparable strength, applying the Positive $P$ representation can lead to incorrect results~\cite{Smith89,Schack91}. At the level of the SDE dynamics, the breakdown of the approximation manifests as diverging trajectories that escape to infinity in a finite time. Analyzing the drift terms of the evolution and identifying diverging trajectories can therefore provide an estimate of when simulations using the Positive $P$ representation become unreliable~\cite{Gilchrist97}.

\subsection{Spin positive $P$ representation}
Here, we discuss the generalization of the positive $P$ representation for spins~\cite{Ng13}. We begin with a single spin-$1/2$, employing spin coherent states defined as
\begin{equation}
    \ket z = e^{-\frac{z}{2}e^z\hat \sigma^+}\ket g = \begin{pmatrix}
        e^{\frac{z}{2}}\\
        e^{-\frac{z}{2}}
    \end{pmatrix},
\end{equation}
with $z\in\mathbb C$. This definition is equivalent to that of Eq.~\eqref{eq:spin_coherent_state}, under the substitution $e^z\mapsto \tan(\theta/2)e^{\ii\phi}$ along with an appropriate change in the normalization factor. Due to the periodicity in $\phi$, the  new variable $z$ exhibits discrete translational symmetry along the imaginary axis. Specifically, all regions in $z$-space of the form $\{x+\ii(y+2k\pi) \ | \ x\in \mathbb R, y\in (-\pi,\pi)\}$ with $k\in\mathbb Z$ are equivalent. Furthermore, the boundary of the parametrization of the sphere [Eq.~\eqref{Boundary sphere}] is mapped to the complex infinity.

\begin{figure}[!ht]
    \begin{center}
    \includegraphics[width = 0.6\textwidth]{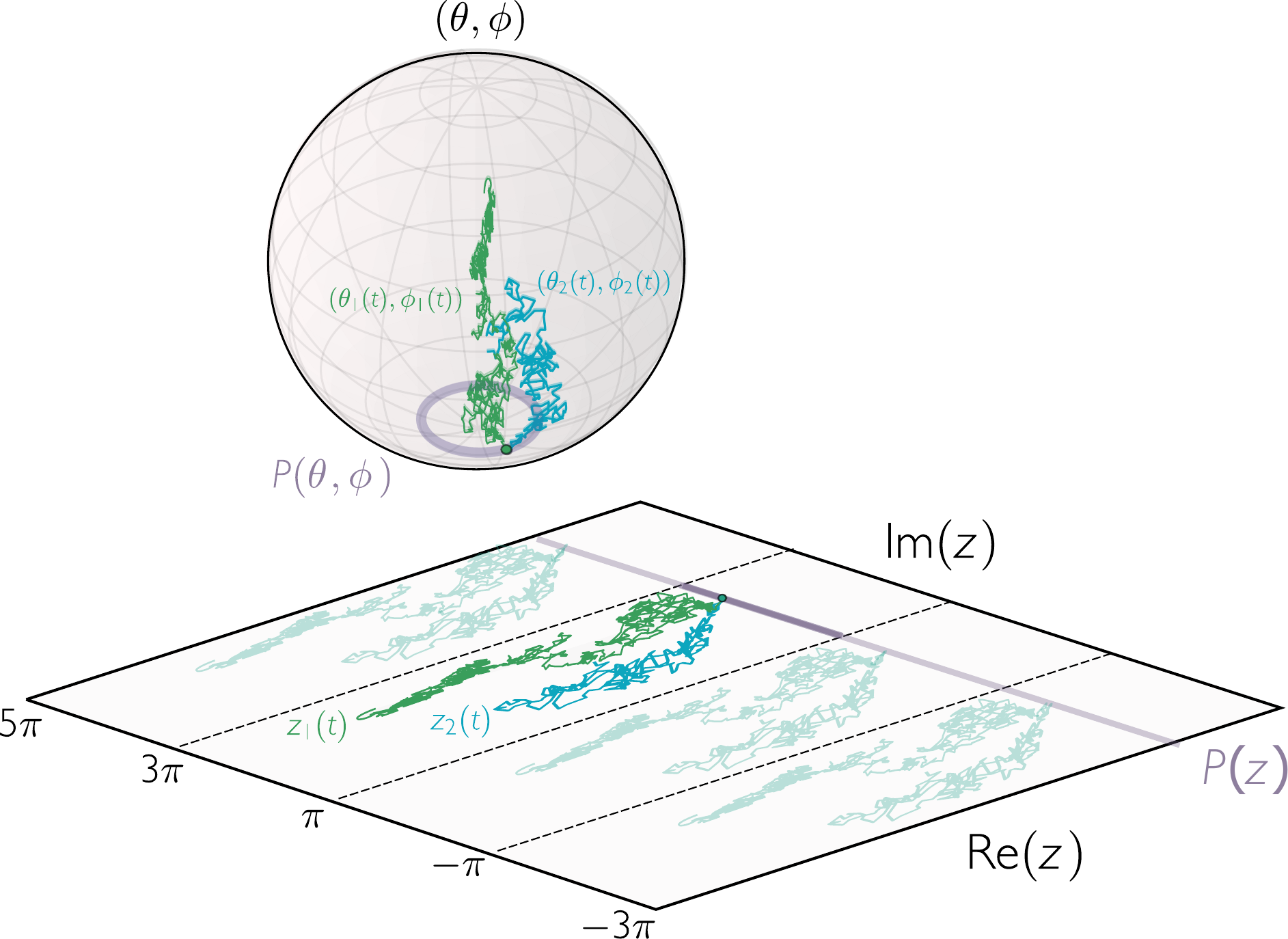}
\caption{\textbf{Schematics of two available phase spaces for the positive P.} A single run of the SDE system simulation produces two trajectories, either 
    $(\theta_1(t),\phi_1(t),\theta_2(t),\phi_2(t))$ on the sphere or $(z_1(t),z_2(t))$ in z-space.}\label{fg:sketch_stereographic_projection}
    \end{center}
\end{figure}

In analogy with the kernel definition for the bosonic Positive $P$ representation [Eq.~\eqref{kernel positive P}], we define the kernel for the spin system as
\begin{equation}\label{KernelPositivePSpin}
    \hat\Delta(z_1,z_2) =\frac{\ket{z_1}\bra{z_2^*}}{\braket{z_2^*|z_1}} = \begin{pmatrix}
        1-\frac{1}{1+e^{z_1+z_2}} & \frac{1}{e^{-z_1}+e^{z_2}}\\
        \frac{1}{e^{-z_2}+e^{z_1}} & \frac{1}{1+e^{z_1+z_2}}
    \end{pmatrix},
\end{equation}
where $z_1,z_2\in \mathbb C$. The positive $P$ representation of any state can now be systematically obtained using the correspondence rule in Eq.~\eqref{eq:CorrespondenceRuleKernel}. Alternatively, harnessing the gauge freedom in the phase-space formalism, the excited and ground states can be expressed as the probability distributions

\begin{equation}\label{eq:Pexcited}
    P(z_1,z_2) = \frac{1}{2\pi}\delta (z_1 - z_2) \delta(\text{Re}\{z_1\}-x_0)\ \mathbb 1_{|\text{Im}\{z_1\}|< \pi}(z_1),
\end{equation}
where $x_0$ is an arbitrary real number satisfying $x_0> 0$ for $\ket e$ and $x_0< 0 $ for $\ket g$.  These representations correspond to sampling $z_1=z_2$ within the segments 
$\{\pm x_0 + \ii y\ | \ y \in (-\pi,\pi) \}$. Although the choice of $x_0$ is irrelevant for the theoretical description of the state, the fidelity of a reconstructed state from a finite sample set improves with increasing $x_0$, which can be easily seen by inspecting the diagonal entries of Eq.~\eqref{KernelPositivePSpin}. Therefore, for numerical simulations, it is advantageous to select a large $x_0$. Besides, we note this positive $P$ representation is one of (infinitely) many that exist for $\ket e$ and $\ket g$ due to the non-uniqueness mentioned earlier in the section. The dipheomorphism $z\longleftrightarrow(\theta,\phi)$ and the sampling of the excited state are illustrated in Fig.~\ref{fg:sketch_stereographic_projection}. Because the positive $P$ effectively doubles the phase space to $X=\mathbb C^2$, a single run of the SDEs corresponds to a pair of trajectories, either $(z_1(t),z_2(t))$ or $(\theta_1(t),\phi_1(t),\theta_2(t),\phi_2(t))$.

The extension to $N$ spins follows straightforwardly by defining the phase space $(\bm z_1,\bm z_2)\in X=\mathbb C^{2N}$ and introducing the kernel operators
\begin{equation}\label{kernel N spins positive P}
    \hat\Delta(\bm z_1,\bm z_2) = \bigotimes_{n=1}^N\hat\Delta(z_1^n,z_2^n),
\end{equation}
where $z_i=(z_i^1,\dots,z_i^N)$ for $i=\{1,2\}$, and $\hat\Delta(z_1^n,z_2^n)$ denotes the kernel operator in Eq.~\eqref{KernelPositivePSpin} for the $n$-th spin.

\subsection{Many-body quantum optics in the Positive $P$ representation}

The mapping of the spin model in Eq.~\eqref{eq:master} to a PDE in phase space $\mathbb C^{2N}$ follows the same steps and assumptions as its bosonic counterpart: products of Pauli matrices and the kernel can be rewritten as differential operators acting on the kernel~\cite{Ng13}, allowing the entire master equation to be expressed as
\begin{equation}
    \mathcal L[\hat\Delta(\bm z_1,\bm z_2)] = D^{(+)}[\hat\Delta(\bm z_1,\bm z_2)],
\end{equation}
where $D^{(+)}$ is a differential operator. By integrating by parts and neglecting boundary terms, $D^{(+)}$ is then transferred to the distribution function $ P(\bm z_1,\bm z_2)$, leading to the final equation governing the dynamics of a system of $N$ atoms coupled to a shared bath,
\begin{align}\label{eq positive P spins}
        \dot P(\bm z_1,\bm z_2) =& \left( -\mathcal L_1 + \frac{1}{2}\mathcal L_2\right)P(\bm z_1,\bm z_2), \text{ with }\notag\\
        \mathcal L_1 =& -\ii\sum_{n\neq m}\left(\frac{\partial}{\partial z_1^n}\frac{J_{nm}\sinh A_{nm}-\ii\frac{\Gamma_{nm}}{2}\cosh A_{nm}}{\cosh C_{m}} + \frac{\partial}{\partial z_2^n}\frac{J_{nm}\sinh B_{nm}-\ii\frac{\Gamma_{nm}}{2}\cosh B_{nm}}{\cosh C_{m}} \right)\notag\\
        &+\ii\sum_n\frac{\partial}{\partial z_1^n}\left(-\Omega_ne^{-z_1^n}+\Omega_n^*e^{z_1^n}\right) +\frac{\partial}{\partial z_2^n}\left(-\Omega_ne^{z_2^n}+\Omega_n^*e^{-z_2^n}\right)\notag \\
        &-\sum_n (\Gamma_{nn}-w)\left(\frac{\partial}{\partial z_1^n}+\frac{\partial}{\partial z_2^n}\right),\notag\\
        \mathcal L_2 =& -2\ii\sum_{n\neq m}\left(g_{nm}\frac{\partial^2}{\partial z_1^n\partial z_1^m}\cosh(z_1^n-z_1^m)-g_{nm}^*\frac{\partial^2}{\partial z_2^n\partial z_2^m}\cosh(z_2^n-z_2^m)\right)\notag\\
        &+\sum_{n,m} \left((\Gamma_{nn}+w)\delta_{nm}\left(\frac{\partial^2}{\partial z_1^n\partial z_1^n}+\frac{\partial^2}{\partial z_2^n\partial z_2^n})\right) + 2\frac{\partial^2}{\partial z_1^n\partial z_2^m}\left(\Gamma_{nm}e^{z_1^n+z_2^m} + w\delta_{nm}e^{-(z_n^1+z_n^2)}\right)\right),
\end{align}
where $A_{nm} = (z_1^m-z_2^m)/2-z_1^n$, $B_{nm} = (z_1^m-z_2^m)/2+z_2^n$, $C_m = (z_1^m + z_2^m)/2$, and $g_{nm} = -J_{nm} + \ii\frac{\Gamma_{nm}}{2}$, $g_{nm}^* = -J_{nm} - \ii\frac{\Gamma_{nm}}{2}$. We have also included in the above equation the terms produced by coherent and incoherent pumping, and individual parasitic decay [i.e., Eqs.~\eqref{CohPumping}-\eqref{InCohPumping}].

Just as in the bosonic Positive $P$, Eq.~\eqref{eq positive P spins} is formally equivalent to the spin model only if the distribution vanishes at the boundary $|z_1^n|,|z_2^n|\rightarrow\infty$, a condition whose validity depends heavily on the system's specific dynamics. Under this assumption, the dynamics is exactly mapped to FPE with diffusion matrix
\begin{equation}
    \bm D = \left(\begin{array}{c|c}
        -2\ii g_{nm}\cosh(z_1^n-z_1^m) + w\delta_{nm} & \Gamma_{nm}e^{z_1^n+z_2^m}+w\delta_{nm}e^{-(z_1^n+z_2^n)} \\
        \hline
        \Gamma_{nm}e^{z_2^n+z_1^m}+w\delta_{nm}e^{-(z_1^n+z_2^n)} & -2\ii g_{nm}\cosh(z_2^n-z_2^m) + w\delta_{nm}
    \end{array}\right).
\end{equation}
Since $\bm D$  is complex symmetric, it admits the decomposition $\bm D  = \bm B \cdot \bm B^T$, ensured by the Takagi factorization~\cite{Horn85}. However, there is no closed-form expression for $\bm B$, meaning that computational simulations based on SDEs require a numerical decomposition at each time step. 

\subsection{Numerical examples}
\begin{figure}[!ht]
    \begin{center}
    \includegraphics[width = 1\textwidth]{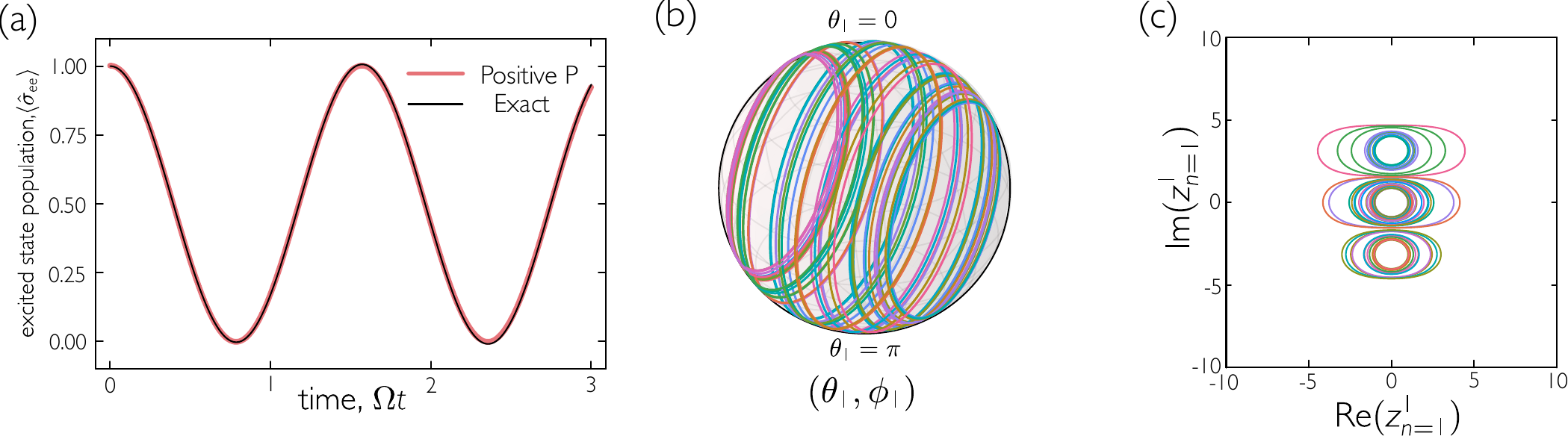}
    \caption{\textbf{Evolution of a coherently-driven single atom in the positive P formalism.} (a) Excited-state population as a function of time, calculated using both the exact master equation and positive $P$ simulations with a total of $M_0=100$ trajectories. Panels~(b) and~(c) show sample trajectories in the Bloch sphere and complex plane for the variables $(\theta_1,\phi_1)$ and $z_1$, respectively [the behavior of the variables  $(\theta_2,\phi_2)$ and $z_2$ is identical].}
    \label{fg:positive_p_N_1}
    \end{center}
\end{figure}

To illustrate the performance of the positive $P$ and gain insight into the conditions under which the approximation remains valid, we present three minimal numerical examples. First, we consider a single atom with zero decay rate that is coherently pumped with Rabi frequency $\Omega$. Figure~\ref{fg:positive_p_N_1}(a) shows that positive $P$ simulations match the exact result perfectly. This agreement can be understood by analyzing the dynamics in $z$-phase space, shown in Figs.~\ref{fg:positive_p_N_1}(b) and~(c). The coherent pumping induces a purely deterministic evolution given by 
\begin{equation}
   \frac{d z_j^1}{dt}=\ii \Omega e^{-z_j^1}-\ii \Omega^* e^{z_j^1}, \quad j=\{1,2\}\implies z_j^1(t)=\ii\arg\Omega + \text{arctanh}\left[e^{c_j\mp \ii|\Omega|t}\right],
\end{equation}    
where $c_j$ are constants determined by the initial conditions. Since the trajectories are periodic and bounded, contributions at the phase-space boundary are prevented. Consequently, the FPE remains an accurate representation of the dynamics described by the master equation.

\begin{figure}[!ht]
    \begin{center}
    \includegraphics[width = 1\textwidth]{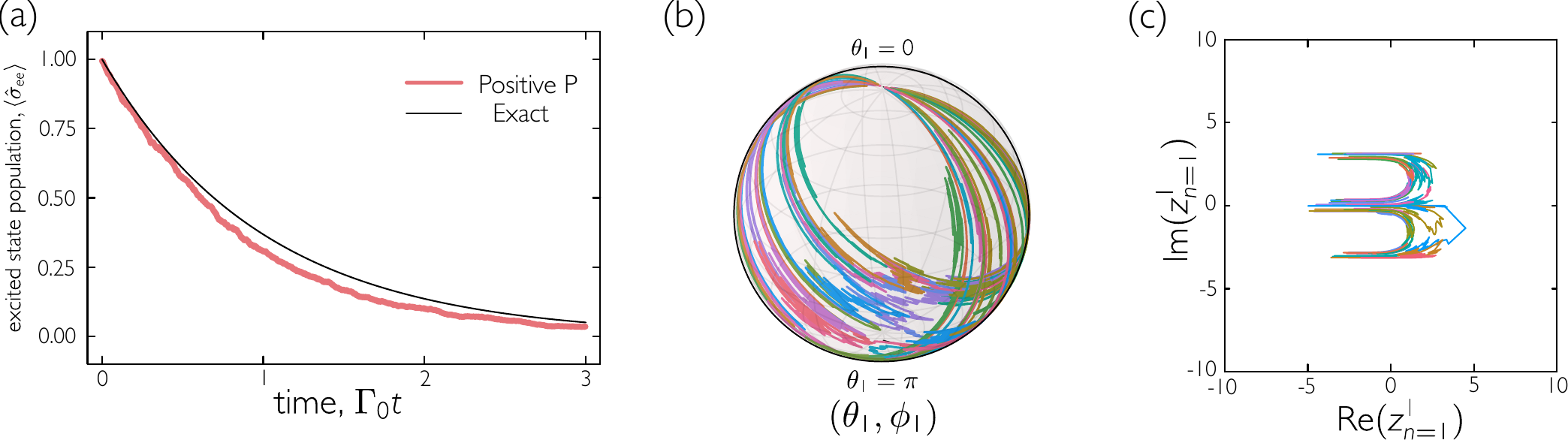}
    \caption{\textbf{Decay of single atom in the positive P formalism.} (a) Evolution of the excited-state population for a single atom decaying with $M_0=200$ trajectories. Pannels (b) and (c) show different sample trajectories in the Bloch sphere and complex plane, respectively.}
\label{fg:positive_p_decay_N_1}
\end{center}
\end{figure}

In the second example, shown in Fig.~\ref{fg:positive_p_decay_N_1}, we consider a single excited atom decaying at a rate $\Gamma_0$ without pumping. The evolution is now stochastic, and the drift term reads
\begin{equation}
    \left(\frac{d z_j}{dt}\right)_{\text{drift}}=-\Gamma_0 t, \quad j=\{1,2\}.
\end{equation}
This causes trajectories in phase space to evolve towards $\text{Re}(z_1) \rightarrow -\infty$ at long times [Fig.~\ref{fg:positive_p_decay_N_1}(c)], which is equivalent to approaching the north pole of $\mathbb{S}^2$ [Fig.~\ref{fg:positive_p_decay_N_1}(b)]. As the stochastic trajectories reach the boundaries, the SDE system is no longer equivalent to the exact master equation, leading to evident discrepancies between the positive $P$ simulation and the exact result, as seen in Fig.~\ref{fg:positive_p_decay_N_1}(a).

\begin{figure}[!ht]
    \begin{center}
    \includegraphics[width = 0.8\textwidth]{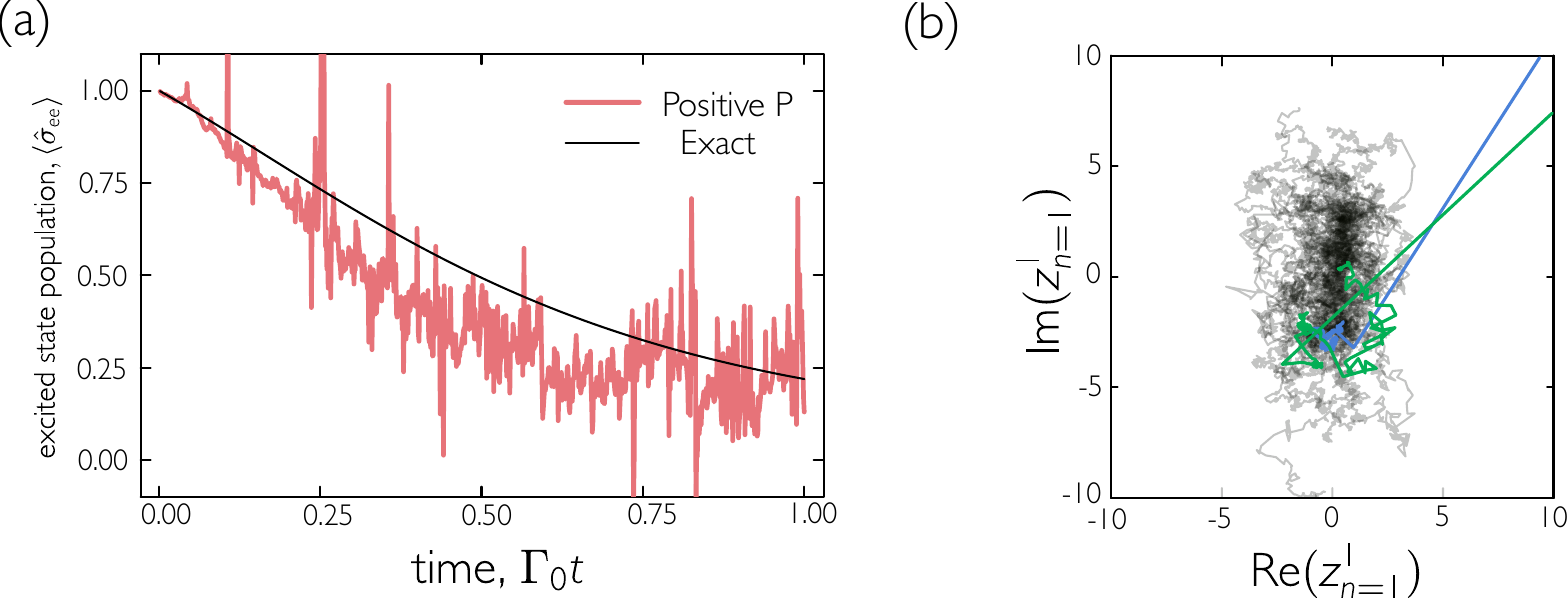}
\caption{\textbf{Collective decay for $N=5$ atoms in the positive P formalism.} (a) Excited-state population and (b) sample trajectories in $z$-space for a 1D array of 5 atoms with polarization perpendicular to the chain and lattice constant $d/\lambda_0 = 0.1$ and $M_0=50$ trajectories. In (b), two diverging trajectories are highlighted in green and blue. }
\label{fg:positive_p_decay_N_5}
\end{center}
\end{figure}

Finally, we consider an example of collective decay. From the previous discussion, we know that the diagonal terms $\Gamma_{nn}$ in Eq.~\eqref{eq positive P spins} induce a drift that eventually leads to population at the boundary. The remaining question is whether the positive $P$ representation can still accurately capture the initial dynamics. To explore this, we consider a 1D array of five atoms decaying collectively in free space. As shown in Fig.~\ref{fg:positive_p_decay_N_5}(a), the method initially captures the decay accurately. However, as time progresses, large oscillations or ``spikes''~\cite{Gilchrist97} emerge, making the discrepancy between the phase-space method and the exact dynamics even more pronounced than for a single atom. The onset of these spikes occurs sooner as the number of atoms increases.

These spikes are not due to statistical or numerical errors but rather indicate that trajectories encounter singular points in phase space, from which they diverge to infinity in a finite time. Unlike for a single atom, where trajectories approach the boundary smoothly, here the simulation fails for two reasons: (1) the dynamics tends to populate the boundary, and (2) the time step is too large to resolve the rapid evolution toward infinity at singular points. The spikes can be partially mitigated by reducing the time step, using higher-order integration schemes, or carefully selecting $x_0\in\mathbb R$ in the initial distribution in Eq.~\eqref{eq:Pexcited}. These measures delay the onset of erratic behavior, but large spikes inevitably appear due to the system instabilities. While discarding diverging trajectories removes the spikes, the resulting statistical averages no longer correspond to the exact dynamics. Diverging trajectories ultimately signal that the description of the exact dynamics by a FPE is invalid. For simple bosonic models, one can identify singularities and estimate the divergence time -- along with the onset of the first spike~\cite{Gilchrist97} -- thus establishing an upper bound on the time range of reliable positive $P$ simulations. However, for Eq.~\eqref{eq positive P spins}, analytical estimation of singularities and divergence times is unfeasible due to the large number of variables.

In an attempt to solve this issue, we discuss in Appendix~\ref{sc:spin gauge P} the gauge $P$ representation for spins~\cite{Deuar02}. This representation exploits the non-uniqueness of phase-space representations to introduce auxiliary drift and diffusion terms in the dynamics to prevent phase-space trajectories from undergoing very rapid dynamics or diverging to infinity. Nevertheless, the construction of these terms is very case-dependent, and a general systematic application of this representation remains an open problem provided the increasing complexity of the equations with atom number.

\section{Summary and Outlook}
In this paper, we have reviewed the phase-space formulation as an equivalent alternative to the standard Hilbert-space formulation of quantum mechanics, with particular emphasis on open quantum many-body dynamics.  In this approach, quantum states are mapped to quasiprobability distributions in phase space, from which observables can be calculated via statistical averages. By using the SW correspondence and the exact Moyal product for spins 1/2, we have derived the exact PDE describing the spin model for many-body quantum optics in phase space, valid for any representation. Although fully equivalent to the master equation, this PDE does not offer a numerical advantage for large systems, as it involves third-order derivatives in a $2N$-dimensional space.

To enable efficient simulation, the exact equation is approximated by a FPE. In the Wigner representation, purely coherent dynamics reduce to the DTWA~\cite{Schachenmayer15}, where quantum fluctuations are incorporated by sampling the initial quasiprobability distribution, while the subsequent evolution follows classical trajectories. For dissipative dynamics, the truncated equation yields the dissipative TWA developed by Mink and Fleischauer~\cite{Mink23}, which captures fluctuations both through stochastic sampling of initial conditions and through noise terms governing the time evolution. By exploiting the linearity of the PDE, we further extend the dissipative TWA to compute multi-time correlation functions, thereby enabling a semiclassical analysis of the light emitted by a many-body system. 

While the dissipative TWA captures many aspects of collective atomic emission, it suffers from unphysical instabilities and breaks down in low-excitation regimes. A major open question is how to quantify its error for general states. We addressed this issue by introducing a geometric interpretation: the exact evolution remains within a subspace $\mathcal{U}$ of functions on $\mathbb{S}^2$ for each atom, while the approximate dynamics leak outside this space. The subsequent projection back to $\mathcal{U}$ required for measurement introduces the error.

We further explored alternative approximations on the exact phase-space equation for the Wigner, $P$, and $Q$ representations that might improve upon existing methods. By imposing reasonable conditions and analyzing differential operators spanned by the first spherical harmonics, we found that the approximation in Ref.~\cite{Mink23} is the only viable alternative yielding a second-order PDE with a positive diffusion matrix, enabling a reformulation in terms of stochastic equations. Future extensions of this analysis could involve the incorporation of higher-order harmonics or the exploration of alternative bases. 

Additionally, we generalized the Positive $P$ formalism to study collective atom emission, an approach successfully used in bosonic systems to circumvent the issue of non-positive diffusion matrices. However, we found that even for a single decaying atom, boundary contributions make the Positive $P$ description deviate from exact quantum dynamics, causing significant errors over time. As the number of atoms increases, singularities emerge, leading to trajectory divergences and ``spikes'' in observable expectation values.

Phase-space methods offer promising avenues for simulating many-body quantum evolution, which remains largely intractable due to Hilbert space's exponential growth. A deeper understanding of approximation errors is crucial to determining their applicability beyond case-by-case validation. Notably, the current approximation is exact for a completely mixed state; an interesting direction would be adapting the approximation dynamically based on the system's current state to minimize errors. Furthermore, a single realization of dissipative TWA evolution does not correspond to a conditioned open-system trajectory based on a specific measurement scheme. Thus, another potentially interesting direction is to extend this framework to conditioned dynamics.

\textbf{Acknowledgments --} The authors would like to thank Eric Sierra for fruitful discussions. We acknowledge support by the National Science Foundation through the CAREER Award (No. 2047380), the Air Force Office of Scientific Research through their Young Investigator Prize (grant No. 21RT0751), as well as by the David and Lucile Packard Foundation.

\appendix
\section{Continuous sampling method}\label{sc:appendix_sampling_method_no_harmonics}
In this Appendix, we introduce a new sampling method applicable to any phase-space representation, offering some advantages over the discrete phase-space sampling method and its generalizations described in the main text. One key advantage is that this method avoids introducing correlations between trajectories due to the spread of initial conditions. Building on our discussion in Section~\ref{sc:beyond_dtwa}, this sampling method also ensures that the initial representation contains no spherical harmonics with $\ell>1$ and thus is fully contained in $\mathcal U$. Furthermore, it faithfully reproduces the Weyl symbol obtained by tracing the density matrix against the kernel. Finally, the method follows the same principles as those in Section~\ref{sc:multi_time_correlation}, leveraging the linearity of the equations and observables to track the evolution of different components of the function.

Let us consider a general Weyl symbol $F_{\rho}(\Omega;s)$ of the density matrix in some representation $s$. By the SW correspondence rules, $\hat \rho$ being Hermitian implies that $F_{\rho}$ is real, but generally not positive everywhere. We define two functions,  $f_+(\Omega)$ and $f_-(\Omega)$, representing the positive and negative parts of $F_{\rho}(\Omega;s)$ and properly normalized, i.e.,
\begin{equation}
    f_{\pm}(\Omega)=\frac{F_{\rho}(\Omega;s)\cdot \mathbb I_{X_{\pm}}(\Omega)}{a_{\pm}},\ \text{ with } a_{\pm}=\int_{X_{\pm}}F_{\rho}(\Omega;s)d\Omega,
\end{equation}
where $X_{\pm}=\{\Omega\in X | F_{\rho}(\Omega;s) \gtrless 0\}$ are the regions where $F_{\rho}(\Omega;s)$ is positive or negative, respectively, and $\mathbb I_{X_{\pm}}(\Omega)$ are the characteristic functions of these regions. It follows that $F_{\rho}(\Omega;s) = a_+f_+(\Omega) + a_-f_-(\Omega)$, where $f_{\pm}(\Omega)$ are positive and normalized distributions. This allows sampling each function as
\begin{equation}
    f_{\pm}(\Omega)\approx\frac{1}{M_{\pm}}\sum_{m=1}^{M_{\pm}}\delta(\Omega-\Omega_{m}^{\pm}).
\end{equation}

By leveraging the linearity of the FPE, if $f_{\pm}(\Omega,t)$ are the solutions to Eq.~\eqref{eq:dtwa_pde} with initial conditions $f_{\pm}(\Omega)$, then the solution with initial condition $F_{\rho}(\Omega;s)$ is
$$F_{\rho}(\Omega,t;s) = a_+f_+(\Omega,t)+a_-f_-(\Omega,t),$$  
and any observable can be computed as
\begin{equation}
\begin{split}
     \langle \hat O(t)\rangle &=\int_XF_O(\Omega;-s)F_{\rho}(\Omega,t;s)d\Omega \\
     &= a_+\int_XF_O(\Omega;-s)f_+(\Omega,t)d\Omega + a_-\int_XF_O(\Omega;-s)f_-(\Omega,t)d\Omega = a_+\langle \hat O(t)\rangle_+ +a_-\langle \hat O(t)\rangle_-.
\end{split}
\end{equation}

This method can be further generalized to non-Hermitian operators, which is useful for computing multi-time correlators, as was done in Section~\ref{sc:multi_time_correlation}, where we have to sample operators such as $\hat O_1(0)\hat \rho$. In the general case, $F_O(\Omega;s)$ is a complex function, so we can redefine the constants
\begin{equation} a_{\pm} = \int_{X_{R\pm}} \text{Re} \{F_O(\Omega; s)\} \,d\Omega, \quad
b_{\pm} = \int_{X_{I\pm}} \text{Im} \{F_O(\Omega; s)\} \,d\Omega, \end{equation} 
where $X_{R\pm}=\{\Omega\in X | \text{Re}\{F_{O}(\Omega;s)\} \gtrless 0\}$ and $X_{I\pm}=\{\Omega\in X | \text{Im}\{F_{O}(\Omega;s)\} \gtrless 0\}$. We then define
\begin{equation}
    f_{\pm}(\Omega)=\frac{\text{Re}\{F_{O}(\Omega;s)\}\cdot \mathbb I_{X_{R\pm}}(\Omega)}{a_{\pm}}, \ g_{\pm}(\Omega)=\frac{\text{Im}\{F_{O}(\Omega;s)\}\cdot \mathbb I_{X_{I\pm}}(\Omega)}{b_{\pm}},
\end{equation}
which are positive and normalized distributions, allowing us to expand the original Weyl symbol as $F_O(\Omega;s)=a_+f_+(\Omega)+a_-f_-(\Omega)+\ii b_+g_+(\Omega)+\ii b_-g_-(\Omega)$. From here, we compute expectation values by evolving each $f_{\pm}$, $g_{\pm}$ in time and then averaging them accounting for the corresponding weights.

Although this sampling method requires more trajectories than other approaches to achieve the same initial state precision, it prevents strong correlations between trajectories due to initial conditions. Moreover, it accurately reproduces the Weyl symbol without needing higher harmonics to construct a Dirac delta expansion. On a more technical note, once the Weyl symbol $F_O(\Omega;s)$ is computed, it cannot be directly sampled. Instead, we must account for the phase-space measure by sampling $F_O(\Omega;s)\frac{\sin\theta}{2\pi}$, so that the sampling points are distributed as $\delta(\Omega-\Omega_m)$ rather than $\delta(\theta-\theta_m)\delta(\phi-\phi_m)$.

\section{Generalized dissipative TWA for other representations}\label{sc:generalized_DTWA}
Here, we provide some of the mathematical details of Sec. \ref{sc:beyond_dtwa}. After defining the operator $\tilde{\bm{\mathcal S}}_s$ in Eq. \eqref{eq:generalized_approx_alpha_beta} and truncating the functional space to $\ell_{max}=1$ with $\bm\alpha_s,\bm\beta_s\in \mathcal U^3=\text{span}\{Y_{00},Y_{1-1},Y_{10},Y_{11}\}^3$, we solve the linear system resulting from imposing the approximation to be exact after projecting back to $\mathcal U$. That is, we require $(P_{|\mathcal U}\circ \tilde{\bm{{\mathcal S}}}_s )[Y_{\ell m}] =  \bm{\mathcal S}_s [Y_{\ell m}]$ for $Y_{\ell m}\in U$. The solutions are given by
\begin{equation}
    \bm\alpha_s = \frac{-8s}{3\pi}\begin{pmatrix}
        0\\
        0\\
        1
    \end{pmatrix},\quad \bm\beta_s=2s\begin{pmatrix}
        -\sin\theta\sin\phi\\
        \sin\theta\cos\phi\\
        0
    \end{pmatrix}+\bm b_0\cos\theta,
\end{equation}
where $\bm b_0\in\mathbb R^3$ is a free parameter.

If we were to generalize the approach in Eqs.~\eqref{eq:fleischauer_approx_moyal_definition} to other representations by simply substituting $\bm s = (-1)^s 3^{\frac{1+s}{2}} \bm m$, the resulting operator would not remain exact after the first projection due to the nonzero contributions of $\bm\alpha_s$ and $\bm\beta_s$ when $s\neq 0$. The corresponding PDE for any choice of $\bm\beta_0$ is given by
\begin{equation}
\begin{split}
    \frac{\partial}{\partial t}F_{\rho}(\bm \Omega,t;s) =& \left(-\mathcal L_1 + \frac{1}{2}\mathcal L_2\right)F_{\rho}(\bm \Omega,t;s),\\
    \mathcal L_1 =& \sum_{n=1}^N \nabla_{\theta_n}\left[\frac{\Gamma_{nn}}{2}(\cot\theta_n+2s\sin\theta_n+b_n^I\cos\theta_n) + (-1)^s3^{\frac{1-s}{2}}\sum_{m=1}^N\sin\theta_m A_{mn}\right] \\
    &+\sum_{n=1}^N \nabla_{\phi_n} \cot\theta_n\left[\frac{\Gamma_{nn}}{2}b_n^R\cos\theta_n+(-1)^s3^{\frac{1-s}{2}}\sum_{m=1}^N\sin\theta_mB_{mn}\right],\\
\mathcal L_2 =& \sum_{n,m=1}\Bigg[ \nabla_{\theta_m}\nabla_{\theta_n}\Gamma_{mn}\cos\phi_{mn} + \nabla_{\theta_m}\nabla_{\phi_n}\Gamma_{mn}\cot\theta_n\sin\phi_{mn} \\
    &+\nabla_{\phi_m}\nabla_{\theta_n}\left(-\Gamma_{mn}\cot\theta_n\sin\phi_{mn} +4s\sin\theta_mB_{mn}+2\cos\theta_m\left(J_{mn}b_n^I-\frac{\Gamma_{mn}}{2}b_n^R\right)\right) \\
    &+ \nabla_{\phi_m}\nabla_{\phi_n}\bigg(\Gamma_{mn} \cot\theta_n\cot\theta_m\cos\phi_{mn} -4s\cot\theta_n\sin\theta_mA_{mn}\\&\quad\quad\quad\quad+2\cos\theta_m\cot\theta_n\left(J_{mn}b_n^R+\frac{\Gamma_{mn}}{2}b_n^I\right)\bigg)\Bigg],
\end{split}\end{equation}
where $A_{mn},B_{mn}$ are defined in Eq. \eqref{eq:exact_pde} and $b_n^R=b_x\cos\phi_n+b_y\sin\phi_n$, $b_n^I=b_x\sin\phi_n-b_y\cos\phi_n$ are respectively the real and imaginary part of $(b_x-\ii b_y)e^{i\phi_n}$.

Since the phase-space representation of spins is invariant under rotations in the angle $\phi_n$, as observed in the fundamental PDE in Eq.~\eqref{eq:exact_pde}, we can set the degrees of freedom in $\bm\beta_0$ as $\beta_x=\beta_y=0$. Additionally, for $s=0$, certain terms in the diffusion matrix vanish, recovering the dissipative TWA equation \eqref{eq:dtwa_pde}. However, for the $P$ and $Q$ representations, numerical diagonalization reveals that the difussion matrix is not positive semidefinite. This highlights the unique symmetry of the Wigner representation, which enables an efficient approximate implementation of the dynamics, unlike other representations and approximations.

A more general approach, involving the absorption of the operator $\bm L$ into the coefficients $\bm\alpha_s$ and $\bm\beta_s$ while introducing an additional coefficient $\bm\gamma_s$, leads to the modified operator
$$\tilde{\bm{\mathcal S}}_s=\bm \gamma_s+\bm\alpha_s\frac{\partial}{\partial\theta}+\bm\beta_s\frac{\partial}{\partial\phi}.$$
However, this results in the nonexistence of solutions to Eq. \eqref{eq:dtwa_moyal_exact_projection}, indicating that higher spherical harmonics contained in the operator $\bm L$ are necessary for the approximation.

This procedure can be systematically extended by increasing the dimension of the truncated functional space of $\bm\alpha_s$ and $\bm\beta_s$, thereby incorporating higher-order corrections to the approximate operator $\tilde{\bm{\mathcal S}}_s$. Nevertheless, numerical analysis suggests that including additional spherical harmonics only introduces more undetermined constants, while the corresponding PDE still lacks a positive semidefinite diffusion matrix.

\section{Spin gauge $P$ representation}\label{sc:spin gauge P}
Our study of the positive $P$ representation was motivated by the pursuit of an efficient framework for simulating many-body dynamics within the spin model~\eqref{eq:master}. However, diverging trajectories -- which are unavoidable for the dynamics we aim to capture --  disrupt the averaging process.  To address this issue, in this Appendix we introduce the gauge $P$ representation for spins~\cite{Deuar02}: the $G$ representation. This approach leverages the non-uniqueness of phase-space representations by introducing an additional degree of freedom,  $\Omega\in\mathbb C$, which enables the construction of an effective potential that confines phase-space trajectories within a bounded region.

Following the definition for bosonic systems~\cite{Deuar02}, we introduce a new phase-space variable $\Omega\in \mathbb C$ and modify the kernel in Eq.~\eqref{kernel N spins positive P} to
\begin{equation}\label{kernelG}
    \hat\Delta(\bm z_1,\bm z_2, \Omega) = \Omega \bigotimes_{n=1}^N\frac{\ket {z_1^n}\bra{z_2^{n*}}}{\braket{z_2^{n*}|z_1^n}} = \Omega \bigotimes_{n=1}^N \begin{pmatrix}
        1-\frac{1}{1+e^{z_1^n+z_2^n}} & \frac{1}{e^{-z_1^n}+e^{z_2^n}}\\
        \frac{1}{e^{-z_2^n}+e^{z_1^n}} & \frac{1}{1+e^{z_1^n+z_2^n}}\end{pmatrix}.
\end{equation}
The $G$ representation of an operator $\hat \rho$ is then defined as
\begin{equation}
    \hat \rho = \iiint d^{2N}\bm z_1\ d^{2N}\bm z_2 \ d\Omega\ G(\bm z_1, \bm z_2, \Omega)\hat\Delta(\bm z_1,\bm z_2, \Omega).
\end{equation}

At first glance, the introduction of $\Omega$ does not seem to significantly alter the previously defined positive $P$ representation. Since $\Omega$ is a scalar, products of Pauli matrices and kernels remain represented by the same differential operators, meaning that the partial differential equation governing the evolution of $G$ remains identical to Eq.~\eqref{eq positive P spins}. Moreover, the ability to sample any state from a positive distribution is preserved: given the positive $P$ function $P(\bm z_1, \bm z_2)$, we can always define a corresponding $G$ representation as
$G(\bm z_1, \bm z_2, \Omega) = P(\bm z_1, \bm z_2)\delta(\Omega-1)$.
 
By defining the set $T=\{z_1^1,\dots,z_1^N,z_2^1,\dots,z_2^N\}$, we can compactly write Eq.~\eqref{eq positive P spins} for the $G$ representation as
\begin{equation}\label{eq for G}
    \frac{\partial}{\partial t} G(\bm z_1, \bm z_2, \Omega) = D^{(-)}[G(\bm z_1, \bm z_2, \Omega)] = \left(-\sum_{a\in T}\frac{\partial}{\partial a}A_a + \frac{1}{2}\sum_{a,b\in T}\frac{\partial^2}{\partial a\partial b}D_{ab} \right)G(\bm z_1, \bm z_2, \Omega),
\end{equation}
where $D^{(-)}$ is the adjoint operator of the action of the Lindbladian $D^{(+)}$ on the kernel in the positive $P$ representation, while $\bm A$ and $\bm D$ correspond to the drift and diffusion terms, respectively. At this stage, the introduction of $\Omega$ has not yet led to any solution to the divergence problem. Equation~\eqref{eq for G} remains susceptible to the same diverging trajectories observed in the positive $P$ formalism. However, the kernel in Eq.~\eqref{kernelG} satisfies
\begin{equation}\label{gauge P sym}
    \left(\Omega\frac{\partial}{\partial\Omega}-1\right)\hat\Delta(\bm z_1,\bm z_2, \Omega) = 0\quad \forall \bm z_1,\bm z_2\in \mathbb C^{2N},\,\Omega\in \mathbb C.
\end{equation}

This property can be exploited to modify Eq.~\eqref{eq for G} without affecting the underlying dynamics of $\ra$. Specifically, we introduce the modified operator~\cite{Deuar02}
\begin{equation}
    D_{\Omega}^{(+)} = \left(\frac{\Omega}{2}\bm g^T\cdot \bm g\frac{\partial}{\partial\Omega} + \sum_{a,b\in T}\bm B_{ab} g_b\frac{\partial}{\partial a}\right)\left(\Omega\frac{\partial}{\partial\Omega}-1\right),
\end{equation}
where  $\bm g(\bm z_1,\bm z_2,\Omega) = (g_a(\bm z_1,\bm z_2,\Omega))_{a\in T}$ are unspecified functions, and $\bm B$ is the decomposition of the diffusion matrix for the positive $P$ representation, $\bm D = \bm B\cdot\bm B^T$. We incorporate the modification into the differential operator by defining $D^{(+)}_G =D^{(+)} + D^{(+)}_{\Omega}$, which takes the form
\begin{equation}
    D^{(+)}_G = \sum_{a\in T}\left(A_a-\sum_{b\in T}g_bB_{ab}\right)\frac{\partial}{\partial a} + \frac{1}{2}\sum_{a,b\in T}\left(D_{ab}\frac{\partial^2}{\partial a\partial b} + \frac{g_bg_b}{2N}\Omega^2\frac{\partial^2}{\partial \Omega^2} + 2\Omega g_b B_{ab}\frac{\partial^2}{\partial a\partial\Omega}\right).
\end{equation}

After integrating by parts, the modified PDE retains the form of a FPE, but with an updated drift vector and diffusion matrix:
\begin{equation}
    \tilde{\bm A} = \left(\begin{array}{c}
        \bm A - \bm B\cdot \bm g\\
        \hline
        0
    \end{array}\right),\quad
    \tilde{\bm D} = \left(\begin{array}{c|c}
        \bm D & \Omega \bm B\cdot \bm g \\
        \hline
         \Omega (\bm B\cdot \bm g)^T& \Omega^2\bm g^T\cdot\bm g   
    \end{array}\right).
\end{equation}
This modified diffusion matrix can be decomposed as $\tilde{\bm D}=\tilde{\bm B}\cdot \tilde{\bm B}^T$ with
\begin{equation}
    \tilde{\bm B} = \left(\begin{array}{c|c}
        \bm B & \bm 0 \\
        \hline
        \Omega \bm g^T & 0
    \end{array}\right).
\end{equation}\\
Finally, the set of SDEs for the gauge $G$ representation is given by
\begin{equation}
    \begin{cases}\label{SDE gauge}
        da = \left(A_a - \sum_{b\in T} B_{ab}g_b\right)dt + \sum_{b\in T}B_{ab}\,dW_b,\\
        d\Omega = \Omega \sum_{b\in B}g_b\,dW_b,
    \end{cases}
\end{equation}
for all $a\in T$, borrowing the expressions for $\bm A$ and $\bm B$ from the positive $P$ representation detailed above. The introduction of the new variable $\Omega$ modifies the drift term in Eq.~\eqref{eq positive P spins}. The functions $\bm g(\bm z_1,\bm z_2,\Omega)$ can be chosen strategically to eliminate the problematic terms in $A_a$ responsible for the divergences~\cite{Deuar02}. To determine an effective choice for $\bm g$, it is necessary to analyze the locations of singular points in the drift dynamics and how trajectories tend to approach them. 

Nevertheless, given the increasing complexity of the system as the number of atoms grows, it remains an open question whether there exists a suitable choice of  $\bm g(\bm z_1, \bm z_2,\Omega)$ that eliminates the spikes.

\bibliography{references}

@article{Dicke54,
  title = {Coherence in Spontaneous Radiation Processes},
  author = {Dicke, R. H.},
  journal = {Phys. Rev.},
  volume = {93},
  issue = {1},
  pages = {99--110},
  numpages = {0},
  year = {1954},
  month = {Jan},
  publisher = {American Physical Society},
  doi = {10.1103/PhysRev.93.99},
  url = {https://link.aps.org/doi/10.1103/PhysRev.93.99}
}

@article{Lehmberg70,
  title = {Radiation from an $N$-Atom System. I. General Formalism},
  author = {Lehmberg, R. H.},
  journal = {Phys. Rev. A},
  volume = {2},
  issue = {3},
  pages = {883--888},
  numpages = {0},
  year = {1970},
  month = {Sep},
  publisher = {American Physical Society},
  doi = {10.1103/PhysRevA.2.883},
  url = {https://link.aps.org/doi/10.1103/PhysRevA.2.883}
}

@Article{Greiner02,
author={Greiner, Markus
and Mandel, Olaf
and Esslinger, Tilman
and H{\"a}nsch, Theodor W.
and Bloch, Immanuel},
title={Quantum phase transition from a superfluid to a {M}ott insulator in a gas of ultracold atoms},
journal={Nature},
year={2002},
month={Jan},
day={01},
volume={415},
number={6867},
pages={39-44},
issn={1476-4687},
doi={10.1038/415039a},
url={https://doi.org/10.1038/415039a}
}

@article{Domokos02,
  title = {Collective Cooling and Self-Organization of Atoms in a Cavity},
  author = {Domokos, Peter and Ritsch, Helmut},
  journal = {Phys. Rev. Lett.},
  volume = {89},
  issue = {25},
  pages = {253003},
  numpages = {4},
  year = {2002},
  month = {Dec},
  publisher = {American Physical Society},
  doi = {10.1103/PhysRevLett.89.253003},
  url = {https://link.aps.org/doi/10.1103/PhysRevLett.89.253003}
}

@Article{Baumann10,
author={Baumann, Kristian
and Guerlin, Christine
and Brennecke, Ferdinand
and Esslinger, Tilman},
title={Dicke quantum phase transition with a superfluid gas in an optical cavity},
journal={Nature},
year={2010},
month={Apr},
day={01},
volume={464},
number={7293},
pages={1301-1306},
issn={1476-4687},
doi={10.1038/nature09009},
url={https://doi.org/10.1038/nature09009}
}

@Article{Bloom14,
author={Bloom, B. J.
and Nicholson, T. L.
and Williams, J. R.
and Campbell, S. L.
and Bishof, M.
and Zhang, X.
and Zhang, W.
and Bromley, S. L.
and Ye, J.},
title={An optical lattice clock with accuracy and stability at the {$10^{-18}$}level},
journal={Nature},
year={2014},
month={Feb},
day={01},
volume={506},
number={7486},
pages={71-75},
issn={1476-4687},
doi={10.1038/nature12941},
url={https://doi.org/10.1038/nature12941}
}

@article{Pichler15,
  title = {Quantum optics of chiral spin networks},
  author = {Pichler, Hannes and Ramos, Tom\'as and Daley, Andrew J. and Zoller, Peter},
  journal = {Phys. Rev. A},
  volume = {91},
  issue = {4},
  pages = {042116},
  numpages = {19},
  year = {2015},
  month = {Apr},
  publisher = {American Physical Society},
  doi = {10.1103/PhysRevA.91.042116},
  url = {https://link.aps.org/doi/10.1103/PhysRevA.91.042116}
}

@article{Asenjo17,
  title = {Exponential Improvement in Photon Storage Fidelities Using Subradiance and ``Selective Radiance'' in Atomic Arrays},
  author = {Asenjo-Garcia, A. and Moreno-Cardoner, M. and Albrecht, A. and Kimble, H. J. and Chang, D. E.},
  journal = {Phys. Rev. X},
  volume = {7},
  issue = {3},
  pages = {031024},
  numpages = {36},
  year = {2017},
  month = {Aug},
  publisher = {American Physical Society},
  doi = {10.1103/PhysRevX.7.031024},
  url = {https://link.aps.org/doi/10.1103/PhysRevX.7.031024}
}

@article{Verstraete08,
author = {F. Verstraete, V. Murg and J.I. Cirac},
title = {Matrix product states, projected entangled pair states, and variational renormalization group methods for quantum spin systems},
journal = {Advances in Physics},
volume = {57},
number = {2},
pages = {143--224},
year = {2008},
publisher = {Taylor \& Francis},
doi = {10.1080/14789940801912366},
}

@article{Meiser09,
  title = {Prospects for a Millihertz-Linewidth Laser},
  author = {Meiser, D. and Ye, Jun and Carlson, D. R. and Holland, M. J.},
  journal = {Phys. Rev. Lett.},
  volume = {102},
  issue = {16},
  pages = {163601},
  numpages = {4},
  year = {2009},
  month = {Apr},
  publisher = {American Physical Society},
  doi = {10.1103/PhysRevLett.102.163601},
  url = {https://link.aps.org/doi/10.1103/PhysRevLett.102.163601}
}

@book{Wiseman09, place={Cambridge}, title={Quantum Measurement and Control}, publisher={Cambridge University Press}, author={Wiseman, Howard M. and Milburn, Gerard J.}, year={2009}}

@article{Leroux10,
  title = {Implementation of Cavity Squeezing of a Collective Atomic Spin},
  author = {Leroux, Ian D. and Schleier-Smith, Monika H. and Vuleti\ifmmode \acute{c}\else \'{c}\fi{}, Vladan},
  journal = {Phys. Rev. Lett.},
  volume = {104},
  issue = {7},
  pages = {073602},
  numpages = {4},
  year = {2010},
  month = {Feb},
  publisher = {American Physical Society},
  doi = {10.1103/PhysRevLett.104.073602},
  url = {https://link.aps.org/doi/10.1103/PhysRevLett.104.073602}
}

@article{Meiser10,
  title = {Steady-state superradiance with alkaline-earth-metal atoms},
  author = {Meiser, D. and Holland, M. J.},
  journal = {Phys. Rev. A},
  volume = {81},
  issue = {3},
  pages = {033847},
  numpages = {4},
  year = {2010},
  month = {Mar},
  publisher = {American Physical Society},
  doi = {10.1103/PhysRevA.81.033847},
  url = {https://link.aps.org/doi/10.1103/PhysRevA.81.033847}
}

@article{Schollwock11,
title = {The density-matrix renormalization group in the age of matrix product states},
journal = {Annals of Physics},
volume = {326},
number = {1},
pages = {96-192},
year = {2011},
issn = {0003-4916},
doi = {https://doi.org/10.1016/j.aop.2010.09.012},
url = {https://www.sciencedirect.com/science/article/pii/S0003491610001752},
author = {Ulrich Schollw{\"o}ck},
}

@Article{Bohnet12,
author={Bohnet, Justin G.
and Chen, Zilong
and Weiner, Joshua M.
and Meiser, Dominic
and Holland, Murray J.
and Thompson, James K.},
title={A steady-state superradiant laser with less than one intracavity photon},
journal={Nature},
year={2012},
month={Apr},
day={01},
volume={484},
number={7392},
pages={78-81},
issn={1476-4687},
doi={10.1038/nature10920},
url={https://doi.org/10.1038/nature10920}
}

@article{Xu13,
  title = {Simulating open quantum systems by applying {SU}(4) to quantum master equations},
  author = {Xu, Minghui and Tieri, D. A. and Holland, M. J.},
  journal = {Phys. Rev. A},
  volume = {87},
  issue = {6},
  pages = {062101},
  numpages = {7},
  year = {2013},
  month = {Jun},
  publisher = {American Physical Society},
  doi = {10.1103/PhysRevA.87.062101},
  url = {https://link.aps.org/doi/10.1103/PhysRevA.87.062101}
}

@article{Daley14,
   title={Quantum trajectories and open many-body quantum systems},
   volume={63},
   ISSN={1460-6976},
   url={http://dx.doi.org/10.1080/00018732.2014.933502},
   DOI={10.1080/00018732.2014.933502},
   number={2},
   journal={Advances in Physics},
   publisher={Informa UK Limited},
   author={Daley, Andrew J.},
   year={2014},
   month=mar, pages={77–149} }

@article{Bolanos15,
	author = {Bola\~{n}os, Marduk and Barberis-Blostein, Pablo},
	date = {2015/10/08},
	doi = {10.1088/1751-8113/48/44/445301},
	isbn = {1751-8121; 1751-8113},
	journal = {J. Phys. A},
	number = {44},
	pages = {445301},
	publisher = {IOP Publishing},
	title = {Algebraic solution of the {L}indblad equation for a collection of multilevel systems coupled to independent environments},
	url = {https://dx.doi.org/10.1088/1751-8113/48/44/445301},
	volume = {48},
	year = {2015}}

@article{Sarkar87,
doi = {10.1088/0305-4470/20/8/028},
url = {https://dx.doi.org/10.1088/0305-4470/20/8/028},
year = {1987},
month = {jun},
publisher = {},
volume = {20},
number = {8},
pages = {2147},
author = {S Sarkar and J S Satchell},
title = {Solution of master equations for small bistable systems},
journal = {Journal of Physics A: Mathematical and General},
}

@article{Bonifacio71,
  title = {Quantum Statistical Theory of Superradiance. I},
  author = {Bonifacio, R. and Schwendimann, P. and Haake, Fritz},
  journal = {Phys. Rev. A},
  volume = {4},
  issue = {1},
  pages = {302--313},
  numpages = {0},
  year = {1971},
  month = {Jul},
  publisher = {American Physical Society},
  doi = {10.1103/PhysRevA.4.302},
  url = {https://link.aps.org/doi/10.1103/PhysRevA.4.302}
}

@article{Glauber76,
  title = {Superradiant pulses and directed angular momentum states},
  author = {Glauber, Roy J. and Haake, Fritz},
  journal = {Phys. Rev. A},
  volume = {13},
  issue = {1},
  pages = {357--366},
  numpages = {0},
  year = {1976},
  month = {Jan},
  publisher = {American Physical Society},
  doi = {10.1103/PhysRevA.13.357},
  url = {https://link.aps.org/doi/10.1103/PhysRevA.13.357}
}

@article{Haake72,
  title = {Quantum Statistics of Superradiant Pulses},
  author = {Haake, Fritz and Glauber, Roy J.},
  journal = {Phys. Rev. A},
  volume = {5},
  issue = {3},
  pages = {1457--1466},
  numpages = {0},
  year = {1972},
  month = {Mar},
  publisher = {American Physical Society},
  doi = {10.1103/PhysRevA.5.1457},
  url = {https://link.aps.org/doi/10.1103/PhysRevA.5.1457}
}

@article{Gordon67,
  title = {Quantum Theory of a Simple Maser Oscillator},
  author = {Gordon, J. P.},
  journal = {Phys. Rev.},
  volume = {161},
  issue = {2},
  pages = {367--386},
  numpages = {0},
  year = {1967},
  month = {Sep},
  publisher = {American Physical Society},
  doi = {10.1103/PhysRev.161.367},
  url = {https://link.aps.org/doi/10.1103/PhysRev.161.367}
}

@Article{Gronchi78,
author={Gronchi, M.
and Lugiato, L. A.},
title={Fokker- {P}lanck equation for optical bistability},
journal={Lettere al Nuovo Cimento (1971-1985)},
year={1978},
month={Dec},
day={16},
volume={23},
number={16},
pages={593-598},
issn={1827-613X},
doi={10.1007/BF02776284},
url={https://doi.org/10.1007/BF02776284}
}

@article{Caneva15,
doi = {10.1088/1367-2630/17/11/113001},
url = {https://dx.doi.org/10.1088/1367-2630/17/11/113001},
year = {2015},
month = {oct},
publisher = {IOP Publishing},
volume = {17},
number = {11},
pages = {113001},
author = {Caneva, Tommaso and Manzoni, Marco T and Shi, Tao and Douglas, James S and Cirac, J Ignacio and Chang, Darrick E},
title = {Quantum dynamics of propagating photons with strong interactions: a generalized input-output formalism},
journal = {New Journal of Physics},
}

@article{Xu15,
  title = {Input-output formalism for few-photon transport: A systematic treatment beyond two photons},
  author = {Xu, Shanshan and Fan, Shanhui},
  journal = {Phys. Rev. A},
  volume = {91},
  issue = {4},
  pages = {043845},
  numpages = {11},
  year = {2015},
  month = {Apr},
  publisher = {American Physical Society},
  doi = {10.1103/PhysRevA.91.043845},
  url = {https://link.aps.org/doi/10.1103/PhysRevA.91.043845}
}

@article{Solano17,
	Author = {Solano, P. and Barberis-Blostein, P. and Fatemi, F. K. and Orozco, L. A. and Rolston, S. L.},
	Da = {2017/11/30},
	Doi = {10.1038/s41467-017-01994-3},
	Id = {Solano2017},
	Isbn = {2041-1723},
	Journal = {Nat. Commun.},
	Number = {1},
	Pages = {1857},
	Title = {Super-radiance reveals infinite-range dipole interactions through a nanofiber},
	Volume = {8},
	Year = {2017}}

@Article{Mirhosseini19,
author={Mirhosseini, Mohammad
and Kim, Eunjong
and Zhang, Xueyue
and Sipahigil, Alp
and Dieterle, Paul B.
and Keller, Andrew J.
and Asenjo-Garcia, Ana
and Chang, Darrick E.
and Painter, Oskar},
title={Cavity quantum electrodynamics with atom-like mirrors},
journal={Nature},
year={2019},
month={May},
day={01},
volume={569},
number={7758},
pages={692-697},
doi={10.1038/s41586-019-1196-1},
url={https://doi.org/10.1038/s41586-019-1196-1}
}

@article{Albrecht19,
doi = {10.1088/1367-2630/ab0134},
url = {https://dx.doi.org/10.1088/1367-2630/ab0134},
year = {2019},
month = {feb},
publisher = {IOP Publishing},
volume = {21},
number = {2},
pages = {025003},
author = {Albrecht, Andreas and Henriet, Loïc and Asenjo-Garcia, Ana and Dieterle, Paul B and Painter, Oskar and Chang, Darrick E},
title = {Subradiant states of quantum bits coupled to a one-dimensional waveguide},
journal = {New Journal of Physics}
}

@Article{Rui2020,
author={Rui, Jun
and Wei, David
and Rubio-Abadal, Antonio
and Hollerith, Simon
and Zeiher, Johannes
and Stamper-Kurn, Dan M.
and Gross, Christian
and Bloch, Immanuel},
title={A subradiant optical mirror formed by a single structured atomic layer},
journal={Nature},
year={2020},
month={Jul},
day={01},
volume={583},
number={7816},
pages={369-374},
issn={1476-4687},
doi={10.1038/s41586-020-2463-x},
url={https://doi.org/10.1038/s41586-020-2463-x}
}

@Article{Gonzalez24,
author={Gonz\'{a}lez-Tudela, Alejandro
and Reiserer, Andreas
and Garc{\'i}a-Ripoll, Juan Jos{\'e}
and Garc{\'i}a-Vidal, Francisco J.},
title={Light--matter interactions in quantum nanophotonic devices},
journal={Nature Reviews Physics},
year={2024},
month={Mar},
day={01},
volume={6},
number={3},
pages={166-179},
issn={2522-5820},
doi={10.1038/s42254-023-00681-1},
url={https://doi.org/10.1038/s42254-023-00681-1}
}

@article{Robicheaux21,
  title = {Beyond lowest order mean-field theory for light interacting with atom arrays},
  author = {Robicheaux, F. and Suresh, Deepak A.},
  journal = {Phys. Rev. A},
  volume = {104},
  issue = {2},
  pages = {023702},
  numpages = {12},
  year = {2021},
  month = {Aug},
  publisher = {American Physical Society},
  doi = {10.1103/PhysRevA.104.023702},
  url = {https://link.aps.org/doi/10.1103/PhysRevA.104.023702}
}

@article{Plankensteiner22,
  doi = {10.22331/q-2022-01-04-617},
  url = {https://doi.org/10.22331/q-2022-01-04-617},
  title = {Quantum{C}umulants.jl: {A} {J}ulia framework for generalized mean-field equations in open quantum systems},
  author = {Plankensteiner, David and Hotter, Christoph and Ritsch, Helmut},
  journal = {{Quantum}},
  issn = {2521-327X},
  publisher = {{Verein zur F{\"{o}}rderung des Open Access Publizierens in den Quantenwissenschaften}},
  volume = {6},
  pages = {617},
  month = jan,
  year = {2022}
}

@Article{Zanner22,
author={Zanner, Maximilian
and Orell, Tuure
and Schneider, Christian M. F.
and Albert, Romain
and Oleschko, Stefan
and Juan, Mathieu L.
and Silveri, Matti
and Kirchmair, Gerhard},
title={Coherent control of a multi-qubit dark state in waveguide quantum electrodynamics},
journal={Nature Physics},
year={2022},
month={May},
day={01},
volume={18},
number={5},
pages={538-543},
issn={1745-2481},
doi={10.1038/s41567-022-01527-w},
url={https://doi.org/10.1038/s41567-022-01527-w}
}

@article{Sierra22,
  title = {Dicke Superradiance in Ordered Lattices: Dimensionality Matters},
  author = {Sierra, Eric and Masson, Stuart J. and Asenjo-Garcia, Ana},
  journal = {Phys. Rev. Res.},
  volume = {4},
  issue = {2},
  pages = {023207},
  numpages = {11},
  year = {2022},
  month = {Jun},
  publisher = {American Physical Society},
  doi = {10.1103/PhysRevResearch.4.023207},
  url = {https://link.aps.org/doi/10.1103/PhysRevResearch.4.023207}
}

@Article{Masson22,
author={Masson, Stuart J.
and Asenjo-Garcia, Ana},
title={Universality of {D}icke superradiance in arrays of quantum emitters},
journal={Nature Communications},
year={2022},
month={Apr},
day={27},
volume={13},
number={1},
pages={2285},
issn={2041-1723},
doi={10.1038/s41467-022-29805-4},
url={https://doi.org/10.1038/s41467-022-29805-4}
}

@article{Holzinger22,
  title = {Control of Localized Single- and Many-Body Dark States in Waveguide {QED}},
  author = {Holzinger, R. and Guti\'errez-J\'auregui, R. and H\"onigl-Decrinis, T. and Kirchmair, G. and Asenjo-Garcia, A. and Ritsch, H.},
  journal = {Phys. Rev. Lett.},
  volume = {129},
  issue = {25},
  pages = {253601},
  numpages = {6},
  year = {2022},
  month = {Dec},
  publisher = {American Physical Society},
  doi = {10.1103/PhysRevLett.129.253601},
  url = {https://link.aps.org/doi/10.1103/PhysRevLett.129.253601}
}

@Article{Ferioli23,
author={Ferioli, Giovanni
and Glicenstein, Antoine
and Ferrier-Barbut, Igor
and Browaeys, Antoine},
title={A non-equilibrium superradiant phase transition in free space},
journal={Nature Physics},
year={2023},
month={Sep},
day={01},
volume={19},
number={9},
pages={1345-1349},
issn={1745-2481},
doi={10.1038/s41567-023-02064-w},
url={https://doi.org/10.1038/s41567-023-02064-w}
}

@article{Ferioli24,
  title = {Non-Gaussian Correlations in the Steady State of Driven-Dissipative Clouds of Two-Level Atoms},
  author = {Ferioli, Giovanni and Pancaldi, Sara and Glicenstein, Antoine and Cl\'ement, David and Browaeys, Antoine and Ferrier-Barbut, Igor},
  journal = {Phys. Rev. Lett.},
  volume = {132},
  issue = {13},
  pages = {133601},
  numpages = {6},
  year = {2024},
  month = {Mar},
  publisher = {American Physical Society},
  doi = {10.1103/PhysRevLett.132.133601},
  url = {https://link.aps.org/doi/10.1103/PhysRevLett.132.133601}
}

@article{Mok23,
  title = {Dicke Superradiance Requires Interactions beyond Nearest Neighbors},
  author = {Mok, Wai-Keong and Asenjo-Garcia, Ana and Sum, Tze Chien and Kwek, Leong-Chuan},
  journal = {Phys. Rev. Lett.},
  volume = {130},
  issue = {21},
  pages = {213605},
  numpages = {7},
  year = {2023},
  month = {May},
  publisher = {American Physical Society},
  doi = {10.1103/PhysRevLett.130.213605},
  url = {https://link.aps.org/doi/10.1103/PhysRevLett.130.213605}
}

@article{Cardenas23,
  title = {Many-Body Superradiance and Dynamical Mirror Symmetry Breaking in Waveguide {QED}},
  author = {Cardenas-Lopez, Silvia and Masson, Stuart J. and Zager, Zoe and Asenjo-Garcia, Ana},
  journal = {Phys. Rev. Lett.},
  volume = {131},
  issue = {3},
  pages = {033605},
  numpages = {7},
  year = {2023},
  month = {Jul},
  publisher = {American Physical Society},
  doi = {10.1103/PhysRevLett.131.033605},
  url = {https://link.aps.org/doi/10.1103/PhysRevLett.131.033605}
}

@article{Masson24,
  title = {Dicke Superradiance in Ordered Arrays of Multilevel Atoms},
  author = {Masson, Stuart J. and Covey, Jacob P. and Will, Sebastian and Asenjo-Garcia, Ana},
  journal = {PRX Quantum},
  volume = {5},
  issue = {1},
  pages = {010344},
  numpages = {19},
  year = {2024},
  month = {Mar},
  publisher = {American Physical Society},
  doi = {10.1103/PRXQuantum.5.010344},
  url = {https://link.aps.org/doi/10.1103/PRXQuantum.5.010344}
}

@article{Goncalves24,
  title = {Driven-Dissipative Phase Separation in Free-Space Atomic Ensembles},
  author = {Goncalves, D. and Bombieri, L. and Ferioli, G. and Pancaldi, S. and Ferrier-Barbut, I. and Browaeys, A. and Shahmoon, E. and Chang, D.E.},
  journal = {PRX Quantum},
  volume = {6},
  issue = {2},
  pages = {020303},
  numpages = {22},
  year = {2025},
  month = {Apr},
  publisher = {American Physical Society},
  doi = {10.1103/PRXQuantum.6.020303},
  url = {https://link.aps.org/doi/10.1103/PRXQuantum.6.020303}
}

@article{Agarwal24,
  title = {Directional Superradiance in a Driven Ultracold Atomic Gas in Free Space},
  author = {Agarwal, Sanaa and Chaparro, Edwin and Barberena, Diego and  Pi\~neiro Orioli, A. and Ferioli, G. and Pancaldi, S. and Ferrier-Barbut, I. and Browaeys, A. and Rey, A. M.},
  journal = {PRX Quantum},
  volume = {5},
  issue = {4},
  pages = {040335},
  numpages = {31},
  year = {2024},
  month = {Dec},
  publisher = {American Physical Society},
  doi = {10.1103/PRXQuantum.5.040335},
  url = {https://link.aps.org/doi/10.1103/PRXQuantum.5.040335}
}

@Article{Bluvstein24,
author={Bluvstein, Dolev
and Evered, Simon J.
and Geim, Alexandra A.
and Li, Sophie H.
and Zhou, Hengyun
and Manovitz, Tom
and Ebadi, Sepehr
and Cain, Madelyn
and Kalinowski, Marcin
and Hangleiter, Dominik
and Bonilla Ataides, J. Pablo
and Maskara, Nishad
and Cong, Iris
and Gao, Xun
and Sales Rodriguez, Pedro
and Karolyshyn, Thomas
and Semeghini, Giulia
and Gullans, Michael J.
and Greiner, Markus
and Vuleti{\'{c}}, Vladan
and Lukin, Mikhail D.},
title={Logical quantum processor based on reconfigurable atom arrays},
journal={Nature},
year={2024},
month={Feb},
day={01},
volume={626},
number={7997},
pages={58-65},
issn={1476-4687},
doi={10.1038/s41586-023-06927-3},
url={https://doi.org/10.1038/s41586-023-06927-3}
}

@misc{Mok24,
      title={Universal scaling laws for correlated decay of many-body quantum systems}, 
      author={Wai-Keong Mok and Avishi Poddar and Eric Sierra and Cosimo C. Rusconi and John Preskill and Ana Asenjo-Garcia},
      year={2024},
      eprint={2406.00722},
      archivePrefix={arXiv},
      primaryClass={quant-ph},
      url={https://arxiv.org/abs/2406.00722}, 
}

@article{Gruner96,
  title = {Green-function approach to the radiation-field quantization for homogeneous and inhomogeneous {K}ramers-{K}ronig dielectrics},
  author = {Gruner, T. and Welsch, D.-G.},
  journal = {Phys. Rev. A},
  volume = {53},
  issue = {3},
  pages = {1818--1829},
  numpages = {0},
  year = {1996},
  month = {Mar},
  publisher = {American Physical Society},
  doi = {10.1103/PhysRevA.53.1818},
  url = {https://link.aps.org/doi/10.1103/PhysRevA.53.1818}
}

@article{Dung02,
  title = {Resonant dipole-dipole interaction in the presence of dispersing and absorbing surroundings},
  author = {Dung, Ho Trung and Kn\"oll, Ludwig and Welsch, Dirk-Gunnar},
  journal = {Phys. Rev. A},
  volume = {66},
  issue = {6},
  pages = {063810},
  numpages = {16},
  year = {2002},
  month = {Dec},
  publisher = {American Physical Society},
  doi = {10.1103/PhysRevA.66.063810},
  url = {https://link.aps.org/doi/10.1103/PhysRevA.66.063810}
}

@article{Wigner32,
  title = {On the Quantum Correction For Thermodynamic Equilibrium},
  author = {Wigner, E.},
  journal = {Phys. Rev.},
  volume = {40},
  issue = {5},
  pages = {749--759},
  numpages = {0},
  year = {1932},
  month = {Jun},
  publisher = {American Physical Society},
  doi = {10.1103/PhysRev.40.749},
  url = {https://link.aps.org/doi/10.1103/PhysRev.40.749}
}

@article{Husimi40,
  title={Some Formal Properties of the Density Matrix},
  author={Husimi, K},
  journal={Proceedings of the Physico-Mathematical Society of Japan. 3rd Series},
  volume={22},
  number={4},
  pages={264-314},
  year={1940},
  doi={10.11429/ppmsj1919.22.4_264}
}

@article{Moyal49, title={Quantum mechanics as a statistical theory}, volume={45}, DOI={10.1017/S0305004100000487}, number={1}, journal={Mathematical Proceedings of the Cambridge Philosophical Society}, author={Moyal, J. E.}, year={1949}, pages={99–124}}

@article{Stratonovich57,
  title = {On Distributions in Representation Space},
  author = {Stratonovich, R. L.},
  journal = {JETP},
  volume = {4},
  issue = {6},
  pages = {981},
  numpages = {0},
  year = {1957},
  month = {June},

}

@article{Sudarshan63,
  title = {Equivalence of Semiclassical and Quantum Mechanical Descriptions of Statistical Light Beams},
  author = {Sudarshan, E. C. G.},
  journal = {Phys. Rev. Lett.},
  volume = {10},
  issue = {7},
  pages = {277--279},
  numpages = {0},
  year = {1963},
  month = {Apr},
  publisher = {American Physical Society},
  doi = {10.1103/PhysRevLett.10.277},
  url = {https://link.aps.org/doi/10.1103/PhysRevLett.10.277}
}

@article{Glauber63,
  title = {Coherent and Incoherent States of the Radiation Field},
  author = {Glauber, Roy J.},
  journal = {Phys. Rev.},
  volume = {131},
  issue = {6},
  pages = {2766--2788},
  numpages = {0},
  year = {1963},
  month = {Sep},
  publisher = {American Physical Society},
  doi = {10.1103/PhysRev.131.2766},
  url = {https://link.aps.org/doi/10.1103/PhysRev.131.2766}
}

@article{Qu19,
  title = {Spin squeezing and many-body dipolar dynamics in optical lattice clocks},
  author = {Qu, Chunlei and Rey, Ana M.},
  journal = {Phys. Rev. A},
  volume = {100},
  issue = {4},
  pages = {041602},
  numpages = {7},
  year = {2019},
  month = {Oct},
  publisher = {American Physical Society},
  doi = {10.1103/PhysRevA.100.041602},
  url = {https://link.aps.org/doi/10.1103/PhysRevA.100.041602}
}

@article{Drummond81,
  title = {Quantum theory of optical bistability. II. Atomic fluorescence in a high-{$Q$} cavity},
  author = {Drummond, P. D. and Walls, D. F.},
  journal = {Phys. Rev. A},
  volume = {23},
  issue = {5},
  pages = {2563--2579},
  numpages = {0},
  year = {1981},
  month = {May},
  publisher = {American Physical Society},
  doi = {10.1103/PhysRevA.23.2563},
  url = {https://link.aps.org/doi/10.1103/PhysRevA.23.2563}
}

@article{Agarwal81,
  title = {Relation between atomic coherent-state representation, state multipoles, and generalized phase-space distributions},
  author = {Agarwal, G. S.},
  journal = {Phys. Rev. A},
  volume = {24},
  issue = {6},
  pages = {2889--2896},
  numpages = {0},
  year = {1981},
  month = {Dec},
  publisher = {American Physical Society},
  doi = {10.1103/PhysRevA.24.2889},
  url = {https://link.aps.org/doi/10.1103/PhysRevA.24.2889}
}

@book{Haken84, place={Berlin}, edition={1}, title={Laser theory}, publisher={Springer-Verlag}, author={Haken, H.}, year={1984}}

@article{Wooters87,
title = {A {W}igner-function formulation of finite-state quantum mechanics},
journal = {Annals of Physics},
volume = {176},
number = {1},
pages = {1-21},
year = {1987},
issn = {0003-4916},
doi = {https://doi.org/10.1016/0003-4916(87)90176-X},
author = {William K Wootters},
}

@article{Varilly89,
title = {The {M}oyal representation for spin},
journal = {Annals of Physics},
volume = {190},
number = {1},
pages = {107-148},
year = {1989},
issn = {0003-4916},
doi = {https://doi.org/10.1016/0003-4916(89)90262-5},
url = {https://www.sciencedirect.com/science/article/pii/0003491689902625},
author = {Joseph C V\'arilly and Jos\e M Gracia-Bond\'ia},
}

@book{Risken96, edition={2}, title={The {F}okker-{P}lanck equation: Methods of solution and applications}, publisher={Springer International Publishing}, author={Risken, H.}, year={1996}}

@book{Buhmann12, edition={1}, title={
Dispersion {F}orces {I}}, publisher={Springer Berlin, Heidelberg}, author={Buhmann, S.}, year={2012}}

@article{Steel98,
  title = {Dynamical quantum noise in trapped {B}ose-{E}instein condensates},
  author = {Steel, M. J. and Olsen, M. K. and Plimak, L. I. and Drummond, P. D. and Tan, S. M. and Collett, M. J. and Walls, D. F. and Graham, R.},
  journal = {Phys. Rev. A},
  volume = {58},
  issue = {6},
  pages = {4824--4835},
  numpages = {0},
  year = {1998},
  month = {Dec},
  publisher = {American Physical Society},
  doi = {10.1103/PhysRevA.58.4824},
  url = {https://link.aps.org/doi/10.1103/PhysRevA.58.4824}
}

@article{Brif98,
doi = {10.1088/0305-4470/31/1/002},
url = {https://dx.doi.org/10.1088/0305-4470/31/1/002},
year = {1998},
month = {jan},
publisher = {},
volume = {31},
number = {1},
pages = {L9},
author = {C Brif and A Mann},
title = {A general theory of phase-space quasiprobability distributions},
journal = {Journal of Physics A: Mathematical and General}
}

@article{Klimov02_2,
    author = {Klimov, A. B.},
    title = {Exact evolution equations for {SU}(2) quasidistribution functions},
    journal = {Journal of Mathematical Physics},
    volume = {43},
    number = {5},
    pages = {2202-2213},
    year = {2002},
    month = {05},
    issn = {0022-2488},
    doi = {10.1063/1.1463711},
}

@article{Sinatra02,
doi = {10.1088/0953-4075/35/17/301},
url = {https://dx.doi.org/10.1088/0953-4075/35/17/301},
year = {2002},
month = {aug},
publisher = {},
volume = {35},
number = {17},
pages = {3599},
author = {Alice Sinatra and Carlos Lobo and Yvan Castin},
title = {The truncated {W}igner method for {B}ose-condensed gases: limits of validity and
applications},
journal = {Journal of Physics B: Atomic, Molecular and Optical Physics},
}

@article{Drummond04,
  title = {Canonical {B}ose Gas Simulations with Stochastic Gauges},
  author = {Drummond, P. D. and Deuar, P. and Kheruntsyan, K. V.},
  journal = {Phys. Rev. Lett.},
  volume = {92},
  issue = {4},
  pages = {040405},
  numpages = {4},
  year = {2004},
  month = {Jan},
  publisher = {American Physical Society},
  url = {https://link.aps.org/doi/10.1103/PhysRevLett.92.040405}
}

@book{Kampen92,
  title={Stochastic processes in physics and chemistry},
  author={Van Kampen, Nicolaas Godfried},
  volume={1},
  year={1992},
  publisher={Elsevier}
}

@article{Blakie08,
author = {P. B. Blakie and A. S. Bradley and M. J. Davis and R. J. Ballagh and C. W. Gardiner},
title = {Dynamics and statistical mechanics of ultra-cold {B}ose gases using c-field techniques},
journal = {Advances in Physics},
volume = {57},
number = {5},
pages = {363--455},
year = {2008},
publisher = {Taylor \& Francis},
doi = {10.1080/00018730802564254},
}

@article{Polkovnikov10,
title = {Phase space representation of quantum dynamics},
journal = {Annals of Physics},
volume = {325},
number = {8},
pages = {1790-1852},
year = {2010},
issn = {0003-4916},
doi = {https://doi.org/10.1016/j.aop.2010.02.006},
author = {Anatoli Polkovnikov},
}

@article{Ng11,
doi = {10.1088/1751-8113/44/6/065305},
url = {https://dx.doi.org/10.1088/1751-8113/44/6/065305},
journal = {Journal of Physics A: Mathematical and Theoretical},
year = {2011},
month = {jan},
publisher = {},
volume = {44},
number = {6},
pages = {065305},
author = {Ng, R and S{\o}rensen, E S},
title = {Exact real-time dynamics of quantum spin systems using the positive-{P} representation},
}

@article{Schachenmayer15,
  title = {Many-Body Quantum Spin Dynamics with {M}onte {C}arlo Trajectories on a Discrete Phase Space},
  author = {Schachenmayer, J. and Pikovski, A. and Rey, A. M.},
  journal = {Phys. Rev. X},
  volume = {5},
  issue = {1},
  pages = {011022},
  numpages = {10},
  year = {2015},
  month = {Feb},
  publisher = {American Physical Society},
  doi = {10.1103/PhysRevX.5.011022},
  url = {https://link.aps.org/doi/10.1103/PhysRevX.5.011022}
}

@BOOK{Carmichael13,
  title     = "Statistical methods in quantum optics 1",
  author    = "Carmichael, Howard J",
  publisher = "Springer",
  series    = "Theoretical and Mathematical Physics",
  month     =  apr,
  year      =  2013,
  address   = "Berlin, Germany",
}

@incollection{Gardiner09,
title = {Stochastic Methods},
author = {Crispin Gardiner},
booktitle = {Stochastic Methods, A Handbook for the Natural and Social Sciences},
publisher = {Springer Berlin, Heidelberg},
edition = {Fourth Edition},
address = {Berlin, Heidelberg},
year = {2009},
series = {Springer Series in Synergetics},
issn = {0172-7389},
doi = {},
url = {https://link.springer.com/book/9783540707127}
}

@article{Schachenmayer15_2,
doi = {10.1088/1367-2630/17/6/065009},
url = {https://dx.doi.org/10.1088/1367-2630/17/6/065009},
year = {2015},
month = {jun},
publisher = {IOP Publishing},
volume = {17},
number = {6},
pages = {065009},
author = {Schachenmayer, J and Pikovski, A and Rey, A M},
title = {Dynamics of correlations in two-dimensional quantum spin models with long-range interactions: a phase-space {M}onte-{C}arlo study},
journal = {New Journal of Physics},
}

@article{Zhu19,
doi = {10.1088/1367-2630/ab354d},
url = {https://dx.doi.org/10.1088/1367-2630/ab354d},
year = {2019},
month = {aug},
publisher = {IOP Publishing},
volume = {21},
number = {8},
pages = {082001},
author = {Zhu, Bihui and Rey, Ana Maria and Schachenmayer, Johannes},
title = {A generalized phase space approach for solving quantum spin dynamics},
journal = {New Journal of Physics},
}

@Article{Huber21,
	title={{Phase-space methods for simulating the dissipative many-body dynamics of collective spin systems}},
	author={Julian Huber and Peter Kirton and Peter Rabl},
	journal={SciPost Phys.},
	volume={10},
	pages={045},
	year={2021},
	publisher={SciPost},
	doi={10.21468/SciPostPhys.10.2.045},
	url={https://scipost.org/10.21468/SciPostPhys.10.2.045},
}

@article{Huber22,
  title = {Realistic simulations of spin squeezing and cooperative coupling effects in large ensembles of interacting two-level systems},
  author = {Huber, Julian and Rey, Ana Maria and Rabl, Peter},
  journal = {Phys. Rev. A},
  volume = {105},
  issue = {1},
  pages = {013716},
  numpages = {14},
  year = {2022},
  month = {Jan},
  publisher = {American Physical Society},
  doi = {10.1103/PhysRevA.105.013716},
  url = {https://link.aps.org/doi/10.1103/PhysRevA.105.013716}
}

@article{Mink22,
  title = {Hybrid discrete-continuous truncated {W}igner approximation for driven, dissipative spin systems},
  author = {Mink, Christopher D. and Petrosyan, David and Fleischhauer, Michael},
  journal = {Phys. Rev. Res.},
  volume = {4},
  issue = {4},
  pages = {043136},
  numpages = {14},
  year = {2022},
  month = {Nov},
  publisher = {American Physical Society},
  doi = {10.1103/PhysRevResearch.4.043136},
  url = {https://link.aps.org/doi/10.1103/PhysRevResearch.4.043136}
}

@Article{Mink23,
	title={{Collective radiative interactions in the discrete truncated {W}igner approximation}},
	author={Christopher D. Mink and Michael Fleischhauer},
	journal={SciPost Phys.},
	volume={15},
	pages={233},
	year={2023},
	publisher={SciPost},
	doi={10.21468/SciPostPhys.15.6.233},
	url={https://scipost.org/10.21468/SciPostPhys.15.6.233},
}

@article{Tebbenjohanns24,
  title = {Predicting correlations in superradiant emission from a cascaded quantum system},
  author = {Tebbenjohanns, Felix and Mink, Christopher D. and Bach, Constanze and Rauschenbeutel, Arno and Fleischhauer, Michael},
  journal = {Phys. Rev. A},
  volume = {110},
  issue = {4},
  pages = {043713},
  numpages = {12},
  year = {2024},
  month = {Oct},
  publisher = {American Physical Society},
  doi = {10.1103/PhysRevA.110.043713},
  url = {https://link.aps.org/doi/10.1103/PhysRevA.110.043713}
}

@misc{Bach24,
      title={Emergence of second-order coherence in superfluorescence}, 
      author={Constanze Bach and Felix Tebbenjohanns and Christian Liedl and Philipp Schneeweiss and Arno Rauschenbeutel},
      year={2024},
      eprint={2407.12549},
      archivePrefix={arXiv},
      primaryClass={quant-ph},
      url={https://arxiv.org/abs/2407.12549}, 
}

@article{Perarnau20,
doi = {10.1088/2058-9565/ab6ce5},
url = {https://dx.doi.org/10.1088/2058-9565/ab6ce5},
year = {2020},
month = {feb},
publisher = {IOP Publishing},
volume = {5},
number = {2},
pages = {025003},
author = {Perarnau-Llobet, M and Gonz\'{a}lez-Tudela, A and Cirac, J I},
title = {Multimode {F}ock states with large photon number: effective descriptions and applications in quantum metrology},
journal = {Quantum Science and Technology},
}

@article{Brif99,
  title = {Phase-space formulation of quantum mechanics and quantum-state reconstruction for physical systems with {L}ie-group symmetries},
  author = {Brif, C. and Mann, A.},
  journal = {Phys. Rev. A},
  volume = {59},
  issue = {2},
  pages = {971--987},
  numpages = {0},
  year = {1999},
  month = {Feb},
  publisher = {American Physical Society},
  doi = {10.1103/PhysRevA.59.971},
  url = {https://link.aps.org/doi/10.1103/PhysRevA.59.971}
}

@article{Zueco07,
doi = {10.1088/1751-8113/40/17/015},
url = {https://dx.doi.org/10.1088/1751-8113/40/17/015},
year = {2007},
month = {apr},
publisher = {},
volume = {40},
number = {17},
pages = {4635},
author = {Zueco, D and Calvo, I},
title = {Bopp operators and phase-space spin dynamics: application to rotational quantum Brownian motion},
journal = {Journal of Physics A: Mathematical and Theoretical},
}

@article{Stratonovich56,
    title = {On Distributions in Representation Space},
    author = {R. L. Stratonovich},
    journal = {J. Exptl. Theoret. Phys. (U.S.S.R.)},
    volume = {74},
    number = {9},
    year = {December, 1956},	
    url = {http://jetp.ras.ru/cgi-bin/e/index/e/4/6/p891?a=list}
}

@article{Drummond80_2,
doi = {10.1088/0305-4470/13/7/018},
url = {https://dx.doi.org/10.1088/0305-4470/13/7/018},
year = {1980},
month = {jul},
publisher = {},
volume = {13},
number = {7},
pages = {2353},
author = {P D Drummond and  C W Gardiner},
title = {Generalised {P}-representations in quantum optics},
journal = {Journal of Physics A: Mathematical and General},
}

@article{Chaturvedi77,
doi = {10.1088/0305-4470/10/11/003},
url = {https://dx.doi.org/10.1088/0305-4470/10/11/003},
year = {1977},
month = {nov},
publisher = {},
volume = {10},
number = {11},
pages = {L187},
author = {S Chaturvedi and  P Drummond and  D F Walls},
title = {Two photon absorption with coherent and partially coherent driving fields},
journal = {Journal of Physics A: Mathematical and General},
}

@article{Gilchrist97,
  title = {Positive {P} representation: Application and validity},
  author = {Gilchrist, A. and Gardiner, C. W. and Drummond, P. D.},
  journal = {Phys. Rev. A},
  volume = {55},
  issue = {4},
  pages = {3014--3032},
  numpages = {0},
  year = {1997},
  month = {Apr},
  publisher = {American Physical Society},
  doi = {10.1103/PhysRevA.55.3014},
  url = {https://link.aps.org/doi/10.1103/PhysRevA.55.3014}
}

@article{Smith89,
  title = {Simulations of nonlinear quantum damping using the positive P representation},
  author = {Smith, A. M. and Gardiner, C. W.},
  journal = {Phys. Rev. A},
  volume = {39},
  issue = {7},
  pages = {3511--3524},
  numpages = {0},
  year = {1989},
  month = {Apr},
  publisher = {American Physical Society},
  doi = {10.1103/PhysRevA.39.3511},
  url = {https://link.aps.org/doi/10.1103/PhysRevA.39.3511}
}

@article{Schack91,
  title = {Positive P representation},
  author = {Schack, R\"udiger and Schenzle, Axel},
  journal = {Phys. Rev. A},
  volume = {44},
  issue = {1},
  pages = {682--687},
  numpages = {0},
  year = {1991},
  month = {Jul},
  publisher = {American Physical Society},
  doi = {10.1103/PhysRevA.44.682},
  url = {https://link.aps.org/doi/10.1103/PhysRevA.44.682}
}

@article{Deuar02,
   title={Gauge {P} representations for quantum-dynamical problems: Removal of boundary terms},
   volume={66},
   ISSN={1094-1622},
   doi={10.1103/physreva.66.033812},
   number={3},
   journal={Physical Review A},
   publisher={American Physical Society (APS)},
   author={Deuar, P. and Drummond, P. D.},
   year={2002},
   month=sep }

@book{Horn85, place={Cambridge}, title={Matrix Analysis}, publisher={Cambridge University Press}, author={Horn, Roger A. and Johnson, Charles R.}, year={1985}}

@article{Drummond99,
  title = {Quantum dynamics of evaporatively cooled {B}ose-{E}instein condensates},
  author = {Drummond, P. D. and Corney, J. F.},
  journal = {Phys. Rev. A},
  volume = {60},
  issue = {4},
  pages = {R2661--R2664},
  numpages = {0},
  year = {1999},
  month = {Oct},
  publisher = {American Physical Society},
  doi = {10.1103/PhysRevA.60.R2661},
  url = {https://link.aps.org/doi/10.1103/PhysRevA.60.R2661}
}

@article{Kheruntsyan05,
  title = {Einstein-{P}odolsky-{R}osen Correlations via Dissociation of a Molecular {B}ose-{E}instein Condensate},
  author = {Kheruntsyan, K. V. and Olsen, M. K. and Drummond, P. D.},
  journal = {Phys. Rev. Lett.},
  volume = {95},
  issue = {15},
  pages = {150405},
  numpages = {4},
  year = {2005},
  month = {Oct},
  publisher = {American Physical Society},
  doi = {10.1103/PhysRevLett.95.150405},
  url = {https://link.aps.org/doi/10.1103/PhysRevLett.95.150405}
}

@article{Rubies-Bigorda23,
  title = {Characterizing superradiant dynamics in atomic arrays via a cumulant expansion approach},
  author = {Rubies-Bigorda, Oriol and Ostermann, Stefan and Yelin, Susanne F.},
  journal = {Phys. Rev. Res.},
  volume = {5},
  issue = {1},
  pages = {013091},
  numpages = {12},
  year = {2023},
  month = {Feb},
  publisher = {American Physical Society},
  doi = {10.1103/PhysRevResearch.5.013091},
  url = {https://link.aps.org/doi/10.1103/PhysRevResearch.5.013091}
}

@article{Shankar21,
  title = {Subradiant-to-Subradiant Phase Transition in the Bad Cavity Laser},
  author = {Shankar, Athreya and Reilly, Jarrod T. and J\"ager, Simon B. and Holland, Murray J.},
  journal = {Phys. Rev. Lett.},
  volume = {127},
  issue = {7},
  pages = {073603},
  numpages = {6},
  year = {2021},
  month = {Aug},
  publisher = {American Physical Society},
  doi = {10.1103/PhysRevLett.127.073603},
  url = {https://link.aps.org/doi/10.1103/PhysRevLett.127.073603}
}

@article{Glicenstein22,
author = {Antoine Glicenstein and Giovanni Ferioli and Antoine Browaeys and Igor Ferrier-Barbut},
journal = {Opt. Lett.},
number = {6},
pages = {1541--1544},
publisher = {Optica Publishing Group},
title = {From superradiance to subradiance: exploring the many-body Dicke ladder},
volume = {47},
month = {Mar},
year = {2022},
url = {https://opg.optica.org/ol/abstract.cfm?URI=ol-47-6-1541},
doi = {10.1364/OL.451903},
}

@Article{Bohnet2012,
author={Bohnet, Justin G.
and Chen, Zilong
and Weiner, Joshua M.
and Meiser, Dominic
and Holland, Murray J.
and Thompson, James K.},
title={A steady-state superradiant laser with less than one intracavity photon},
journal={Nature},
year={2012},
month={Apr},
day={01},
volume={484},
number={7392},
pages={78-81},
issn={1476-4687},
doi={10.1038/nature10920},
url={https://doi.org/10.1038/nature10920}
}

@article{Arecchi72,
  title = {Atomic Coherent States in Quantum Optics},
  author = {Arecchi, F. T. and Courtens, Eric and Gilmore, Robert and Thomas, Harry},
  journal = {Phys. Rev. A},
  volume = {6},
  issue = {6},
  pages = {2211--2237},
  numpages = {0},
  year = {1972},
  month = {Dec},
  publisher = {American Physical Society},
  doi = {10.1103/PhysRevA.6.2211},
  url = {https://link.aps.org/doi/10.1103/PhysRevA.6.2211}
}

@Article{Bromley2016,
author={Bromley, S. L.
and Zhu, B.
and Bishof, M.
and Zhang, X.
and Bothwell, T.
and Schachenmayer, J.
and Nicholson, T. L.
and Kaiser, R.
and Yelin, S. F.
and Lukin, M. D.
and Rey, A. M.
and Ye, J.},
title={Collective atomic scattering and motional effects in a dense coherent medium},
journal={Nature Communications},
year={2016},
month={Mar},
day={17},
volume={7},
number={1},
pages={11039},
issn={2041-1723},
doi={10.1038/ncomms11039},
url={https://doi.org/10.1038/ncomms11039}
}

@article{Chang04,
  title = {Controlling dipole-dipole frequency shifts in a lattice-based optical atomic clock},
  author = {Chang, D. E. and Ye, Jun and Lukin, M. D.},
  journal = {Phys. Rev. A},
  volume = {69},
  issue = {2},
  pages = {023810},
  numpages = {10},
  year = {2004},
  month = {Feb},
  publisher = {American Physical Society},
  doi = {10.1103/PhysRevA.69.023810},
  url = {https://link.aps.org/doi/10.1103/PhysRevA.69.023810}
}

@article{Glicenstein20,
  title = {Collective Shift in Resonant Light Scattering by a One-Dimensional Atomic Chain},
  author = {Glicenstein, Antoine and Ferioli, Giovanni and \ifmmode \check{S}\else \v{S}\fi{}ibali\ifmmode \acute{c}\else \'{c}\fi{}, Nikola and Brossard, Ludovic and Ferrier-Barbut, Igor and Browaeys, Antoine},
  journal = {Phys. Rev. Lett.},
  volume = {124},
  issue = {25},
  pages = {253602},
  numpages = {6},
  year = {2020},
  month = {Jun},
  publisher = {American Physical Society},
  doi = {10.1103/PhysRevLett.124.253602},
  url = {https://link.aps.org/doi/10.1103/PhysRevLett.124.253602}
}

@article{Guerin16,
  title = {Subradiance in a Large Cloud of Cold Atoms},
  author = {Guerin, William and Ara\'ujo, Michelle O. and Kaiser, Robin},
  journal = {Phys. Rev. Lett.},
  volume = {116},
  issue = {8},
  pages = {083601},
  numpages = {5},
  year = {2016},
  month = {Feb},
  publisher = {American Physical Society},
  doi = {10.1103/PhysRevLett.116.083601},
  url = {https://link.aps.org/doi/10.1103/PhysRevLett.116.083601}
}

@article{Rubies23-2,
  title = {Dynamic population of multiexcitation subradiant states in incoherently excited atomic arrays},
  author = {Rubies-Bigorda, Oriol and Ostermann, Stefan and Yelin, Susanne F.},
  journal = {Phys. Rev. A},
  volume = {107},
  issue = {5},
  pages = {L051701},
  numpages = {6},
  year = {2023},
  month = {May},
  publisher = {American Physical Society},
  doi = {10.1103/PhysRevA.107.L051701},
  url = {https://link.aps.org/doi/10.1103/PhysRevA.107.L051701}
}

@article{Ferioli21,
  title = {Storage and Release of Subradiant Excitations in a Dense Atomic Cloud},
  author = {Ferioli, Giovanni and Glicenstein, Antoine and Henriet, Loic and Ferrier-Barbut, Igor and Browaeys, Antoine},
  journal = {Phys. Rev. X},
  volume = {11},
  issue = {2},
  pages = {021031},
  numpages = {12},
  year = {2021},
  month = {May},
  publisher = {American Physical Society},
  doi = {10.1103/PhysRevX.11.021031},
  url = {https://link.aps.org/doi/10.1103/PhysRevX.11.021031}
}

@article{Toth10,
doi = {10.1088/1367-2630/12/5/053007},
url = {https://dx.doi.org/10.1088/1367-2630/12/5/053007},
year = {2010},
month = {may},
publisher = {},
volume = {12},
number = {5},
pages = {053007},
author = {T\'oth, G\'eza and W Mitchell, Morgan},
title = {Generation of macroscopic singlet states in atomic ensembles},
journal = {New Journal of Physics},
}

@article{Endres16,
author = {Manuel Endres  and Hannes Bernien  and Alexander Keesling  and Harry Levine  and Eric R. Anschuetz  and Alexandre Krajenbrink  and Crystal Senko  and Vladan Vuleti\'{c}  and Markus Greiner  and Mikhail D. Lukin },
title = {Atom-by-atom assembly of defect-free one-dimensional cold atom arrays},
journal = {Science},
volume = {354},
number = {6315},
pages = {1024-1027},
year = {2016},
doi = {10.1126/science.aah3752},
URL = {https://www.science.org/doi/abs/10.1126/science.aah3752},
eprint = {https://www.science.org/doi/pdf/10.1126/science.aah3752},}

@article{Barredo16,
author = { Daniel Barredo and Sylvain de L{\'e}s{\'e}leuc and Vincent Lienhard and Thierry Lahaye and Antoine Browaeys},
title = {An atom-by-atom assembler of defect-free arbitrary two-dimensional atomic arrays},
journal = {Science},
volume = {354},
number = {6315},
pages = {1021-1023},
year = {2016},
doi = {10.1126/science.aah3778},
URL = {https://www.science.org/doi/abs/10.1126/science.aah3778},
eprint = {https://www.science.org/doi/pdf/10.1126/science.aah3778},}

@Article{Kim16,
author={Kim, Hyosub
and Lee, Woojun
and Lee, Han-gyeol
and Jo, Hanlae
and Song, Yunheung
and Ahn, Jaewook},
title={In situ single-atom array synthesis using dynamic holographic optical tweezers},
journal={Nature Communications},
year={2016},
month={Oct},
day={31},
volume={7},
number={1},
pages={13317},
issn={2041-1723},
doi={10.1038/ncomms13317},
url={https://doi.org/10.1038/ncomms13317}
}

@Article{Kaufman21,
author={Kaufman, Adam M.
and Ni, Kang-Kuen},
title={Quantum science with optical tweezer arrays of ultracold atoms and molecules},
journal={Nature Physics},
year={2021},
month={Dec},
day={01},
volume={17},
number={12},
pages={1324-1333},
issn={1745-2481},
doi={10.1038/s41567-021-01357-2},
url={https://doi.org/10.1038/s41567-021-01357-2}
}

@Article{Kumar18,
author={Kumar, Aishwarya
and Wu, Tsung-Yao
and Giraldo, Felipe
and Weiss, David S.},
title={Sorting ultracold atoms in a three-dimensional optical lattice in a realization of {M}axwell's demon},
journal={Nature},
year={2018},
month={Sep},
day={01},
volume={561},
number={7721},
pages={83-87},
issn={1476-4687},
doi={10.1038/s41586-018-0458-7},
url={https://doi.org/10.1038/s41586-018-0458-7}
}

@Article{Bornet23,
author={Bornet, Guillaume
and Emperauger, Gabriel
and Chen, Cheng
and Ye, Bingtian
and Block, Maxwell
and Bintz, Marcus
and Boyd, Jamie A.
and Barredo, Daniel
and Comparin, Tommaso
and Mezzacapo, Fabio
and Roscilde, Tommaso
and Lahaye, Thierry
and Yao, Norman Y.
and Browaeys, Antoine},
title={Scalable spin squeezing in a dipolar {R}ydberg atom array},
journal={Nature},
year={2023},
month={Sep},
day={01},
volume={621},
number={7980},
pages={728-733},
issn={1476-4687},
doi={10.1038/s41586-023-06414-9},
url={https://doi.org/10.1038/s41586-023-06414-9}
}

@Article{Eckner23,
author={Eckner, William J.
and Darkwah Oppong, Nelson
and Cao, Alec
and Young, Aaron W.
and Milner, William R.
and Robinson, John M.
and Ye, Jun
and Kaufman, Adam M.},
title={Realizing spin squeezing with {R}ydberg interactions in an optical clock},
journal={Nature},
year={2023},
month={Sep},
day={01},
volume={621},
number={7980},
pages={734-739},
doi={10.1038/s41586-023-06360-6},
url={https://doi.org/10.1038/s41586-023-06360-6}
}

@article{Bakr10,
author = {W. S. Bakr  and A. Peng  and M. E. Tai  and R. Ma  and J. Simon  and J. I. Gillen  and S. F\"{o}lling  and L. Pollet  and M. Greiner },
title = {Probing the Superfluid-to-{M}ott Insulator Transition at the Single-Atom Level},
journal = {Science},
volume = {329},
number = {5991},
pages = {547-550},
year = {2010},
doi = {10.1126/science.1192368},
URL = {https://www.science.org/doi/abs/10.1126/science.1192368},
eprint = {https://www.science.org/doi/pdf/10.1126/science.1192368},}

@article{Hutson24,
author = {Ross B. Hutson  and William R. Milner  and Lingfeng Yan  and Jun Ye  and Christian Sanner },
title = {Observation of millihertz-level cooperative {L}amb shifts in an optical atomic clock},
journal = {Science},
volume = {383},
number = {6681},
pages = {384-387},
year = {2024},
doi = {10.1126/science.adh4477},
URL = {https://www.science.org/doi/abs/10.1126/science.adh4477},
eprint = {https://www.science.org/doi/pdf/10.1126/science.adh4477},}

@article{Safronova18,
  title = {Search for new physics with atoms and molecules},
  author = {Safronova, M. S. and Budker, D. and DeMille, D. and Kimball, Derek F. Jackson and Derevianko, A. and Clark, Charles W.},
  journal = {Rev. Mod. Phys.},
  volume = {90},
  issue = {2},
  pages = {025008},
  numpages = {106},
  year = {2018},
  month = {Jun},
  publisher = {American Physical Society},
  doi = {10.1103/RevModPhys.90.025008},
  url = {https://link.aps.org/doi/10.1103/RevModPhys.90.025008}
}

@article{Norcia19,
author = {Matthew A. Norcia  and Aaron W. Young  and William J. Eckner  and Eric Oelker  and Jun Ye  and Adam M. Kaufman },
title = {Seconds-scale coherence on an optical clock transition in a tweezer array},
journal = {Science},
volume = {366},
number = {6461},
pages = {93-97},
year = {2019},
doi = {10.1126/science.aay0644},
URL = {https://www.science.org/doi/abs/10.1126/science.aay0644},
eprint = {https://www.science.org/doi/pdf/10.1126/science.aay0644},}

@Article{Bothwell22,
author={Bothwell, Tobias
and Kennedy, Colin J.
and Aeppli, Alexander
and Kedar, Dhruv
and Robinson, John M.
and Oelker, Eric
and Staron, Alexander
and Ye, Jun},
title={Resolving the gravitational redshift across a millimetre-scale atomic sample},
journal={Nature},
year={2022},
month={Feb},
day={01},
volume={602},
number={7897},
pages={420-424},
issn={1476-4687},
doi={10.1038/s41586-021-04349-7},
url={https://doi.org/10.1038/s41586-021-04349-7}
}

@book{Sudarshan2016, place={Hackensack, NJ}, title={Classical dynamics: A modern perspective}, publisher={World Scientific Publishing Co. Pte. Ltd}, author={Sudarshan, E. C. G. and Mukunda, N.}, year={2016}}

@article{Semeghini21,
author = {G. Semeghini  and H. Levine  and A. Keesling  and S. Ebadi  and T. T. Wang  and D. Bluvstein  and R. Verresen  and H. Pichler  and M. Kalinowski  and R. Samajdar  and A. Omran  and S. Sachdev  and A. Vishwanath  and M. Greiner  and V. Vuleti\'{c}  and M. D. Lukin },
title = {Probing topological spin liquids on a programmable quantum simulator},
journal = {Science},
volume = {374},
number = {6572},
pages = {1242-1247},
year = {2021},
doi = {10.1126/science.abi8794},
URL = {https://www.science.org/doi/abs/10.1126/science.abi8794},
eprint = {https://www.science.org/doi/pdf/10.1126/science.abi8794},
}

@Article{Bernien17,
author={Bernien, Hannes
and Schwartz, Sylvain
and Keesling, Alexander
and Levine, Harry
and Omran, Ahmed
and Pichler, Hannes
and Choi, Soonwon
and Zibrov, Alexander S.
and Endres, Manuel
and Greiner, Markus
and Vuleti{\'{c}}, Vladan
and Lukin, Mikhail D.},
title={Probing many-body dynamics on a 51-atom quantum simulator},
journal={Nature},
year={2017},
month={Nov},
day={01},
volume={551},
number={7682},
pages={579-584},
issn={1476-4687},
doi={10.1038/nature24622},
url={https://doi.org/10.1038/nature24622}
}

@misc{Hosseinabadi25,
      title={Making Truncated {W}igner for dissipative spins 'plain easy'}, 
      author={Hossein Hosseinabadi and Oksana Chelpanova and Jamir Marino},
      year={2025},
      eprint={2503.17443},
      archivePrefix={arXiv},
      primaryClass={quant-ph},
      url={https://arxiv.org/abs/2503.17443}, 
}

@misc{Ho24,
      title={Optomechanical self-organization in a mesoscopic atom array}, 
      author={Jacquelyn Ho and Yue-Hui Lu and Tai Xiang and Cosimo C. Rusconi and Stuart J. Masson and Ana Asenjo-Garcia and Zhenjie Yan and Dan M. Stamper-Kurn},
      year={2024},
      eprint={2410.12754},
      archivePrefix={arXiv},
      primaryClass={quant-ph},
      url={https://arxiv.org/abs/2410.12754}, 
}

@article{Black03,
  title = {Observation of Collective Friction Forces due to Spatial Self-Organization of Atoms: From {R}ayleigh to {B}ragg Scattering},
  author = {Black, Adam T. and Chan, Hilton W. and Vuleti\ifmmode \acute{c}\else \'{c}\fi{}, Vladan},
  journal = {Phys. Rev. Lett.},
  volume = {91},
  issue = {20},
  pages = {203001},
  numpages = {4},
  year = {2003},
  month = {Nov},
  publisher = {American Physical Society},
  doi = {10.1103/PhysRevLett.91.203001},
  url = {https://link.aps.org/doi/10.1103/PhysRevLett.91.203001}
}

@article{Rubies23,
  title = {Characterizing superradiant dynamics in atomic arrays via a cumulant expansion approach},
  author = {Rubies-Bigorda, Oriol and Ostermann, Stefan and Yelin, Susanne F.},
  journal = {Phys. Rev. Res.},
  volume = {5},
  issue = {1},
  pages = {013091},
  numpages = {12},
  year = {2023},
  month = {Feb},
  publisher = {American Physical Society},
  doi = {10.1103/PhysRevResearch.5.013091},
  url = {https://link.aps.org/doi/10.1103/PhysRevResearch.5.013091}
}

@article{Rubies22,
  title = {Superradiance and subradiance in inverted atomic arrays},
  author = {Rubies-Bigorda, Oriol and Yelin, Susanne F.},
  journal = {Phys. Rev. A},
  volume = {106},
  issue = {5},
  pages = {053717},
  numpages = {12},
  year = {2022},
  month = {Nov},
  publisher = {American Physical Society},
  doi = {10.1103/PhysRevA.106.053717},
  url = {https://link.aps.org/doi/10.1103/PhysRevA.106.053717}
}

@article{Robicheaux21-2,
  title = {Theoretical study of early-time superradiance for atom clouds and arrays},
  author = {Robicheaux, F.},
  journal = {Phys. Rev. A},
  volume = {104},
  issue = {6},
  pages = {063706},
  numpages = {10},
  year = {2021},
  month = {Dec},
  publisher = {American Physical Society},
  doi = {10.1103/PhysRevA.104.063706},
  url = {https://link.aps.org/doi/10.1103/PhysRevA.104.063706}
}

@article{Carmichael00,
title = {A quantum trajectory unraveling of the superradiance master equation},
journal = {Optics Communications},
volume = {179},
number = {1},
pages = {417-427},
year = {2000},
issn = {0030-4018},
doi = {https://doi.org/10.1016/S0030-4018(99)00694-X},
url = {https://www.sciencedirect.com/science/article/pii/S003040189900694X},
author = {H.J. Carmichael and Kisik Kim},
}

@article{Ng13,
  title = {Simulation of the dynamics of many-body quantum spin systems using phase-space techniques},
  author = {Ng, Ray and S\o{}rensen, Erik S. and Deuar, Piotr},
  journal = {Phys. Rev. B},
  volume = {88},
  issue = {14},
  pages = {144304},
  numpages = {14},
  year = {2013},
  month = {Oct},
  publisher = {American Physical Society},
  doi = {10.1103/PhysRevB.88.144304},
  url = {https://link.aps.org/doi/10.1103/PhysRevB.88.144304}
}
\end{document}